\documentclass[12pt]{article}

\usepackage{color,amsmath,amssymb,amsfonts,graphicx}
\usepackage{varioref}
\newtheorem{theorem}{Theorem}[section]

\newtheorem{remark}[theorem]{Remark}

\newtheorem{proposition}[theorem]{Proposition}
\newtheorem{corollary}[theorem]{Corollary}
\newtheorem{conjecture}[theorem]{Conjecture}

\definecolor{refkey}{rgb}{1,0.5,0.5}
\definecolor{labelkey}{rgb}{0.5,1,0.5}
\include{abbrevs}
 \numberwithin{equation}{section}
\begin{document}

\title{ A saddle in a corner - a model of collinear triatomic chemical reactions}
\author{L. Lerman$^1$, V. Rom-Kedar$^2$ \\
\normalsize $^1$ Dept. of Diff. Equat. \& Math. Analysis,\\
\normalsize The University of Nizhny Novgorod, Russia,\\
\normalsize $^2$ The Estrin Family Chair of Computer Science and Applied Mathematics, \\ \normalsize The Weizmann Institute of Science, Rehovot, Israel
}
\date{\today}

\maketitle

\begin{abstract}
    A geometrical model which captures  the  main ingredients governing
    atom-diatom collinear chemical reactions is proposed. This model is neither
    near-integrable nor hyperbolic, yet it is amenable to analysis using
    a combination of the recently developed tools for studying systems with steep
    potentials and the study of the phase space structure near a center-saddle equilibrium. The nontrivial
    dependence of the reaction rates on parameters, initial conditions and  energy is thus qualitatively explained.
Conditions under which the phase space transition state theory assumptions are satisfied
and conditions under which these fail are derived.    \end{abstract}

%\tableofcontents

\section{Introduction}

The study of classical, semi-classical and quantum chemical reactions on a molecular level has a rich history \cite{levinbook,tannor,henhans,smithPhysRev69,polak0,polak1,polak,davis,bugas94}.  In these models, the  full Hamiltonian is averaged over the fast motion of the
 electrons, where each electron is assumed to be fixed at a specific quantum energy level (the adiabatic approximation)\cite{levinbook,tannor,henhans}. Such computations produce effective potential energy surfaces (PES) that govern the slow motion of the nuclei.  The resulting Hamiltonians correspond to the {\em ``Born-Oppenheimer"} approximations. Quasi-classical computations\footnote{In these computations the atomic motion is found from the classical dynamics dictated by the PES. The quantization enters through choosing initial quantised ensembles (usually only in vibrational and rotational energies) and through binning of the product states to quantized energies.}  that employ these  Hamiltonians provide surprisingly good approximations to the quantum calculations, hence classical models are extensively studied by chemists, see \cite{levinbook,tannor,henhans} and references therein.

  Dishearteningly, the resulting nuclei motion of even the most basic,  classical ``Elementary" (bimolecular) reactions, that are ``at the heart of chemistry" \cite{henhans}, is not well understood. Numerical simulations of  the nuclei dynamics exhibit sensitive dependence of the trajectories on initial conditions and parameters. Moreover, one finds that  macroscopic observables,  like reaction rates,  have an intricate dependence on the parameters and energy, see e.g.   \cite{levinbook,tannor,henhans,polak0,davis}.

In contrast, the popular transition state theory provides an appealing  intuitive view of this motion and leads to explicit formulae relating the microscopic nuclei kinetics  to the macroscopic  reaction rates. This theory assumes  that the PES is separable to a one dimensional potential (``along the reaction path") and  a potential well  in all the other modes of motion (the bath of oscillators). Hence, the reaction   according to this theory is  described by a product of a one degree of freedom (thus integrable) system and oscillators. This appealing phenomenological model  is  too simple: it does not describe the observed  complex dependence of the nuclei motion on initial conditions and energy  \cite{levinbook,tannor,henhans,smithPhysRev69,polak0,polak1,polak,davis,bugas94}.

The next level of approximation, by which the non oscillatory  part of the potential is two-dimensional, is the subject of our paper.   Here, the translational and vibrational energies of the atom and diatom   are coupled to each other yet are separable from all the other modes of motion (e.g. of those associated with  the bending  and the rotational energies). Such a decoupling occurs, for example, when   the initial configuration is collinear  and the angular momentum is zero \cite{smithPhysRev69,polak0}. This separability assumption is widely used in theoretical investigations of chemical reactions  \cite{levinbook,tannor,henhans,smithPhysRev69,polak0,polak1,polak,davis,bugas94}. Indeed, the first examinations of the principles  underlying the transition state theory were conducted by investigating such models  \cite{smithPhysRev69,polak0}.

Employing such a restrictive reduction\footnote{By which the bath of oscillators is considered to be separable from the reaction dynamics.} still leaves us with a two degree of freedom system which may admit chaotic behavior (in contrast with the transition state theory reduction which leaves us with integrable dynamics). One source for such complicated behavior is associated with the existence of a reaction barrier - a saddle point of the PES\footnote{Some reactions have  potentials that also admit stable triatomic states (indirect reactions), many reactions have a single steady unstable triatomic configuration (direct reactions), and in some exceptional cases there are no steady triatomic configurations at all \cite{polak0,henhans}. Most recently, models with rank-k saddles  have been considered as well  \cite{HaUzPY11,HUPYC10}.
}.      The existence of such a barrier is the main ingredient needed for relating the classical  transition state theory to the analysis of the phase space structure of  two degrees of freedom systems  \cite{polak0,ujpyw02,Davi85,DaSk92,TGKrev96,KoBe01}. Normal form analysis of the local phase-space structure near the PES bottlenecks (local unstable extremal points of the PES) relates these two approaches and leads to  accurate calculations of the minimal flux through them \cite{ujpyw02,HaUzPY11,HUPYC10}.
The local picture near the bottlenecks does not reveal the complexity of the motion. This complexity is revealed only when the global     structure of  the reaction tubes - the phase space regions that pass through the bottlenecks - is  calculated  \cite{polak0,polak1,ujpyw02,Davi85,DaSk92,WaBuWi05,WaWi10,IPPS11}.

Previous works on the global features of these tubes have either employed near-integrable techniques  \cite{davis,DaGr86,DaSk92,IPPS07} or   numerical integrations  \cite{polak0,polak1,ujpyw02,Davi85,DaSk92,WaBuWi05,WaWi10,IPPS11}. In these works it is demonstrated that the structure of the stable and unstable manifolds of unstable periodic orbits (the Periodic Orbits Dividing Surfaces PODS \cite{polak0}  or, similarly, the stable and unstable manifolds of  the Normally Hyperbolic Invariant Manifolds NHIM \cite{ujpyw02,WaWi10}) determines the reaction rates.  In particular, when these asymptotic surfaces intersect each other the    reaction rates depend on both the intersection pattern  and the  local flux near the saddles. In such cases the predictions of  the transition state theory (and its   local minimal flux variants) must be modified. These insights are employed to study numerically the asymptotic surfaces to  NHIMs of high dimensional  problems \cite{ujpyw02,WaBuWi05}.

Here we introduce a new geometrical model for the two dimensional PES. The PES in the reaction region is a approximated by a sum of a quadratic form with a saddle point and a smooth potential which is close to zero within a corner region and increases sharply at the corners' boundaries (see Figs \ref{fig:potcon}b,\ref{fig:config}). We analyze this model by combining the  theory of smooth Hamiltonians with impacts
\cite{KZbook,zhar00,GoNei08,GS}, the recent  generalization of   \cite{TuRK98,RaRkTu07} to the impact case \cite{KlRk11}, and the theory of homoclinic loops to a saddle-center  \cite{Conley,Lerman,LerKol}. The analysis provides qualitative  understanding regarding the global structure of the dividing surfaces in  reactions with one rank-1 saddle point. Indeed, we identify a basic mechanism which explains why and when the dividing surfaces of the periodic orbits that have energies slightly above the barrier energy are especially complicated or especially simple (namely do not intersect each other). Moreover, we provide a mechanism for the emergence of stable triatomic cyclic motion. The applicability of these geometrical insights to more accurate models of the PES and to higher dimensional settings (e.g. when the two degrees of freedom dynamics are weakly coupled to a bath of oscillators) is under current study.

The paper is ordered as follows.  In section 2 we recall that mass-scaled  Jacobi coordinates bring the collinear triatomic reaction's Hamiltonian to a two-degree of freedom Hamiltonian in the standard mechanical form. In section 3 we introduce the geometrical potential function and in section 4 we provide some basic observations regarding its properties and its relation to other potentials. In sections \ref{sec:theta0} we analyze the dynamics in this model under specific geometrical conditions, proving that both chaotic  and stable periodic triatomic motions emerge. In section 6 we show that  complicated dynamics appear at some parameter ranges and simple dynamics at others. In section 7 we discuss the relation of these results to  transition state theory and propose some possible extensions.
\section{Collinear atom-diatom reactions}
 Some of the geometrical characteristics of the PES describing collinear triatomic reactions may be inferred from  general considerations that are common to all such reactions. These characteristics, as described next, motivate our construction of the simplified geometrical model for the PES.

\begin{figure}[t]
\begin{center}
(a)\includegraphics[width=0.45\textwidth]{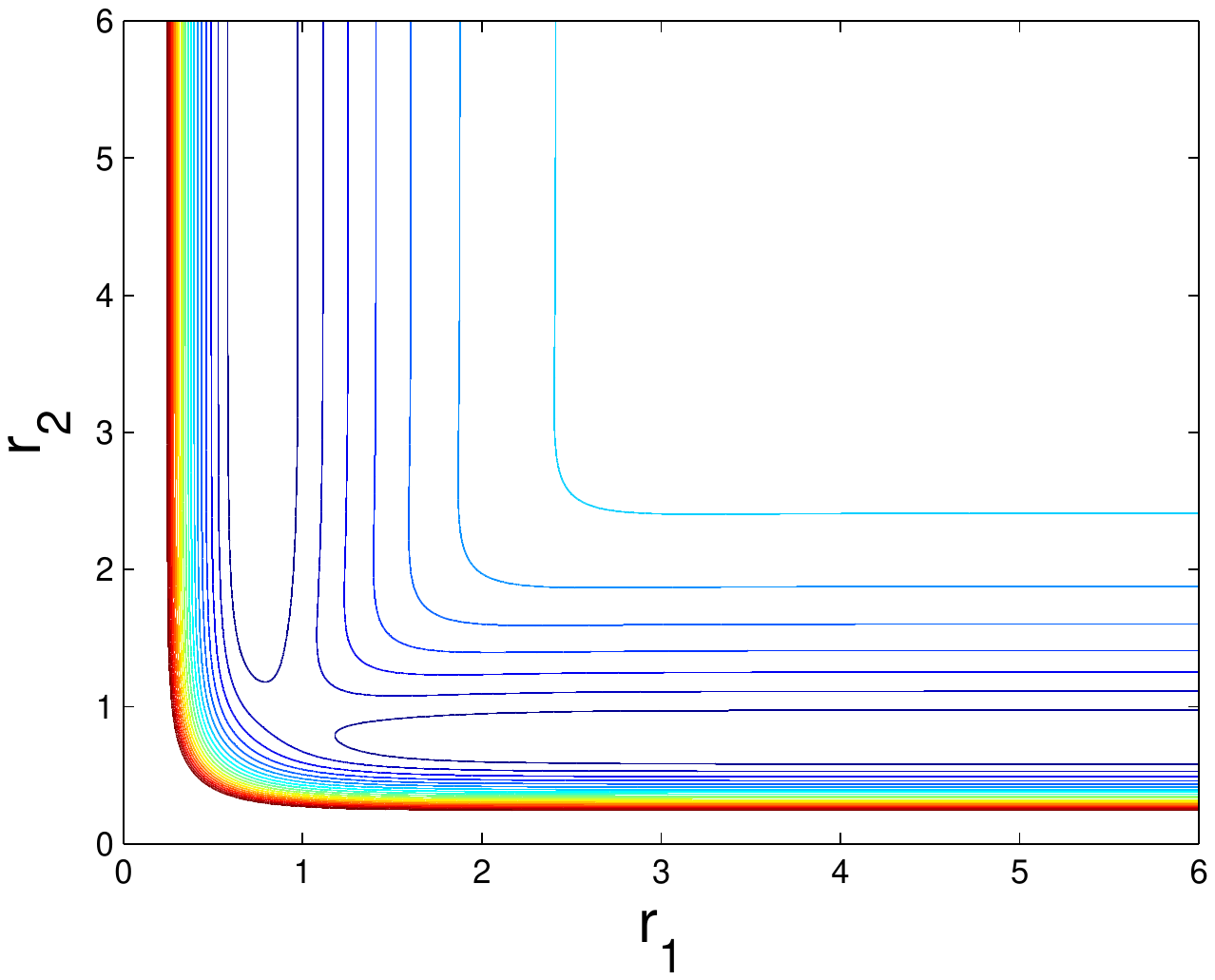}
(b)\includegraphics[width=0.45\textwidth]{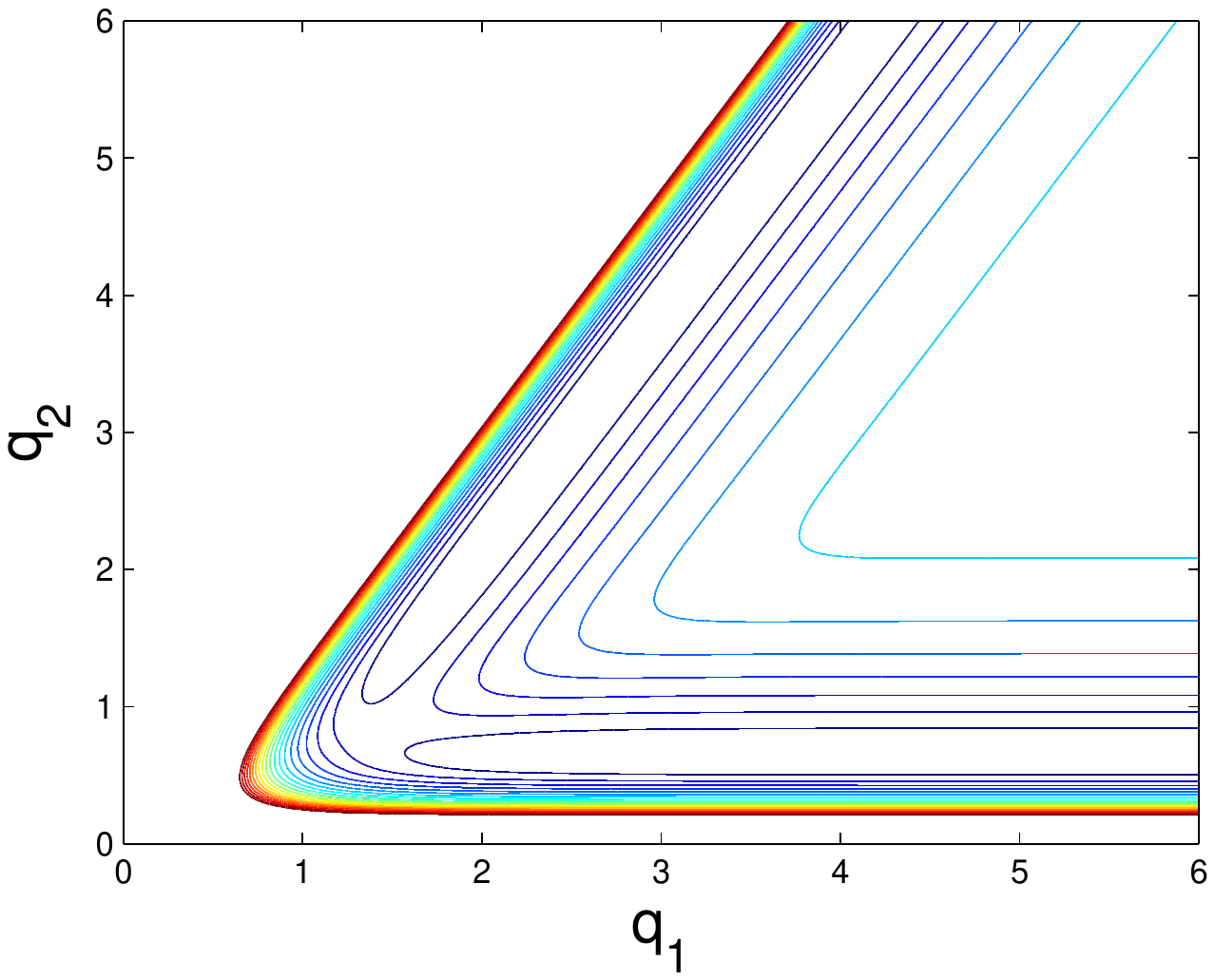}
\end{center}
\caption{\label{fig:potcon}
%\begin{footnotesize}
Contour lines of the effective potential for the \(H_2+H\)
reaction in the  (a) relative positions coordinates \(V(r_1,r_2)\)   (b)   mass scaled coordinates \(V_r(q_1,q_2)\). The  allowed region of motion   lies within  the dense contour   lines (red lines) that correspond to the strong diatomic repulsion at small distances. In the \((r_1,r_2)\) plane this region is essentially the positive quadrant whereas in  the  \((q_1,q_2)\) plane it is confined to a \(\beta\)-wedge/corner: a two-dimensional wedge with a corner angle \(\beta=60^{o}\).  The appearance of a single saddle point in the corner region is apparent. The plotted potential is of the LEPS  form, see   \cite{levinbook,tannor,henhans} and references therein.  Notably, the  \(\beta\)-wedge  feature always appears in the mass scaled coordinates, independent of the exact form of the potential.
%\end{footnotesize}
}
\end{figure}
Consider the triatomic  reaction $A+BC\rightarrow AB+C.$ Namely, here we always consider the atom-diatom case. We denote this reaction by the standard shorthand notation  \(AB+C\). An
effective Hamiltonian for the molecular interaction is of the
form
\[
H(r,p)=\sum_{i\in\{A,B,C\}}\frac{1}{M_{i}}\frac{p_{i}^{2}}{2}+V(r)
\]
where $r=(r_{A},r_{B},r_{C})$ denotes the positions of the atoms \(A,B,C \) in a given inertial frame
and $\left\{  M_{i}\right\}  _{i\in\{A,B,C\}}$  denote the masses of the atoms. Under standard conditions $V $
depends only on the relative positions of these atoms. This $9$ d.o.f. system
simplifies when collinearity is assumed \cite{smithPhysRev69,polak0}. This assumption implies that the
relative positions may be expressed in terms of two scalars $r_{1}%
=(r_{A}-r_{B})\cdot\widehat{e},r_{2}=\left(  r_{B}-r_{C}\right)  \cdot
\widehat{e}$ where $\widehat{e}$ is a unit vector aligned with the molecules,
namely $V(r)=V(r_{1},r_{2})$. Moreover, since at small distances the atoms are strongly repelling, \(V\) becomes large along the rays \(r_i=0\), see Fig. \ref{fig:potcon}a. The kinetic energy term in these new coordinates is
non-diagonal and has a mass dependent quadratic form. A mass-weighted coordinate system (the Jacobi coordinates,
see \cite{tannor} for formulation and references) brings the system to the
standard mechanical Hamiltonian form of a unit mass particle moving in the potential
field $V_{r}(q_{1},q_{2})$:
\begin{equation}
H(q_{i},p_{i})=\frac{p_{1}^{2}}{2}+\frac{p_{2}^{2}}{2}+V_{r}(q_{1},q_{2}).
\label{eq:reactham}\end{equation}
Here, \((q_{1},q_2)\) are the reaction coordinates:
\begin{equation}
q_{1}(r_{1},r_{2})=\hat a\  r_{1}+ \hat b \ r_{2}\cos\beta,\quad q_{2}(r_{2})=\hat b \ r_{2}\sin\beta,
\label{eq:reactcoord}%
\end{equation}
the scaling coefficients \(\hat a,\hat b    \) depend only on the mass of the atoms:
\begin{equation}
\hat a \ =\sqrt{\frac{M_{A}(M_{B}+M_{C})}{M_{A}+M_{B}+M_{C}}},\quad
\hat b \ =\sqrt
{\frac{M_{C}(M_{B}+M_{A})}{M_{A}+M_{B}+M_{C}}};
\end{equation}
and the mass dependent ``skew-angle'' \(\beta\) is  defined by:
\begin{equation}
\beta=\arccos\sqrt{\frac{M_{A}M_{C}}{(M_{A}+M_{B})(M_{B}+M_{C})}}.
\label{eq:beta}%
\end{equation}
For the $H_{2}+H$ reaction, $\beta=60^{\circ}$ (see Fig \ref{fig:potcon}b), for heavy-light-heavy
interactions $\beta$ is small (e.g. for $IH+I$ one finds $\beta=7^{\circ}$)
whereas light-heavy-light interactions lead to $\beta\approx90^{\circ}$,
 see \cite{polak,polak0,polak1,tannor} for the corresponding figures and references therein.

\smallskip

%Notice that $q_{2}=const$ and $q_{1}\rightarrow\infty$ corresponds to the %case
%$r_{2}=const$ and $r_{1}\rightarrow\infty$ namely to a $BC$ molecule and %a free $A$ atom. Thus, these coordinates are called the reactant coordinates %(similarly one may define product coordinates \((Q_{1},Q_2)\), see )

 The short range repulsion of the atoms implies that the motion in the configuration space  is confined by the rays \(r_{1}=0,r_{2}>0\) (namely \(q_{2}=q_1 \tan \beta, q_{2}>0\)), and  \(r_{2}=0,r_{1}>0\) (namely \(q_{2}=0,q_1>0\)). This region, a  two-dimensional wedge in the \((q_{1},q_{2})\) plane, is called hereafter the corner region or the  \(\beta\)-wedge, see Fig \ref{fig:potcon}b. The dynamics within the \(\beta\)-wedge depend on the particular form of the potential \(V_{r}(q_{1},q_2)\).
We will be mostly considering   potentials that have  a single extremal point which is a saddle. \bigskip

In the scaled coordinates,  chemical
reactions are represented by trajectories  of the Hamiltonian (\ref{eq:reactham}).
Reactant states ($BC$ molecules and far away \(A\) atoms) correspond to  configurations with a large  $q_{1}$
and bounded $q_{2}$. We thus say that such configurations belong to the reactants channel. Similarly,  product states ($AB$ molecules and  far away \(C\) atoms) correspond to configurations in the products channel with bounded \((q_1 \tan\beta-q_{2})\) and large \(q_{1},q_2\). In the reactants (products) channel the potential is well approximated by a one dimensional BC (AB) diatomic potential.
The corner region, where both \(q_{1}\) and \(q_{2}\) are bounded, and where all the potential saddle points are located, is called the reaction region.
In symmetric cases with a single saddle point the saddle point is located on  the  bisector - the potential symmetry line. In non-symmetric cases, the barrier location is called `early'  (respectively `late') if it is closer to the reactants (respectively products) channel.

\section{The geometrical potential function}

The above description of the geometrical properties of the potential \(V_{r}(q_1,q_2)\) is independent of the details of the reaction. Below, we propose a specific form for  a potential  \(V_{r}(q_{1},q_{2})\)  with parameters that have a transparent geometrical meaning. This potential may be viewed as a local geometrical approximation in the interaction zone to any other potential surface. The advantage of using this new formulation becomes apparent - it allows for    rigorous analysis of the  model and for qualitative understanding of the dynamics.

Consider Hamiltonians with  geometrical potentials  of the form:
\begin{equation}
H(q,p;a,b,c,\varepsilon)=\frac{p^{2}}{2}+aV_{a}(q)+bV_{b}(q,\varepsilon)+cV_{farfield}(q).
\label{eq:hgeometricalgen}
\end{equation}
Here $H_{b}=H(q,p;b,a=c=0)$ is a billiard-like system limiting to a billiard in a \(\beta\)-wedge (see more details below).
$H_{int}=H(q,p;a,$ $b=c=0)$ is an integrable system where the potential \(V_{a}(q)\) has a single saddle point in the corner region (we will soon fix \(V_{a}(q)\) to be a quadratic potential and remark on the effects of higher order terms when applicable). The far-field potential \(cV_{farfield}(q)\) and its derivatives are small in the reaction region (yet are large away from the corner region).  All together, \(aV_{a}(q)\) corresponds to the normal form of the potential near the saddle point, \(bV_b(q,\varepsilon)\) corresponds to the diatomic repulsion terms and  \(cV_{farfield}(q)\) handles  the remainder terms near the saddle and the reactants and products channels away from the corner region.

By billiard-like system we mean that the potential \(V_{b}(q,\varepsilon)\) is a steep potential   \cite{RKTu98}:  the level set of this potential at, say, $V_b=1/2$, limits, as  \(\varepsilon\rightarrow 0\) to some billiard-like domain and $V_b(q,\varepsilon) \rightarrow 0$ for all $q$ in the interior of this domain (see \cite{RaRkTu07} for a precise definition and Eqs. (\ref{eq:exppot}),(\ref{eq:colpot}) for examples).
Here,   the \(\beta\)-wedge is the  billiard domain. For example, in Fig \ref{fig:potcon}b,  the red level curves are identified with the level curves of $V_{b}(q,\varepsilon)$. For small $a$ values (or equivalently, at energy levels that are much larger than \(a\)), the smooth part of the potential at the corner region may be neglected and the motion in the \(\beta-\)wedge is billiard-like  \cite{TuRK98,RaRkTu07}.

For non-negligible\footnote{Yet bounded, so that along the corner boundaries the diatomic repulsion dominates.} \(a\), for sufficiently small $(\varepsilon,c)$,  at a fixed positive energy level, the system
$H_{\varepsilon}=H(q,p;a,b,c,\varepsilon)$ has trajectories that closely follow the smooth integrable dynamics \(H_{int}\) until they reach the walls defined by
$V_{b}(q,\varepsilon)$, reflect according to the billiard law, and
continue with the smooth dynamics. Indeed, in the limit \(\varepsilon\rightarrow0\), the motion is described by a Hamiltonian system with impacts
\cite{KZbook,zhar00,GoNei08,GS}.
Here, we study the dynamics of such impact systems when their smooth integrable part ($H_{int}$) has  invariant hyperbolic
subsets (here, Lyapunov periodic orbits near a saddle-center \cite{Lyap,Conley,Lerman}). We prove that  for a range of parameter values the reflections of  the stable and unstable manifolds of this set from the billiards' boundary  give rise to
complicated homoclinic behavior (Theorem \ref{thm:bifurc}) and to stable, recurrent triatomic states (Theorem \ref{thm:homtangen}). We also show that at a different range of  parameter values the manifolds reflect to infinity via the reactants and products channels and no recurrent motion near the saddle is possible (Theorem \ref{thm:sd1} and Conjecture \ref{thm:sd2}).
For this latter range of parameters transition state theory is expected to be valid.

  Finally, the results  established for the limiting impact flow at \(\epsilon=c=0\) are shown to be valid at small \(\epsilon,c\). The persistence for small   \(c\)  values follows, as usual, from the robust character of these results under smooth perturbations. The persistence for small \(\epsilon\) follows from the  recent extension of    \cite{RaRkTu07} to smooth Hamiltonians that limit to impact systems \cite{KlRk11}.

\subsection{The simplest form of the geometrical model}%

The assumption that the potential has a single saddle point in the corner region implies that the potential $aV_a$ of  (\ref{eq:hgeometricalgen}) is of the form  \begin{equation}
aV_{a}(q)=\frac{1}{2}(q-q_{s})^{T}A(q-q_{s})+O((q-q_{s})^{3})\label{eq:quadpot}
\end{equation}
where $A$ is   a symmetric $2\times 2 $ matrix ($A^T = A$ ) with eigenvalues  (\(\omega^2,-\lambda^2).\)   Additionally, since the potential saddle point  $q_s=(q_{1,s},q_{2,s}) $ is assumed to be located within the corner region, it satisfies $q_{1,s}>0,q_{2,s}
 < q_{1,s}\tan\beta$.  Consequently (see section \ref{sec:phasspace}), the Hamiltonian flow has a saddle-center equilibrium at $P=(q,p)=(q_s,0)$.

Replacing  \(aV_{a}(q)\)  in Eq. (\ref{eq:hgeometricalgen}) by its quadratic approximation  we obtain the simplest form of the geometrical model:\begin{equation}
H(q,p;b,c,\varepsilon)=\frac{p^{2}}{2}+\frac{1}{2}(q-q_{s})^{T}A(q-q_{s})+
bV_{b}(q;\varepsilon)+cV_{farfield}(q).%
\label{eq:linbilham}
\end{equation}
Recall that the \(bV_b\) term corresponds to a billiard-like potential in the corner, namely it satisfies conditions I-IV of \cite{RaRkTu07}. We may, for example, consider here either an exponential potential (as in the repulsion associated with a Morse potential)  or a power-law potential (as in the   Pauli repulsion term of a Lennard-Jones interaction):
\begin{equation}
bV_b(q,\varepsilon)=bV_{exp}(q,\varepsilon)=b\exp(-q_{2}/\varepsilon)+b\exp((q_{2}-q_{1}\tan\beta )/\varepsilon)
\label{eq:exppot}
\end{equation}
or%
\begin{equation}
bV_b(q,\varepsilon)=bV_{pow}(q,\varepsilon/b)=\frac{\varepsilon}{q_{2}^k}+\frac{\varepsilon}{(q_{1}\tan\beta-q_{2}%
)^k },\ \  k\in\mathbb{Z}^+.
\label{eq:colpot}
\end{equation}

We study Eq. (\ref{eq:linbilham})  in the small  \(c\) regime, namely, we assume that the far-field term is small. We expect this approximation to be valid only in the reaction region. We thus study the dynamics in a  bounded corner region of the configuration space.
The simplest model provides adequate approximation to  (\ref{eq:hgeometricalgen}) in this bounded region if  the far-field potential and all its derivatives are small there and, additionally, \(aV_{a}(q)\) is well approximated by its quadratic approximation. This latter part of the assumption may be relaxed in the future by including higher order terms of the normal form near the saddle-center point of the Hamiltonian.

The motion for $a,b\ne0$  is found by analyzing
 first the singular impact limit (i.e. when $\varepsilon \rightarrow +0$), hereafter called the limit system.
 Then,  the recent  persistence results  of \cite{KlRk11} (generalizing    \cite{TuRK98,RaRkTu07}) and standard perturbation theory allow us to establish that similar
 behavior persists for sufficiently small $\varepsilon$ and \(c\).

\section{The phase-space structure of the limit system and its dependence on parameters\label{sec:phasspace}}

In the limit $\varepsilon \rightarrow +0$ the behavior of (\ref{eq:linbilham}) in the interior of the corner region is
governed by the linear integrable system corresponding to the quadratic  Hamiltonian. The limit system   is defined as this smooth integrable motion inside the corner together with  reflections from the  corners' boundaries.

In section  \ref{subsec:integr} we  recall the integrable phase space structure of the quadratic Hamiltonian in the saddle-center case. In section \ref{subsec:impacts} we recall the reflection law from the corners' boundaries. In section \ref{subsec:geompar} we explain how the parameters of the simplest geometrical model (\ref{eq:linbilham}) may be extracted from a general  PES and motivate our choice of particular ranges of geometrical parameter values  in sections \ref{sec:theta0}-\ref{sec:upperstable}.
\subsection{\label{subsec:integr}The integrable structure near the saddle-center}

\begin{figure}[t]
\begin{center}
(a)\includegraphics[width=0.4\textwidth]{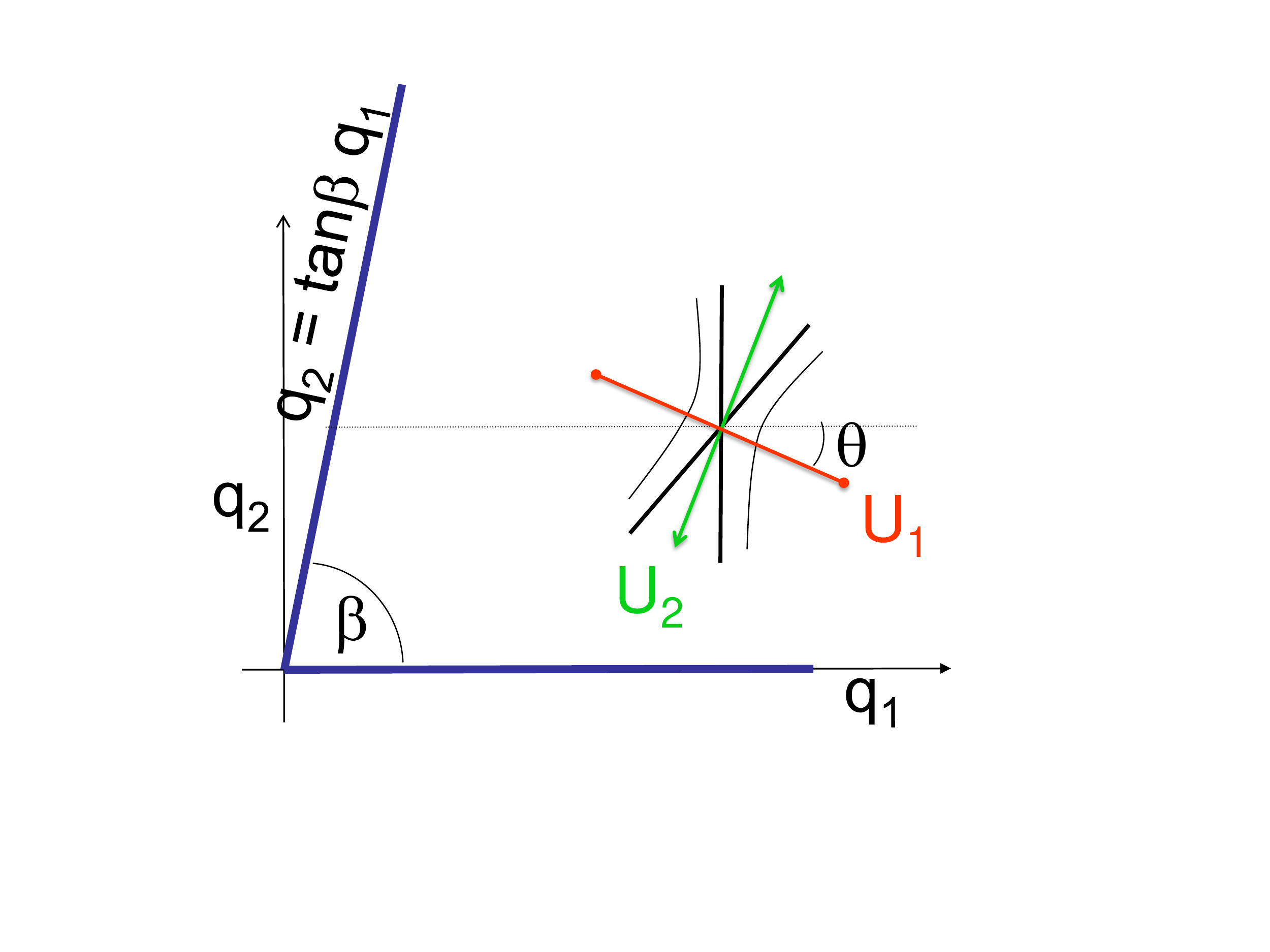}
(b)\includegraphics[width=0.4\textwidth]{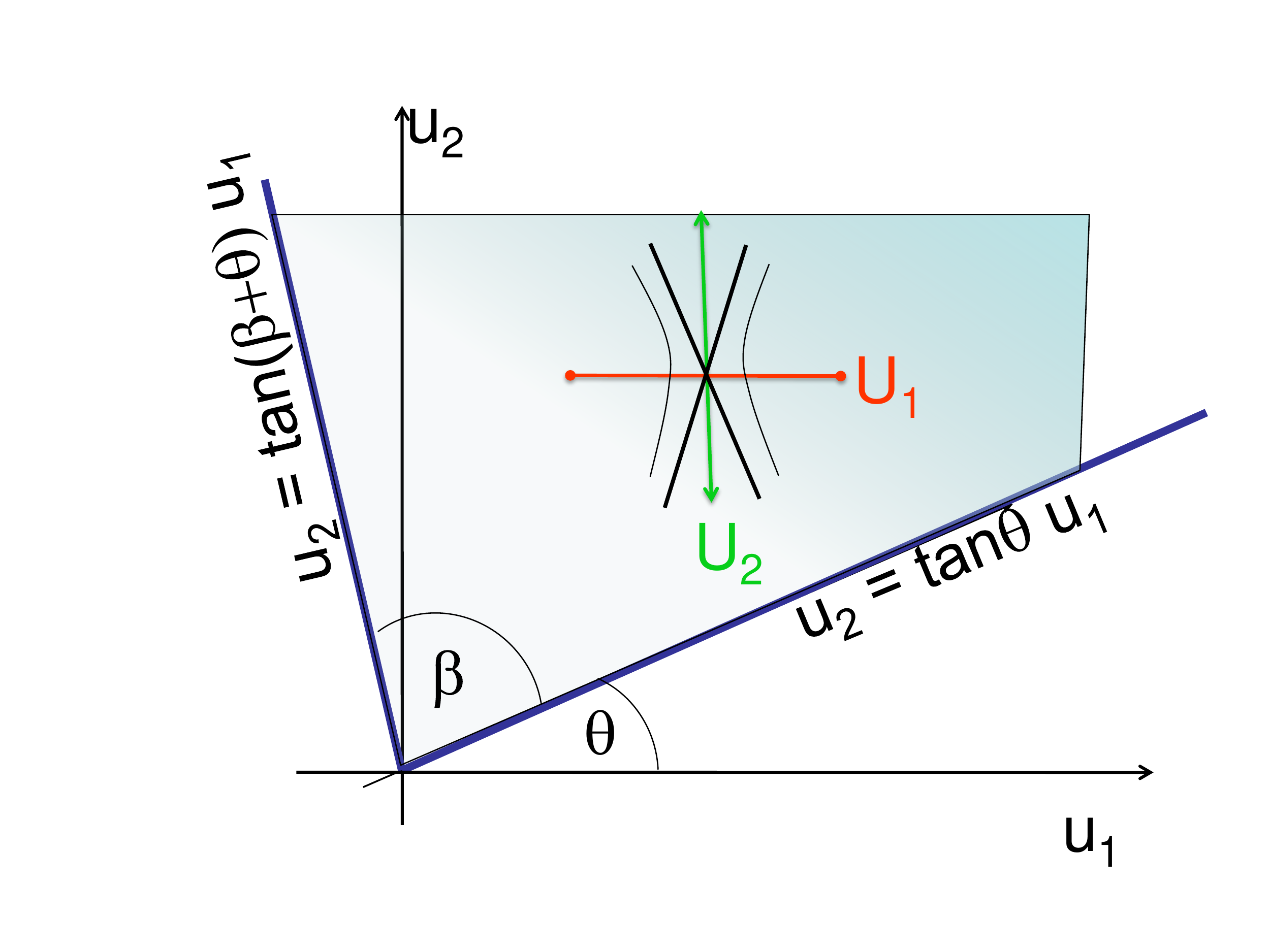}
\end{center}
\begin{center}
(c)\includegraphics[width=0.4\textwidth]{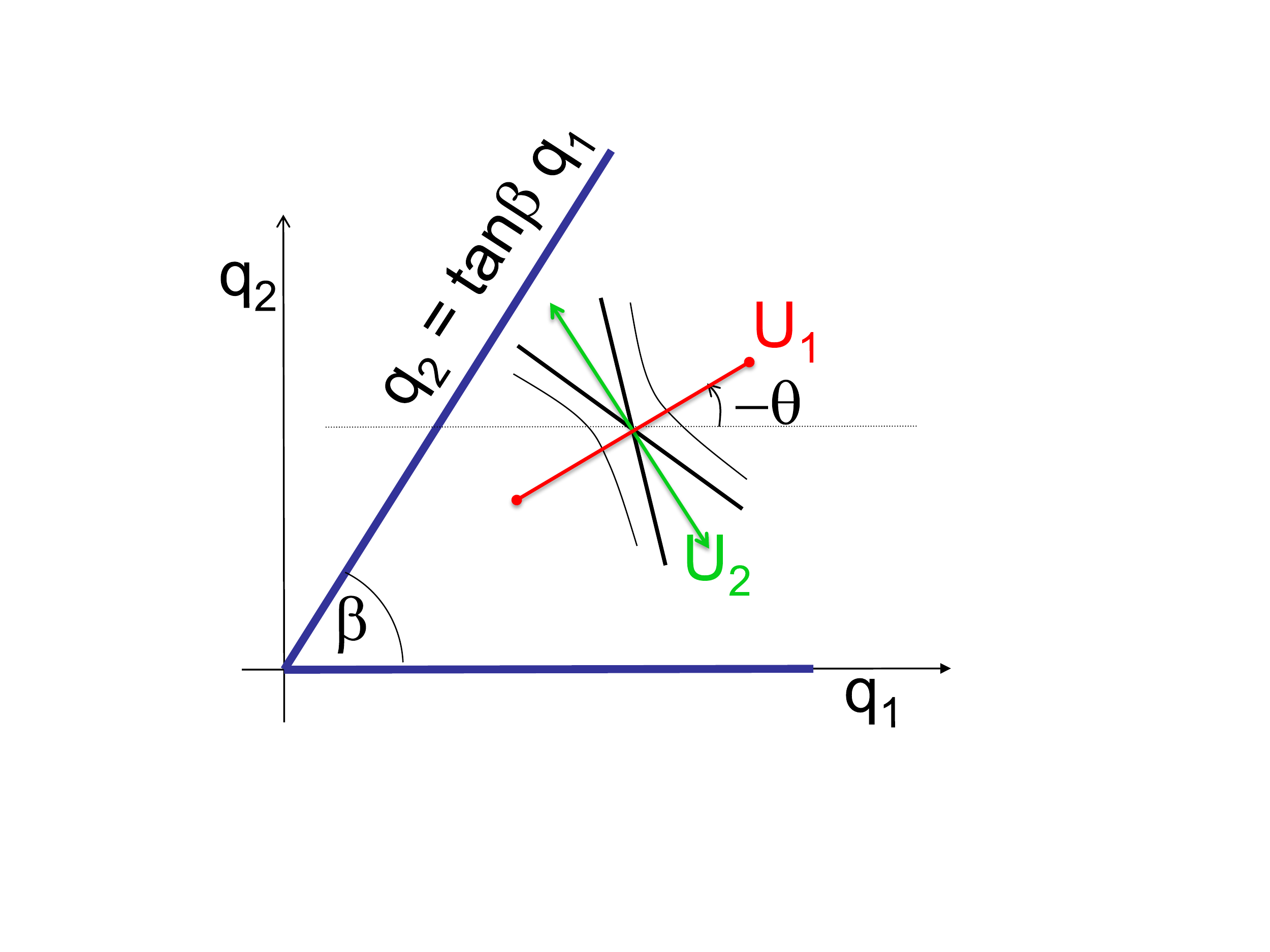}
(d)\includegraphics[width=0.4\textwidth]{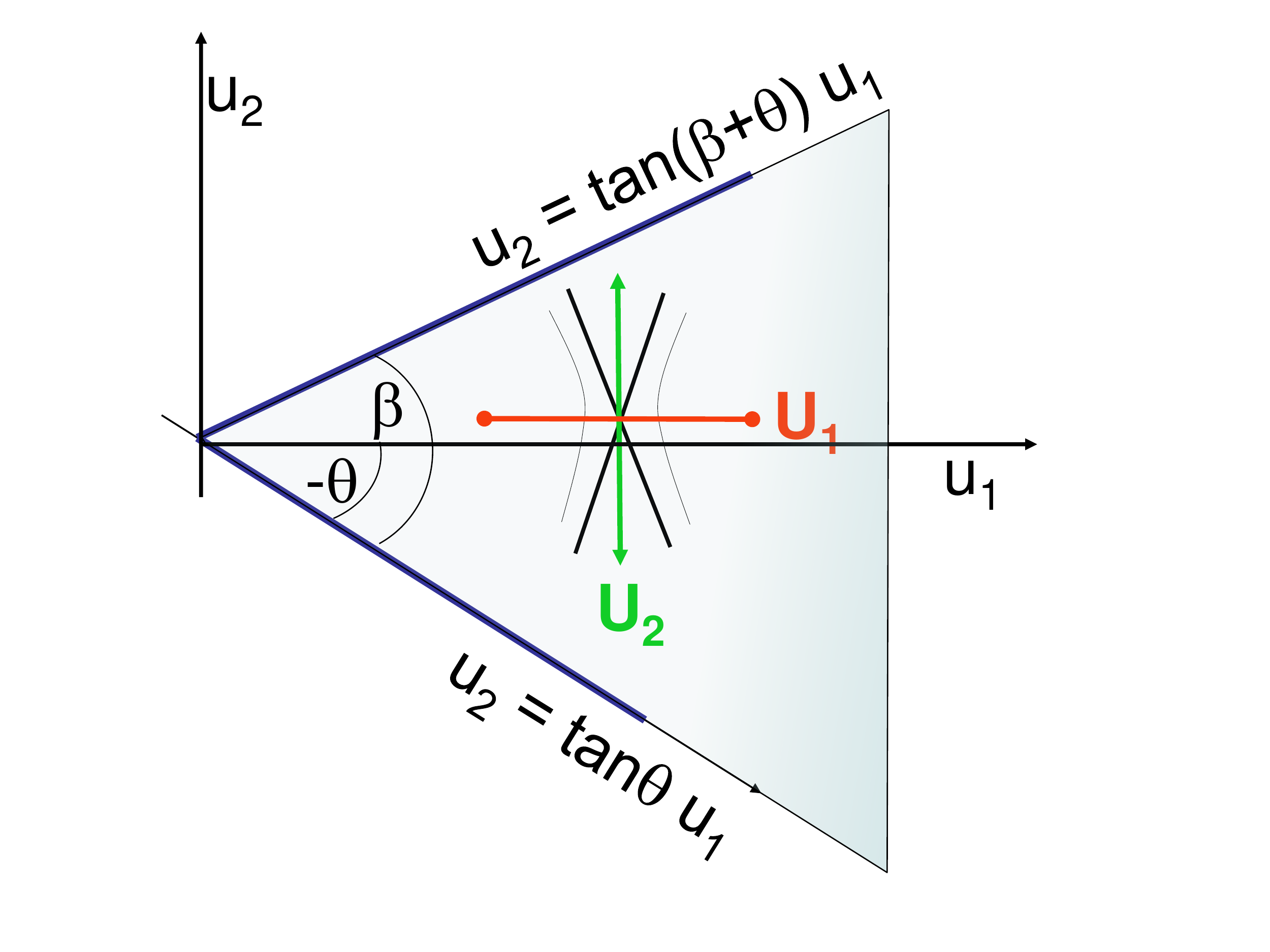}
\end{center}
\caption{
%\begin{footnotesize}
The saddle in a corner geometry in the configuration space. The potential level sets near the saddle are schematically drawn; The zero level lines are the thick black lines whereas the thin black curves are the nearly- zero potential level lines. The projection of the center eigen space is denoted by the red line and that of the stable and unstable eigen spaces  is denoted by the green line. The top (bottom) panels show the geometry when  \(\theta\) is positive (negative). The left (right) panels show the geometry in the mass scaled  \((q_1,q_2)\)  (the rotated \((u_1,u_2))\) coordinates. The  region \(\mathcal{A}_L\) is indicated by grey shading.
%\end{footnotesize}
}
\label{fig:config}%
\end{figure}

Below, we first  rotate the \((q,p) \)  coordinate system so that the quadratic Hamiltonian in (\ref{eq:linbilham}) becomes separable (more generally, one brings the integrable part of (\ref{eq:hgeometricalgen}) to its normal form near the saddle-center point). We then define the corner region in the rotated coordinate system \( (u,v)\), see Fig \ref{fig:config}.
We show that in the rotated system the projections of the stable and unstable manifolds of the saddle-center fixed point and of the nearby periodic orbits are easily found. We end this subsection by defining  two constants of motion and identifying reacting and non reacting trajectories near the saddle-center point in terms of the values of these constants of motion.

\begin{figure}[t]
\begin{center}
\includegraphics[width=0.7\textwidth]{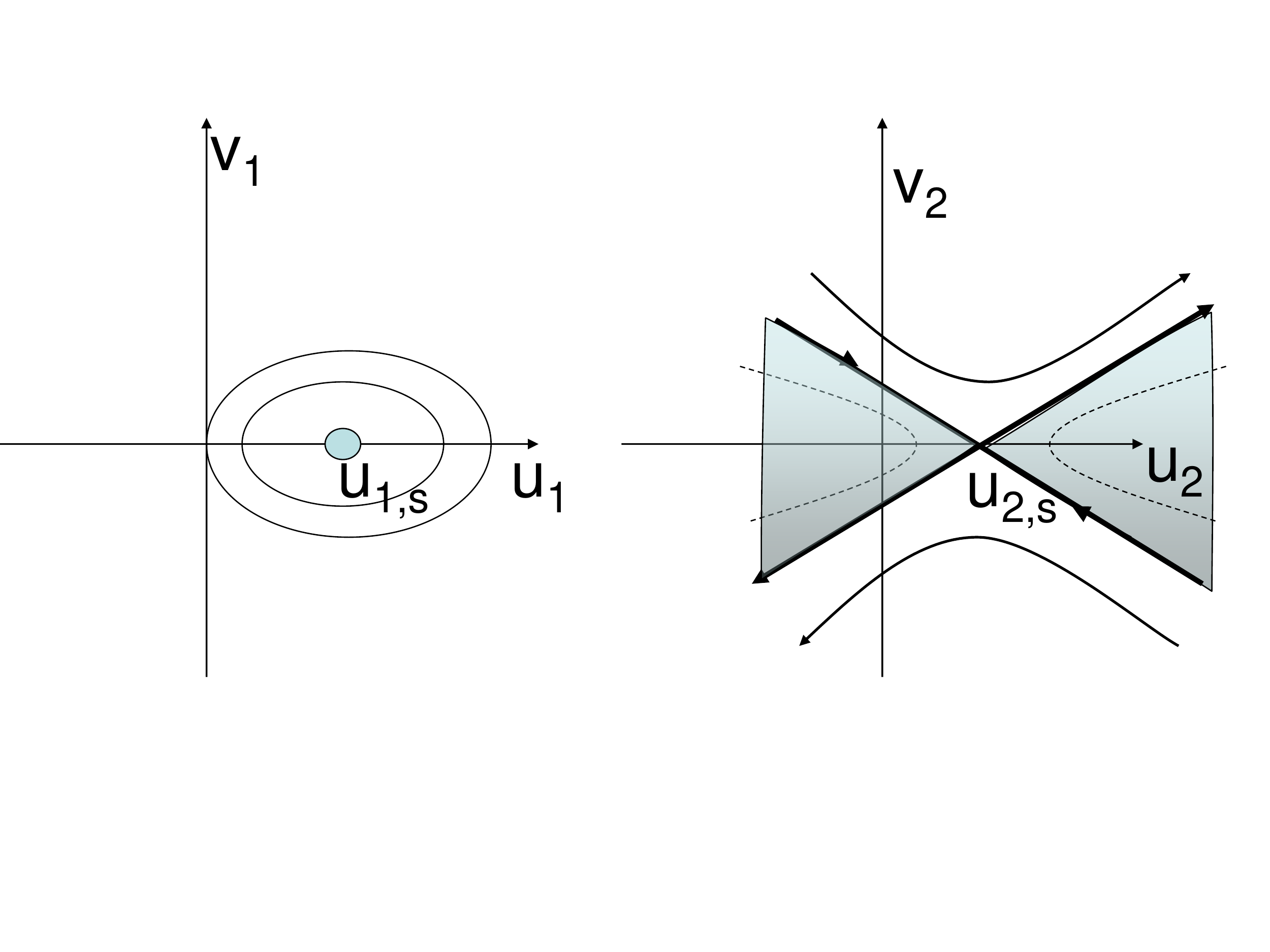}
\end{center}
\caption{
%\begin{footnotesize}
The phase space structure of the linear system. Shaded region with dashed
level curves correspond to non-reacting orbits
having $D_2<0$ (these trajectories cannot pass the dividing configuration-space section $u_2=u_{2,s}$). %\end{footnotesize}
}
\label{fig:phasespace}
\end{figure}

Denote the  normalized eigenvector of \(A\) which corresponds to $\omega^2$ (respectively $(-\lambda^2)$)  by $U_{1}$ (respectively $U_{2} $). Recall that \(U_1\bot U_2\).  A
small computation shows that the eigenvalues of the saddle-center
point of the Hamiltonian flow, $P=(q_{s},0)$, are $\
\pm i\omega, \pm \lambda$. The Hamiltonian flow has a two-dimensional real
invariant subspace $E^c$ (center plane) corresponding to the eigenvalues $\pm
i\omega : E^{c}    =\{(\widehat{ q},\widehat p)|(\widehat q,\widehat p)= ( rU_{1},sU_{1}),\; r,s \in \mathbb R \}$. The center
plane projection onto the configuration space is one dimensional, along the  eigenvector\footnote{If higher order terms of the normal form are included in \(aV_{a}\), these statements apply to the tangent planes of the corresponding manifolds.}
$U_1$. Similarly, the Hamiltonian stable and unstable subspaces corresponding to the real eigenvalues
$\pm \lambda$ are expressed in terms of the matrix eigenvector  $U_{2}$:
 $ E^{u}   =\{(\widehat q,\widehat p)|(\widehat q,\widehat p)=r(U_{2},\lambda U_{2}),\; r \in \mathbb R\},
 E^{s}  =\{(\widehat q,\widehat p)|(\widehat q,\widehat p)=r(U_{2},-\lambda U_{2}),\; r \in \mathbb R\}.
$
Thus, the projection of both the stable and unstable subspaces onto the
configuration space is {along} $U_{2}$ -- the eigenvector of $A$
corresponding to the negative eigenvalue $(-\lambda^2)$.  The eigenvectors directions should not be
confused\footnote{These simple observations are the
source of much confusion as we wrongly tend to confuse the level sets
of $V_a$ in the configuration space $q$ with the phase space plots
in the space spanned by the stable and unstable  directions (see
Figs. \ref{fig:config},\ref{fig:phasespace}).} with the zero
level-lines of $V_a$. Since the potential $V_a$ has a saddle point
at $q=q_s$, there are two directions, $Q^\pm = U_1+\alpha^\pm
U_2,$ along which $V_a$ vanishes $V_a(q_s+tQ^\pm)=0$ for all $t$,
and $\alpha^\pm$ are finite and non-vanishing (see Fig \ref{fig:config}).

To simplify further calculations it is convenient to transform the
Hamiltonian to its linear normal form. Let us denote by $(\cos\theta,
-\sin\theta)^T$ the unitary eigenvector $U_1$, assuming, with no loss of generality, that $-\pi/2 <
\theta \leq\ \pi/2.$ Then $(\sin\theta, \cos\theta)^T$ is the unitary
eigenvector $U_2$. A useful standard observation for natural
mechanical Hamiltonians is that rotations of the configuration
space can be easily incorporated into the Hamiltonian. Indeed,
defining the standard rotation matrix:
\begin{equation}
R_{\theta} = \begin{pmatrix}\cos(\theta)\ & -\sin (\theta ) \\
\sin (\theta ) & \cos (\theta )
\end{pmatrix}\label{eq:matrthet}
\end{equation}
and  making the symplectic transformation:
$(q,p)\rightarrow (u,v)=(R_{\theta}q, R_{\theta}p)$,  the Hamiltonian (\ref{eq:linbilham}) becomes:
\begin{equation}
H(u,v;b,c,\varepsilon)=\frac{v^{2}}{2}+\frac{1}{2}(u-R_{\theta}
q_{s})^{T}R_{\theta}
AR_{-\theta} (u - R_{\theta}q_{s})+bV_{b}%
(R_{-\theta} u,\varepsilon)+cV_{farfield} (R_{-\theta} u)\label{eq:rotateh}
\end{equation}
where  the quadratic part is diagonal:
\begin{equation}
R_{\theta} AR_{-\theta} =
\begin{pmatrix}\omega^2 & 0 \\0 & -\lambda^2 \\
\end{pmatrix}.
\end{equation}The
integrable part of the Hamiltonian (\ref{eq:linbilham}) becomes (where $u_s=R_{\theta} q_s$):
\begin{equation}
H_{lin}=\frac{v_1^{2} + v_2^2}{2} +
\frac{\omega^2  }{2}(u_1-u_{1,s})^2-\frac{\lambda^2}{2}(u_2-u_{2,s})^2.
\label{eq:linhamil}
\end{equation}

Recall that the quadratic (or more generally the integrable) approximation inside the corner is expected to hold only in the reaction zone where the far-field contribution is small. We thus define a bounded corner region in the \((u_{1},u_{2})\) plane,  \(\mathcal A_{L} \), and study the dynamics in this region (see Fig \ref{fig:config}). The lower and upper boundaries of  \(\mathcal A_{L} \)  are the two rays that emanate from the origin  and are aligned with the vectors
$(\cos\theta,\sin\theta)$ (lower boundary), and
$(\cos(\beta+\theta), \sin(\beta+\theta))$ (upper boundary). This \(\beta-\)wedge is then intersected by the square \([-L,L]\times[-L,L]\), where
\(L>\max(u_{1,s},u_{2,s})\).
    Since  $\beta \in(0, \pi/2)$, we have   $\beta + \theta \in
(-\pi/2,\pi)$ and two cases appear. For \(\beta+\theta<\pi/2 \) (see Fig. \ref{fig:config}d):
\begin{eqnarray}\mathcal A_{L} &=&
\{ u_1 \in[0,L], u_2
\in[u_{1}\tan\theta,\min \{u_{1}\tan(\beta+\theta),L\}]\}
 \end{eqnarray}
 whereas for \(\beta+\theta>\pi/2 \) (see Fig. \ref{fig:config}b):
 \begin{equation}
 \mathcal A_{L}=\{ u_1 \in[0,L], u_2
\in[u_{1}\tan\theta,L]\}\cup\{ u_1 \in[\text{cotan}(\beta+\theta),L], u_2 \in[0,L]\}.
 \end{equation}

Finally, we  describe in geometrical terms the well
known structure of the linear flow in  \(\mathcal A_{L}\). The motion under this linear flow occurs on surfaces
defined by the joint levels of $H_{lin}$ and the action $I_1$:
\begin{equation} I_1(u_1,v_1)=\frac{v_1^{2}}{2} +
\frac{\omega^2  }{2}(u_1-u_{1,s})^2.
\label{eq:i1def}\end{equation}
The action $I_1$ is
 the constant of motion associated with the oscillatory motion. The other constant of motion:

\begin{equation}
D_2(u_2,v_2)=\frac{v_2^{2}}{2} -
\frac{\lambda^2}{2}(u_2-u_{2,s})^2=H_{lin}-I_1
\label{eq:d2def}\end{equation}
determines the hyperbolic motion in the $(u_2,v_2)$-plane.   The surface on which
 the motion occurs,  $\{(u,v)| H_{lin}(u,v)=h, I_1(u,v)=k\},$
 is composed, for $k>0$ and $h\ne k$,  of two
 disconnected  2-dimensional cylinders: the direct product
 of an ellipse in the $(u_1,v_1)$ plane and two branches of a hyperbola
 in the $(u_2, v_2)$ plane. The sign of $D_2=h-k$
 determines the nature of the hyperbolic  motion. For negative $D_2$
 the hyperbola branches are directed sideways, one having
positive $u_2 - u_{2,s}$ and the other having  negative $u_2 - u_{2,s}$, see
the shaded region in Fig. \ref{fig:phasespace}. Trajectories belonging to level
sets with $D_2<0$ do not cross the 3-plane $u_2=u_{2,s}$: they
approach it and then return to the same side. Keeping in mind the
chemical origin of our model,  we shall say that the motion
corresponding to such trajectories occurs ``without reaction". On the
other hand, if $D_2$ is positive, branches of the hyperbola extend
horizontally along the full $u_2$ axis. All the trajectories that
belong to level sets with $D_2>0$  cross the surface $u_2=u_{2,s}$.
The upper branch of the hyperbola corresponds to orbits with a
monotonically increasing $u_2$ (``reactants to products"), whereas the
lower one to monotonically decreasing $u_2$ (``products to
reactants").  We thus say that such trajectories ``realize the
reaction".

The level sets on which $H_{lin}=h=k=I_1$ (so $D_2=0$), are the
singular level sets. These sets separate  the  two types
of motion (with vs. without reaction). Each such singular level set contains a normally
hyperbolic Lyapunov periodic orbit $\gamma_h=\{I_1=h>0,
u_2=u_{2,s},v_2=0\}$ belonging to the center plane $E^c$ along
with its stable and unstable manifolds \(W^{s,u}(\gamma_h)\)(each being a straight
cylinder). At $h=0$ this singular level set contains only the
saddle-center point $P$ and its stable and unstable manifolds \(W^{s,u}({P=\gamma_0})\)(each being
a straight-line). The projection of these
local stable and unstable manifolds (\(W^{s,u}({P})\)) onto the configuration space is a straight line, aligned with the vector $U_2$.
This projected line is divided by the saddle point $u_s$  into two rays: the extension of the lower ray, the projection of
 \(W^{s,u}_{-}(P)\),
 intersects the lower boundary of the corner, whereas the extension
of the upper ray, the projection of \(W^{s,u}_{+}(P) \), intersects (if $\tan(\theta+\beta)>0$) the upper
boundary of the corner (see right panels of Fig. \ref{fig:config}). The projection
onto the configuration space of the stable and unstable
manifolds of a Lyapunov orbit $\gamma_h$ with $h>0$ appears as a
collection of many oscillatory orbits that are centered around this
line.
The projection of the Lyapunov periodic orbit   $\gamma_h$ lies within the corner region provided \(h\) is smaller than \(h_{crit-\gamma}\):\begin{equation}
h_{crit-\gamma}= \min\left\{\frac{\omega^2  }{2}(u_{2,s}/\tan(\beta+\theta)-u_{1,s})^2,\frac{\omega^2  }{2}(u_{2,s}/\tan\theta-u_{1,s})^2\right\}.
\label{eq:hcritgam}\end{equation}
For large \(h \) values the impacts destroy these periodic orbits. Hereafter we always consider energies that are close to the saddle point energy and are thus  strictly smaller than \(h_{crit-\gamma}\).

Summarizing, for any given $h$, the energy surface $H_{lin}=h$ is
foliated by the levels of $I_1$ into the level sets on which the
motion occurs. For $h<0,$ all these level sets have  negative
$D_2,$  hence, these do not cross the surface $u_2=u_{2,s}$
and  no reaction may occur there. When $h>0$ the energy surface
contains, additionally, a Lyapunov saddle periodic orbit with
action $I_1(\gamma_h)=h$ and  level sets with $0\le
I_1<I_1(\gamma_h)$. These level sets have positive $D_2$ and thus
orbits belonging to these correspond to reacting trajectories, namely, these orbits belong to the reaction tubes.

See
\cite{WaScWi08} for a detailed explanation of the very similar analogous geometry of the
energy surfaces near saddle-center-center-..-center points in the higher dimensional
settings and when higher order terms of the non-resonant normal forms are incorporated. {Here we concentrate on the two degrees-of-freedom case: the higher dimensional saddle-multi-center case with impacts may be studied similarly, leading to more complicated dynamics (involving, for example, homoclinic orbits to invariant tori as in \cite{KL1})}.

\subsection{\label{subsec:impacts}The  impacts }

In the rest of this paper we study how the standard integrable behavior changes when the
trajectories of the linear system are reflected from the walls of
the billiard corner.
Here
we recall the reflection law from the  lower and upper boundaries of the corner region.

The unit vector defining the upper  boundary of the corner in the  $(u_1, u_2)$ plane
is  ${\bf s}_{\beta+\theta} = (\cos(\beta+\theta), \sin(\beta+\theta))^T$
and its inward normal is  ${\bf n}_{\beta+\theta}=(\sin(\beta+\theta),$ $ -\cos(\beta+\theta))^T$.  The resulting reflection law is:
\begin{equation}
\begin{array}{l}
{\bf v}=(v_1, v_2)\mapsto \\
((v_1\cos2(\beta+\theta)+v_2\sin2(\beta+\theta),
v_1\sin2(\beta+\theta)- v_2\cos2(\beta+\theta))).
\end{array}\label{eq:reflectionupper}
\end{equation}
It is defined for velocities that exit the corner region, namely those satisfying \(\left\langle{\bf v}, {\bf n}_{\beta+\theta}\right\rangle <0\). Similarly, for the lower ray, ${\bf s}_{\theta} = (\cos\theta, \sin\theta)^T$ and
${\bf n}_{\theta}=(-\sin\theta, \cos\theta)^T$, so the reflection law becomes:
\begin{equation}\label{lrl}
\begin{array}{l}
{\bf v}=(v_1, v_2)\mapsto \\
(v_1\cos(2\theta)+v_2\sin(2\theta),
v_1\sin(2\theta)- v_2\cos(2\theta)).
\end{array}
\end{equation}
\begin{remark}
The reflection law preserves energy, namely the integral $H $. However, reflections from the lower boundary (respectively upper boundary) do \textbf{not} preserve the integrals $I_1$ and $D_2 $    whenever \(\theta\neq0,\pi/2\) (respectively
whenever \(\beta+\theta\neq0,\pi/2\)).\end{remark}
The change in the integrals by the reflections leads to the non-trivial behavior of the impact system.
  \subsection{\label{subsec:geompar}The geometrical parameters }
We show in sections \ref{sec:theta0}-\ref{sec:upperstable} that the dynamics of the limiting
system depend in an essential way on the location and orientation
of the saddle point with respect to the corner. The coordinates of
the saddle point $u_{s}=R_\theta q_s$, the angle $\beta$ of the corner, the ratio \(\omega/\lambda\)  and the angle $\theta$ between the eigenvector
$U_1$  and the $q_1$-axis  all matter in an essential way in determining the
dynamics.

These geometrical parameters, \((\beta,u_{s},L,\theta,{\omega},{\lambda,b,\varepsilon})\), may be  extracted numerically from any
potential surface describing triatomic reactions. First, one finds the saddle point location \(q_{s}\), and linearizes  the vector field at this point to obtain the matrix \(A\). The angle $\theta$  may be calculated via \(A'\)s entries
$a_{ij}$:
$$\tan\theta = -\frac{2a_{12}}{\sqrt{(a_{11}-a_{22})^2 +
4a_{12}^2}+a_{11}-a_{22}}.$$
The location of the saddle point in the normal coordinates is then found via \(u_{s}=R_{\theta}q_s\) (see \ref{eq:matrthet}).
   The estimates of \(b,\varepsilon\)  may be extracted from the diatomic potentials: these parameters are determined by the form of the strong atomic repulsion at short distances. Estimating the linear zone range \(L\) may be more delicate. In principle, comparing the approximation (\ref{eq:linhamil}) to the numerical potential energy surface provides the range of validity of the linear approximation. If it appears to be too small, it is possible to extend our theory by including  higher order terms of the integrable normal form near the saddle center (as explained in detail, in another context, for the high dimensional chemical reaction settings in \cite{WaScWi08}).
The application of this scheme to concrete reactions is under current study.

A detailed classification of all possible dynamical behaviors for various
\((\beta,u_{s},L,\theta,{\omega},{\lambda,b,\varepsilon})\) and \(h\) in the symmetric and asymmetric cases is beyond the scope of the current
paper (moreover, it is hardly possible at all). In section \ref{sec:theta0} we analyze  the nearly perpendicular behavior ($\theta $ nearly zero or equivalently,   $\theta+\beta$ close to zero). In section \ref{sec:upperstable} we provide rough classification for the dependence of the manifolds' geometry
on \({\omega}/{\lambda}\).

\section{\label{sec:theta0}The nearly perpendicular dynamics}

Recall that the projections of  \(W^{s,u}_{-}(P)\), the lower branches of the local stable and unstable manifolds of the saddle-center point \(P\), onto the configuration space are directed downward along the vector $U_2$. We show next that when  $U_2$ is close to being
perpendicular to the lower boundary of the corner
 (so $\theta  $ is nearly zero),  near integrable behavior occurs.
We establish first that when $\theta= \varepsilon=c=0$ the limit motion in some region
containing the saddle-center point is integrable.  We then prove that when $\theta
\ne 0 $ the picture changes dramatically, leading sometimes to chaotic dynamics and sometimes to stable triatomic periodic motion.
The same results apply to the upper branches of the manifolds when $(\beta+\theta)$ is small. These two cases may arise, for example, in light-heavy-light reactions (\(\beta\) is close to \(\pi/2\)) with late (small  $\theta  $) or early (small $(\beta+\theta)  $) barriers.
\subsection{Integrable behavior of the perpendicular limit system}

\begin{figure}[t]
\begin{center}
\includegraphics[width=0.8\textwidth]{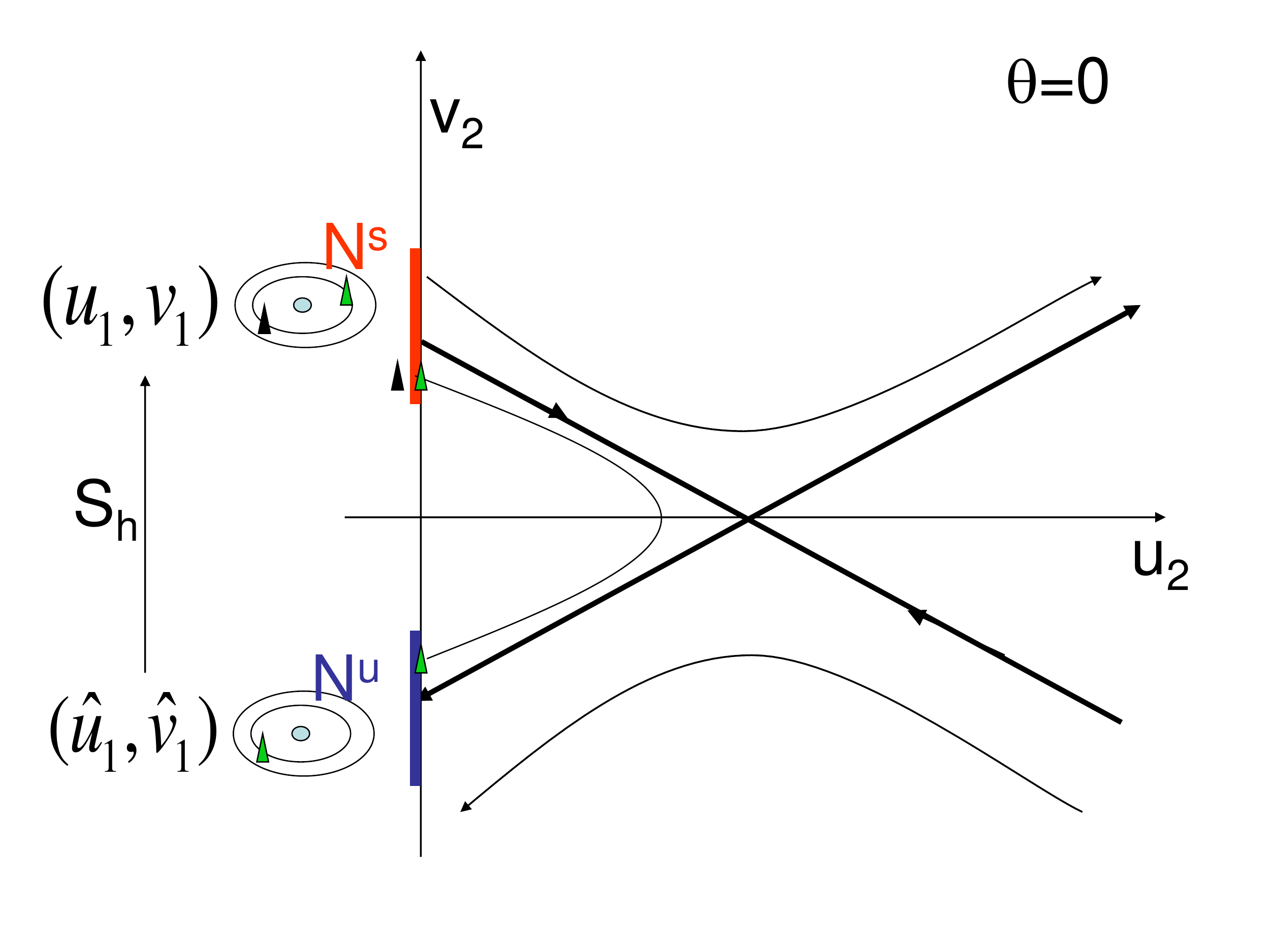}
\end{center}
\caption{%\begin{footnotesize}
The return map for $\theta=0$. The green triangle on $N^s$ is mapped by the smooth  flow to the green triangle on $N^u$. Its reflection (the \(S_{h}\) image)  is the black triangle belonging to $N^s$. Here, it belongs to the same level sets \(I_{1},D_2\) yet its velocity \(v_{2}\) changes sign. %\end{footnotesize}
}%
\label{fig:return0}%
\end{figure}

 When $\theta= \varepsilon=c=0$ and the energy \(h\) is smaller than \(h_{crit-\gamma}\) (see (\ref{eq:hcritgam})),
homoclinic loops are created by  \(W^{s,u}_{-}(\gamma_h)\).   Indeed, as shown below, the
projection of the stable and unstable manifolds of the
center-saddle point $P$    (respectively of the Lyapunov orbit \(\gamma_{h}\)) is a straight line (a cylinder) which is perpendicular to the lower boundary of the billiard corner.  Thus,  after one reflection these manifolds coincide, see Fig. \ref{fig:return0}.
\begin{proposition}\label{prop:theta0} Consider the limit system (\( \varepsilon =c=0\)) at  $\theta=0$ at an energy level \(h\in[0, h_{crit-\gamma}).\) The lower branch of the unstable manifold of $\gamma _{h}$
coincides after reflection with the lower branch of its stable manifold,
forming a family of homoclinic orbits to $\gamma _{h}$. The flow of this limit
system near the surface of homoclinic orbits is locally integrable: all
nearby orbits belonging to  the same energy surface \(h\) either belong to invariant tori or leave the
homoclinic loop region after one round to the $u_2>u_{2,s}$ region.
\end{proposition}
{\bf Proof}. We  first prove the existence of a homoclinic
orbit for \(h=0 \), where \(\gamma_{h=0}=P\). For the case $\theta =0$ the projection of the
stable/unstable manifolds of \(P \) onto the configuration space is a line perpendicular to the line $u_2 =0.$ Indeed, the linear system
\begin{equation}
\dot u_1 = v_1,\; \dot u_2 = v_2, \;\dot v_1 = -\omega^2   (u_1 -
u_{1,s}), \;\dot v_2 = \lambda^{2}(u_2 - u_{2,s})
\label{eq:equationsuv}\end{equation}
has the flow
\begin{equation}
\begin{array}{l} u_1(t)-u_{1,s} = (u_1^0 -
u_{1,s})\cos(\omega t)+
\displaystyle{\frac{v_1^0}{\omega}\sin(\omega t)},\\
v_1(t)= -\omega(u_1^0 - u_{1,s})\sin(\omega t)+
v_1^0\cos(\omega t),\\
u_2(t)-u_{2,s} = (u_2^0 - u_{2,s})\cosh(\lambda t)+
\displaystyle{\frac{v_2^0}{\lambda }\sinh(\lambda t)},\\
v_2(t)= \lambda (u_2^0 - u_{2,s})\sinh(\lambda t)+
v_2^0\cosh(\lambda t).
\end{array}\label{eq:explicittraj}
\end{equation}
So, the stable $W^{s}(P)$ and unstable $W^u(P)$ one-dimensional manifolds of the equilibrium $P$ are the straight lines $\{u_1 = u_{1,s},$ $v_1 =0,$ $u_2 = u_{2,s} -
v_2/\lambda $ \} and \{$u_1 = u_{1,s},$ $v_1 =0,$ $u_2 = u_{2,s}
+ v_2/\lambda \},$ respectively. Each straight line is divided
by the point $P$ into two rays  $W^{u,s}_+(P)$ and  $W^{u,s}_-(P)$. The lower rays  $W^{u,s}_-(P)$ intersect the wall
$u_2 =0$ (and the other rays intersect either the upper box boundary
or the upper corner boundary). The stable and unstable
rays that hit the lower wall intersect it at two different phase
space points $m_s = (u_{1,s}, 0, 0,\lambda u_{2,s})$ and $m_u =
(u_{1,s}, 0, 0, -\lambda u_{2,s}),$ respectively (recall that
$u_{1,s}, u_{2,s}$ are both positive for sufficiently small $\theta $). We
choose, near each of these two points,  sufficiently small
3-dimensional cross-sections to trajectories in the phase space:
$N^{s,u}=\{(u,v)|u_2=0,||(u,v)- m_{s,u}||<\delta\}$. Each of these
cross-sections is foliated into 2-disks $N^s_h = N^s\cap \{H=h\},$
$N^u_h = N^u\cap \{H=h\}.$ As coordinates on the disk $N^s_h$ we
take $(u_1, v_1)$, since the third coordinate $v_2$ on $N^s$ is
expressed from $H:$ $v_2 = \sqrt{2h - (v_1^2 + \omega^2  (u_1 -
u_{1,s})^2) +\lambda^{2}(u_{2,s})^2} = \sqrt{2h - 2I_1 +\lambda^{2}(u_{2,s})^2}.$ The
same coordinates $(\hat u_1, \hat v_1)$ work for $N^u_h,$ where the
$\hat v_2$-coordinate has the same form as $v_2$ but with a
$``-"$  sign in front of the root. If a trajectory of the linear flow hits
$N^u_h,$ it is transformed to $N^s_h$ due to the reflection law (see
(\ref{lrl})): the coordinates $u_1, u_2, v_1$ at the incidence point
remain the same, but $v_2=\dot u_2$ changes its sign. This means
that the reflection law defines the symplectic global map (gluing
map) $S_h: N^u_h \to N^s_h$ as follows: $u_1 =\hat u_1, v_1 = \hat
v_1$. In particular, $m_u$ is transformed to $m_s:$ we get a
homoclinic orbit $\Gamma$ to $P $.

More generally, since \(\theta=0\), for any trajectory hitting\footnote{i.e. trajectories arriving to this section with \(\hat v_{2}<0\).} the 3-plane \(u_{2}=0\),    $S_h$ simply changes the sign of \(v_2\). In particular,   $S_h$ preserves the energy and the  integrals of motion $\hat
I_1=I_1(\hat u_1,\hat v_1)=I_1(u_1,v_1)=k$ and  $D_2(u_2, v_2)=h - k=\hat D_2$.   Thus, trajectories belonging to the level set $\hat
I_1(\hat u,\hat v)= k,\;H_{lin}(\hat u,\hat v)=h$ that hit the lower
boundary remain, after reflection, on the same level set with
$I_1(u,v)=\hat I_1=k$ and $D_2=h - k=\hat D_2$: the reflection just
changes their relative position along the same level set of the
integrable linear system (see Fig \ref{fig:return0}).

Thus, for \(h\in(0,h_{crit-\gamma})\),
 the  stable (\(I_1=h,D_{2}=0\)) and unstable (\(\hat I_1=h,\hat D_{2}=0\)) manifolds of the Lyapunov periodic orbit \(\gamma_h\) are cylinders that coincide after one reflection. Moreover, since \(\beta<\pi/2\), these cylinders hit the 3-plane \(u_{2}=0\)  at \(u_1\) values that are bounded away from 0, namely bounded away from the corner.

The dynamics for trajectories near the homoclinic loops are also fairly simple.
 In particular, trajectories lying on the
given level set $H=h, I_1=k$ are projected onto the plane $(u_2, v_2)$
into one of the two hyperbola branches of $D_2 = h-k$. If $h-k <0$
and $u_2 <u_{2,s}$, then trajectories starting on $N^s$ move towards $P$ and then return to the cross-section $N^u$. In other words, the reflection law
glues this hyperbola-like branch into a closed curve (see  Fig. \ref{fig:return0}: the green triangle on $N^s$ is mapped by the flow to the green triangle on $N^u$ and then reflects  to the black triangle, namely, back to $N^s$). Since such orbits also belong, in the  $(u_1,
v_1)$ plane, to the closed curve $I_1 =k$, these orbits belong, topologically, to an invariant torus in the phase space. Similarly, trajectories
for which $D_2=h-k>0$ belong to two disjoint cylinders (direct
product of the circle $I_1=k$ in the $(u_1, v_1)$-plane and the two branches
of the corresponding  hyperbola in  the $ (u_2, v_2)$-plane). Here, the gluing map defined by the
reflection law glues the two end circles (those intersecting
the section $u_2=0$) of these cylinders.    Trajectories move along one of the cylinders towards
the section $u_2=0$ and after the first reflection escape along the second cylinder  to the region
$u_2>u_{2,s}$. Their global dynamics, after they pass the saddle point, depend on
($\beta,u_s,\omega,\lambda$) as discussed in section \ref{sec:upperstable}. $\blacksquare$

We conclude that for $\theta=\varepsilon=c=0$ the motion {near} $\Gamma$ in
the region $u_2<u_{2,s}$ is indeed simple. For negative energy \(h\), the motion near\footnote{Namely for small enough $I^*_1,-h$.} the
homoclinic loop occurs within the solid torus $0\leq I_1\le I_1^*,  D_{2}=h-I_1\leq0$. The motion is quasi-periodic if the rotation number on the related
invariant 2-torus is irrational and it is periodic if this number
is rational\footnote{The same behavior extends to all\ energies in  \(-
\frac{\lambda^2}{2}(u_{2,s})^2<h\leq0\), but here we are concerned with the near saddle-center behavior.}. Inside this solid torus there is also a unique
elliptic periodic orbit $I_1 =0,$ $D_2 = h$. For $h=0$ this periodic orbit is
replaced by the homoclinic orbit $\Gamma.$
When \(h>0, \) the orbits with negative \(D_{2}\) are again periodic or quasiperiodic. On the other hand, the orbits with positive \(D_{2}\)   cross into the $u_2>u_{2,s}$
region. Their behavior may be rather complicated since these  may hit the upper
boundary of the billiard which is slanted. For sufficiently small \(h\),     such trajectories
  closely follow  $W^{u,s}_+(P)$, the upper
branches of the stable and unstable manifolds of $P$. Hence,  the global structure of $W^{u,s}_+(P)$ determines their behavior.

We say that $W^{u,s}_+(P)$ exhibit simple dynamics (SD) if these manifolds intersect the upper boundary at a
monotonically increasing sequence of \(u\) until  exiting the corner
region, see Fig. \ref{fig:smtheta} and first column of  Fig \ref{fig:bildynam}.   Otherwise, we say that their behavior is
complicated, see second and third columns of Fig  \ref{fig:bildynam}. We see that in the complicated cases the
manifolds  return back to the corner
region, possibly hitting both boundaries, possibly hitting the
boundaries  at some \(u_{1}<u_{1s}\).
We describe next the behavior of  $W^{u,s}_-(P)$ and of trajectories in their neighborhood near a slightly slanted lower
boundary, and discuss the behavior of the
trajectories at $u_2>u_{2,s}$  in
section \ref{sec:upperstable}.

 We should note that the behavior of trajectories with a larger oscillatory component (e.g. trajectories starting near \(u_{2,s}\) with $I_1>h_{crit-\gamma}$)
is expected to be complicated as well: almost always such trajectories eventually hit the upper ray and we expect  that this reflection would, almost always, destroy the integrability. Notice that such reflections, induced by the geometry, supply an alternative route for the creation of reactions which is unrelated to the local structure near $P$. We will return to this point in the discussion.

\subsection{Non-integrable behavior}

We establish next that for small enough nonzero $\theta$    the motion in the limit system is chaotic for a range of energies.  We then prove that similar behavior occurs for sufficiently small \(\varepsilon,c\).
 More precisely, we prove below the existence of transverse homoclinic orbits to a saddle periodic orbit (Poincar\'e homoclinic orbits). To this aim we use some of the ideas developed in \cite{Lerman,LerKol} where the center-saddle case was analyzed (the first discussion of the problem was in \cite{Conley}).  The chaotic nature of the motion follows from this result: it is well known that near such homoclinic orbits there is an invariant subset which is described on some cross-section by a transitive Markov chain, so in particular, this set contains a countable set of saddle periodic orbits, almost periodic orbits, etc. \cite{Smale65,Shil67}. \begin{theorem}[Complicated Dynamics I] \label{thm:bifurc}
If \(|\theta|\) is nonzero and sufficiently small, then, for \(\varepsilon=c=0\), the  system
(\ref{eq:linbilham}) admits the following properties:
\begin{enumerate}
\item The lower branches of the stable and unstable separatrices of $P$ are split;
\item There is a critical energy value \(h^{0}\), depending on the geometrical parameters \((\theta,u_s,{\lambda},\omega)\), satisfying  $0<h^{0}<h_{crit-\gamma}$, such that at energies \(h\in (h^{0}, h_{crit-\gamma})\)
the Lyapunov periodic orbit
$\gamma_h$ has two {\bf transverse} homoclinic orbits. \item At the energy level $H=h^{0}(\theta,u_s,{\lambda},\omega)$ the flow has a tangent homoclinic
orbit to the related Lyapunov periodic orbit.

\item For $0<h< h^{0}$ the lower branches of the separatrices do not admit simple\footnote{Orbits that reflect only once from the corner boundaries.} homoclinic orbits  to $\gamma_h$.
\end{enumerate}
\end{theorem}

\begin{figure}[t]
\begin{center}
\includegraphics[width=0.8\textwidth]{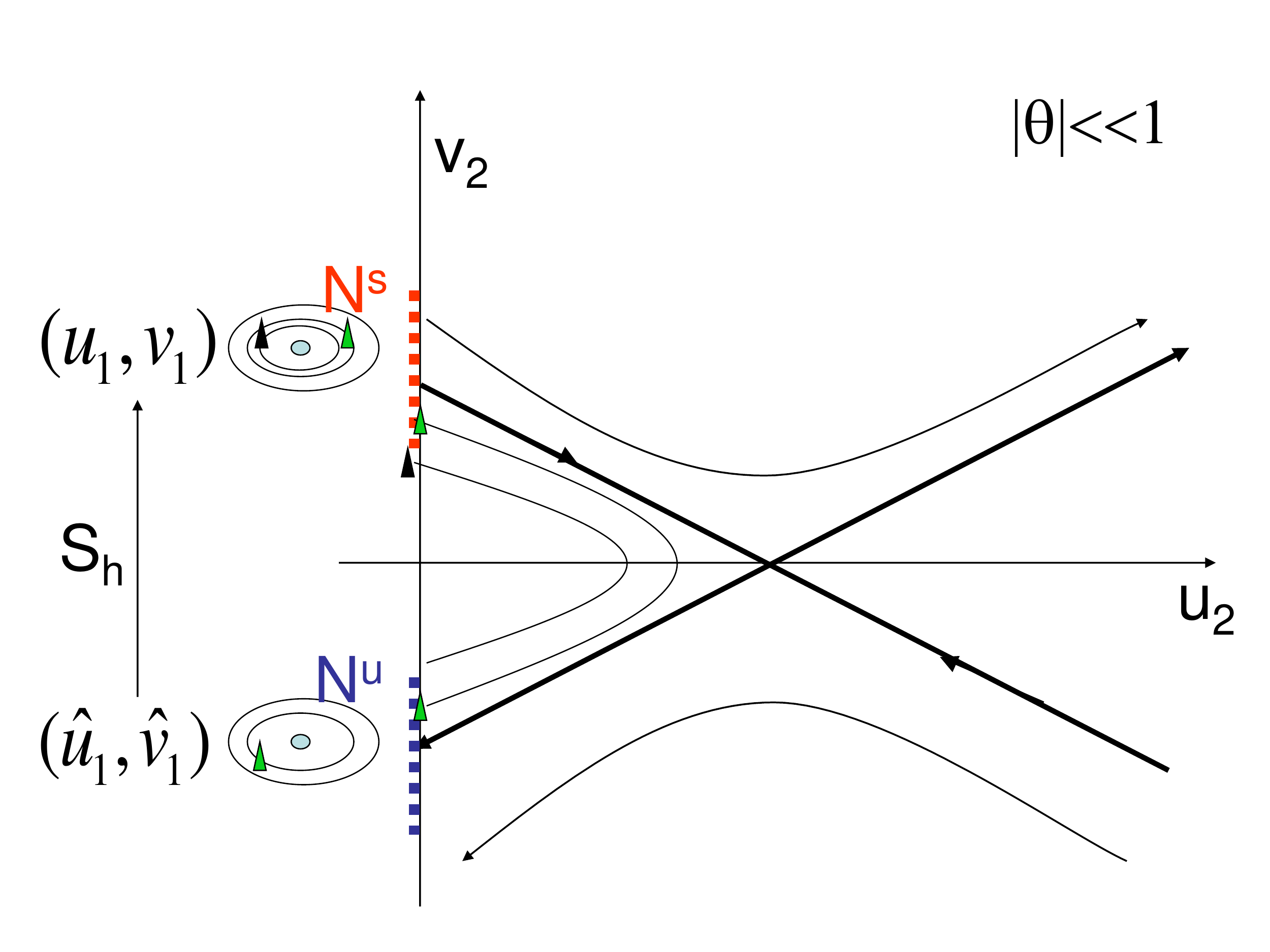}
\end{center}
\caption{%\begin{footnotesize}
The return map for small $\theta$. As in Fig 4, the image of the green triangle of \(N^{s}\) under the flow and the gluing map is the black triangle. Here, the velocities $(v_1,v_2)$ and the corresponding level sets $I_1,D_2$ are changed due to the reflection. %\end{footnotesize}
}%
\label{fig:returnnot0}%
\end{figure}

{\bf Proof}.
1. We construct, as in the proof of Proposition \ref{prop:theta0}, the
global map that is defined by the reflection law from the slightly
slanted bottom wall. Let $\mu = \tan\theta$ and  ${\bf s} =
(1/\sqrt{1+\mu^2} =\cos\theta, \mu/\sqrt{1+\mu^2}=\sin\theta)^T$ be,
as above, a unit vector defining the lower boundary of the corner.
Then, for sufficiently small $\theta$, the 3-plane given by $u_2 =
\mu u_1$ provides a cross-section to the linear flow. Recall that  the projections of the lower branches of the local stable and
unstable manifolds of $P$,   $W^{u,s}_-(P)$, are  straight lines. Their extensions
intersect the bottom wall\footnote{Recall that since
$\theta$ is small, $u_{1,s}$ and $u_{2,s}$ are positive.} at the
points: $m_s = (u_{1,s}, \mu u_{1,s}, 0, \lambda (u_{2,s} - \mu
u_{1,s}))$ and $m_u = (u_{1,s}, \mu u_{1,s}, 0, -\lambda (u_{2,s} - \mu
u_{1,s})). $   As before, we consider some small cross-sections
$N^s,$ $N^u$ near these points. They are again foliated into
2-disks. The coordinates on these cross-sections are $(u_1,
v_1),$ where here $u_2 = \mu u_1$, and $v_2$ is expressed from $H:$ $v_2 =
\pm\sqrt{2h - 2I_1+\lambda^{2}(\mu u_1 - u_{2,s})^2}$. The sign $``+"$
corresponds to $N^s_h$ and the sign $``-"$  to $N^u_h.$ The restriction of the
2-form $dv_1\wedge du_1 + dv_2 \wedge du_2$ to $N^s_h$ (and similarly for $N^u_h$) is now given as $dv_1\wedge du_1 + \mu dv_2 \wedge
du_1,$ where $v_2$ is taken from $H$. Applying the reflection law
from the slanted wall boundary (\ref{lrl}) to values \((\hat u_1, \hat v_1)\in N^u_h\), we obtain the gluing symplectic map $S_h:N^u_h \to N^s_h$ (w.r.t.
the non-trivial symplectic form, see Appendix): \begin{equation}
S_h: (\hat u_1, \hat v_1)\mapsto (\hat u_1,
\frac{(1-\mu^2)\hat{v}_1+ 2\mu \hat{v}_2}{1+\mu^2})= (\hat u_1,
\hat v_1\cos2\theta + \hat v_2 \sin2\theta).\label{eq:globalmap}
\end{equation}
This gluing map is defined for trajectories pointing down so that $-\mu \hat{v}_1+\hat{v}_2 < 0$.
Namely, for sufficiently small $|\mu|$     we require that $\hat{v}_2 < 0,$ so
the sign in front of the radical for $\hat{v}_2$ should be
negative.

We now prove the first assertion of the
theorem. We show that the reflection of  the unstable branch  $(W^{u}_-(P))$ from the lower boundary (the  $S_0$-image of the point $m_u$) does not
coincide with the stable branch  $(W^{s}_-(P))$ intersection with this boundary (the point $m_s$).  For $S_0$ we should set $\hat{u}_1 = u_{1,s}, \hat{v}_1 =
0,$ $\hat v_2 = - \lambda (u_{2,s} - \mu u_{1,s}).$ Then we get for
$\mu \ne 0:$
$$
u_1 = u_{1,s}, \;\;v_1 = \frac{-2\mu \lambda(u_{2,s}-\mu u_{1,s})}{1+\mu^2}= -2\mu\lambda u_{2,s}(1+O(\mu))\ne 0,
$$
and therefore the stable and unstable manifolds of $P$ are split - they are of  the order $\lambda|\mu|$ apart.

2-3. Now, let us fix $h>0$. We first show that the intersection of
$W^s_{-}(\gamma_h)$ (the lower branch of the stable manifold of the Lyapunov periodic orbit
$\gamma_h$) with $N^s_h$ is an ellipse. We then show that the $S_h$-image of the intersection of the unstable manifold branch $W_{-}^u(\gamma_h) $ with
$N^u_h$ is another ellipse. We then prove that these two ellipses intersect as the energy is increased beyond a critical energy.
Recall that for $0<h<h_{crit-\gamma}$
the intersections of $W^{s,u}_{-}(\gamma_h)$  with $N^{s,u}$ are bounded away from the corner\footnote{more precisely, these are bounded away from the 2-plane $(u_1=0,u_{2}=\mu u_{1}=0,v_1,v_2)$ which corresponds to the corner in the phase space.}   and occur, for sufficiently small \(|\mu|\), with a  vertical velocity which is
strictly bounded away from $0$. We consider here only such \(h\) values.

  First, note that
the 2-dimensional cylinders $W^{s,u}_{-}(\gamma_h)$ are given by
solutions of the two equations: $I_1(u_1,v_1)=v_1^2/2 + \omega^2   (u_1 -
u_{1,s})^2/2=h,$ $v_2 =\pm \lambda (u_{2,s} - u_2)$, where the
$+/-$ signs correspond to the stable/unstable manifolds. The
intersection of $W^s_{-}(\gamma_h)$ with $N^s$ (with coordinates
$(u_1,v_1,v_2)$) is the closed curve $ I_1(u_1,v_1)= h,$ $v_2 =
\lambda (u_{2,s} - \mu u_1),$ its projection onto the $(u_1,v_1)$-plane
is an ellipse. Similarly, $W^u(\gamma_h)$ intersects $N^u$
(with coordinates $(\hat u_1, \hat v_1, \hat v_2)$) along the closed
curve $I_1(\hat{u}_1, \hat v_1)= h,$ $\hat{v}_2 =
-\lambda (u_{2,s} - \mu \hat{u}_1),$ and its projection onto the
$(\hat{u}_1,\hat{v}_1)$-plane is an ellipse as well.

To find the intersection of the
$S_h$-image of the trace of $W^u_{-}(\gamma_h)$ with the trace of
$W^s_{-}(\gamma_h)$, it is convenient to use the action-angle (here
simply polar) coordinates $(I_1,\varphi)$ on $N^s_h$ and $(\hat I_1,
\psi)$ on $N^u_h:$ $v_1 = \sqrt{2I_1}\cos\varphi,$ $u_1 = u_{1,s} +
\sqrt{2I_1/\omega^2  }\sin\varphi,$ $\hat v_1 = \sqrt{2\hat
I_1}\cos\psi,$ $\hat u_1 = u_{1,s} + \sqrt{2\hat I_1/\omega^2  }\sin\psi.$
Using eq. (\ref{eq:globalmap}), the equation $S_h(\hat u_1,\hat
v_1)=(u_1,v_1)$ becomes:
$$
\begin{array}{l}
u_{1,s} + \sqrt{2h/\omega^2  }\sin\varphi = u_{1,s} +
\sqrt{2h/\omega^2  }\sin\psi,\\
\sqrt{2h}\cos\varphi =
\displaystyle{\frac{(1-\mu^2)\sqrt{2h}\cos\psi -
2\mu\lambda [u_{2,s} - \mu (u_{1,s} +
\sqrt{2h/\omega^2  }\sin\psi)]}{1+\mu^2}}.
\end{array}
$$
The first equation implies that either $\varphi = \psi$ or
$\varphi = \pi - \psi.$ In the first case we obtain the following
equation for $\varphi$:
$$
\mu\sqrt{2h}(\cos\varphi -
\frac{\lambda}{\omega}\sin\varphi)= -\lambda (u_{2,s} -
\mu u_{1,s}),
$$
which has no solutions for sufficiently small $|\mu|$. In the
second case the equation for $\varphi $ becomes:
\begin{equation}
\sqrt{2h}\left(\cos\varphi -
\mu^2\frac{\lambda}{\omega}\sin\varphi\right) =
-\mu\lambda (u_{2,s} - \mu u_{1,s}).\label{eq:secondphi}
\end{equation}
Defining $$\sigma=\frac{\omega^2  \lambda^{2}(u_{2,s} - \mu u_{1,s})^2}{2(\omega^2
+\mu^4\lambda^{2})}=\frac{  (\lambda u_{2,s} )^2}{2}+O(\mu),$$ we see that for a fixed $\mu$ eq.
(\ref{eq:secondphi}) has no solutions  for $0< h < \sigma\mu^2$, has a
unique solution at $\sigma\mu^2,$ and has two solutions  for
$h\in (\sigma\mu^2,h_{crit-\gamma})$. Indeed, notice that the  $S_h$-image of the trace of
$W_{-}^u(\gamma_h)$ is  given in the   $(u_1,v_1)$ coordinates by
\begin{equation}
\begin{array}{l}
u_1 = \hat{u}_1, \\v_1=\hat{v_1}\cos2\theta
+\hat{v}_2\sin2\theta=-\lambda u_{2,s}\sin2\theta
+\hat{v_1}\cos2\theta + \mu\lambda \hat{u}_{1}\sin2\theta,
\end{array}\label{eq:u1v1reflec}
\end{equation}
that is, it is a linear transformation of the ellipse, so it is also an
ellipse. It follows from the geometry of intersecting ellipses
that for $h > \sigma\mu^2$ these two solutions correspond to two
transversal homoclinic orbits to the Lyapunov periodic orbit
$\gamma_h,$ and that at the intermediate case:
\begin{equation}
h^{0}=h^{0}(\theta,u_s,{\lambda},\omega)=\sigma\mu^2=\frac{(\lambda u_{2,s} \tan \theta)^2}{2}+O(\tan \theta^3)
\label{eq:h0}\end{equation}
the
unique solution corresponds to a tangent homoclinic orbit of
$\gamma_h$ (it corresponds to the outer tangency of the ellipses).

4. Since for \(h<h^{0}\) the two ellipses do not intersect, and since in the corner interior the flow is  integrable,  the lower branches do not admit simple homoclinic orbits at such energies. In general, it is still possible that consequent reflections  of the extensions of  \(W^{s,u}_-(\gamma_h)\) from the corner boundaries will produce homoclinic orbits.    $\blacksquare$

Next, we assert that similar behavior appears when \(c   \) and \(\epsilon\) are  sufficiently small and the billiard-like potential indeed limits to the impact flow. More precisely, in \cite{KlRk11} we prove that away from a small boundary layer near the billiard boundary, regular reflections of impact flows are \(C^{r}\) close to the corresponding reflection-like segments of the smooth Hamiltonians. The smooth Hamiltonians are assumed to be in the standard mechanical form with  potentials that are a sum of a smooth potential and a family of  billiard-like potentials (satisfying assumptions I-IV of \cite{RaRkTu07}). Additionally, to obtain the correct impact limit, it is assumed that the value of the full potential  along the billiard boundary is strictly positive. Then, for positive \(\epsilon\) the Hill region of the smooth flow lies within the billiard region and, near regular impacts, limits to it as \(\epsilon\rightarrow0^{+}\). Under these conditions it is proved  in \cite{KlRk11} that for sufficiently small \(\epsilon\) the smooth reflection from the Hill region boundary limits to the billiard reflection law.  Here, to comply with this latter  condition, we assume hereafter that the potential along the corner rays in   \(\mathcal A_{L}\): \begin{equation}
R_{\beta,L}=\left\{(u_{1},u_{2})|\left( u_2=u_1\tan\theta\right) \cup\left( u_2=u_1\tan(\beta+\theta)\right)\cap(|u_{1,2}|\leq L)\right\}
\end{equation} is strictly positive. Namely, we assume that there exist positive constants \(\epsilon_{0}, B_1\) such that for all \(0<\epsilon<\epsilon_{0}\):\begin{equation}
\left(bV_{b}%
(R_{-\theta} u,\varepsilon)+\frac{\omega^2  }{2}(u_1-u_{1,s})^2
-\frac{\lambda^2}{2}(u_2-u_{2,s})^2\right)_{(u_1,u_2)\in R_{\beta,L}}>B_{1}.
\label{eq:condonb}\end{equation}
 For the power law repulsion law (\ref{eq:colpot}), the first term is infinite, so the inequality is satisfied for any \(b>0\). When the diatomic repulsion is modeled by a bounded potential (e.g. (\ref{eq:exppot})), \(b\) should be taken to be sufficiently large so that (\ref{eq:condonb}) holds. Such an assumption is  natural in the chemical-reaction context:   the nuclear potential energy associated with small diatomic distances is  much larger than the barrier energy. Thus, such an assumption is satisfied by adequate models of the PES of triatomic reactions. Then, by \cite{KlRk11}, we can establish:
\begin{theorem}[Complicated Dynamics - I - smooth case] \label{conj:bifurcsm}
 Assume that \(V_{b}(q;\epsilon)\)   is a billiard-like potential family limiting to the  billiard in the \(\beta\)-wedge, \(V_{farfield}(q)\) is  bounded in the \(C^{r}\) topology in the corner region, and (\ref{eq:condonb}) is satisfied. Then,  for sufficiently small \(\varepsilon,c\) and \(\theta_{max}\), for all \(0<|\theta|<\theta_{max}\) the  system
(\ref{eq:linbilham}) admits the same properties as listed in Theorem \ref{thm:bifurc}. There, the critical energy levels \(h^{0},h_{crit-\gamma}\) need to be replaced by a  family of \((\varepsilon,c)\) dependent critical energies \(h^{\epsilon,c}\) and by \(h^{\theta_{max}}_{crit-\gamma}\). Moreover, the critical  values at which the tangent homoclinic bifurcation occurs depend smoothly on \((\varepsilon,c)\) and approach the limiting bifurcation energy as  \((\varepsilon,c)\)  are decreased to zero:  \(h^{\varepsilon,c}\rightarrow h^0\).
\end{theorem}

\textbf{Proof:} Notice that in the proof of  Theorem \ref{thm:bifurc}  for the limit system we may replace the cross sections \(N^{s }\) with any other locally transverse cross-section to \(W^u_-(P)  \) in the interior of the domain. Consider for example such a local cross section  at \(\Sigma ^{\bar u_{2}}=\{(u,v):u_2=\bar u_{2}<(u_{2,s}+u_{1,s}\tan\theta)/2\}\).  The first intersection of the extensions of \(W^{u,s}_-(\gamma_{h})  \) with  \(\Sigma ^{\bar u_{2}}\)  are always transverse for the limit system.   \(\Sigma ^{\bar u_{2}}\)  also provides, under specified conditions, a locally transverse cross section to the image of the first reflection of the extension of  \(W^{u}_-(\gamma_{h})  \) from the lower boundary. Indeed, for any fixed \(\bar u_{2}\) one can choose sufficiently small  \(|\theta|\)    and fix an energy level \(h^{\bar u_{2}}(\theta)\) so that for all \(h\in[0,h^{\bar u_{2}}(\theta)] \) this section is transverse to these orbits. Moreover, \(h^{\bar u_{2}}(\theta)\rightarrow h_{crit-\gamma}\) monotonically as \(|\theta|\rightarrow0\).  Thus, the proof of  Theorem \ref{thm:bifurc}  applies, in the limit system, to the traces of  \(W^{u,s}_-(\gamma_{h}) \) on \(\Sigma ^{\bar u_{2}}\) for all \(h\in[0,h^{\bar u_{2}}(\theta)]\). Namely, for small \(|\theta|\), the trace which corresponds to the first intersection of the extension of     \(W^{s}_-(\gamma_{h}) \) with \(\Sigma ^{\bar u_{2}}\) and the image of the corresponding trace of      \(W^{u}_-(\gamma_{h}) \) after its first reflection      from the lower boundary intersect transversely at \(h>h^{0}\), do not intersect at \(h<h^{0}\) and are tangent at \(h=h^{0}\).   Notice that for sufficiently small \(|\theta|\), \(h^{\bar u_{2}}(\theta)\) is of order one whereas  the \(\theta\)-dependent homoclinic bifurcation value  \(h^{0}\) is small (see \ref{eq:h0}), namely \( h^{0}\in (0,h^{\bar u_{2}}(\theta)). \)  Also, recall that in the proof of  \ref{thm:bifurc}    \(|\theta|\) is assumed to be sufficiently small so that the reflection of  \(W^{u}_-(\gamma_{h}) \) from the lower boundary is a regular reflection: it occurs with strictly negative \(v_{2}\).   Let  \(\theta_{max}\) be sufficiently small so that a) Theorem  \ref{thm:bifurc}   applies, b)  at  \(\theta_{max}\) the homoclinic bifurcation occurs at an energy at which   \(\Sigma ^{\bar u_{2}}\) is transverse as explained above, and c)  this energy is smaller than the nuclear diatomic repulsion energy. Namely, at  \(\pm\theta_{max}\),   \(h^{0}< h^{\theta_{max}}_{crit-\gamma}:=\min (h^{\bar u_{2}}(\pm\theta_{max}),B_{1})\). This last condition implies that for sufficiently small \((c,\epsilon)\), the Hill region of
(\ref{eq:linbilham}) near the impact point \((u_{1,s},u_{1,s}\tan\theta)\) limits to the lower ray boundary of the corner for all    \(h\in[0,h^{\theta_{max}}_{crit-\gamma}] \).  Thus,  by \cite{KlRk11}, for  all \(|\theta|<\theta_{max}\) and    \(h\in[0,h^{\theta_{max}}_{crit-\gamma}], \)    for sufficiently small \((c,\epsilon) \),  the stable trace and the image of the unstable trace under  the smooth flow are   \(C^{r}\) close to the corresponding traces of the impact flow.  Hence, the result is established. $\blacksquare$

\begin{figure}[t]
\begin{center}
(a)\includegraphics[width=0.4\textwidth]{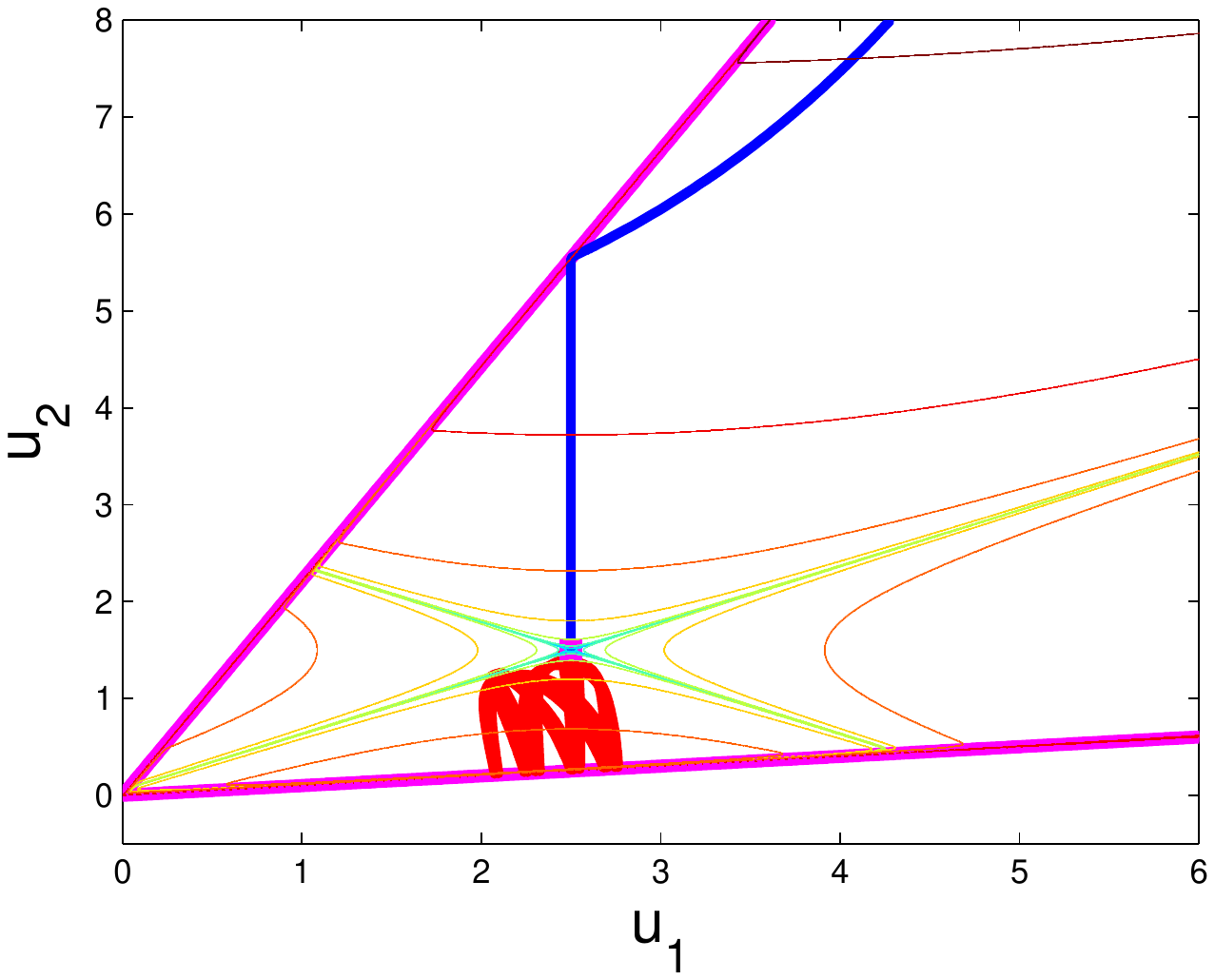}
(b)\includegraphics[width=0.4\textwidth]{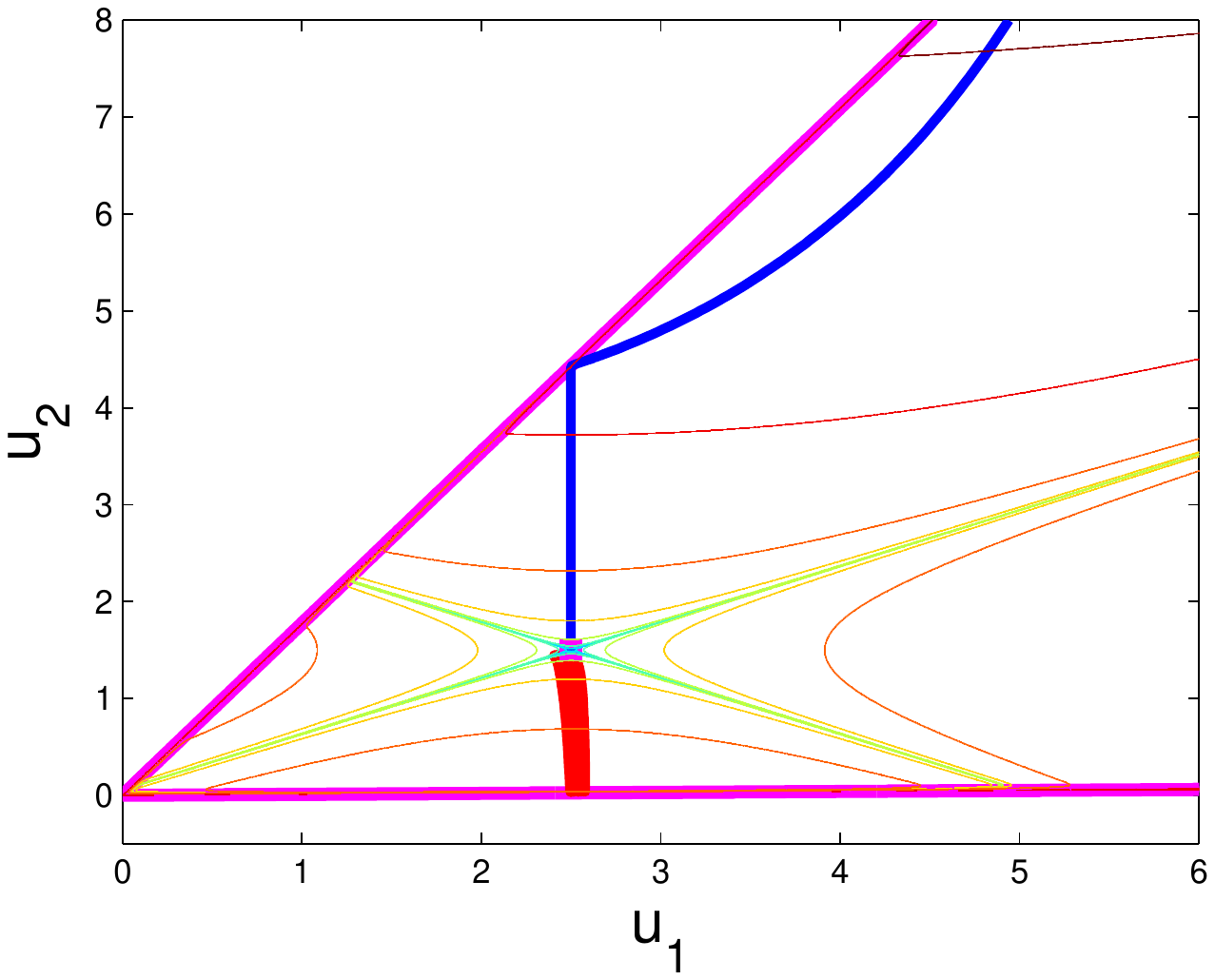}
\end{center}
\begin{center}
(c)\includegraphics[width=0.4\textwidth]{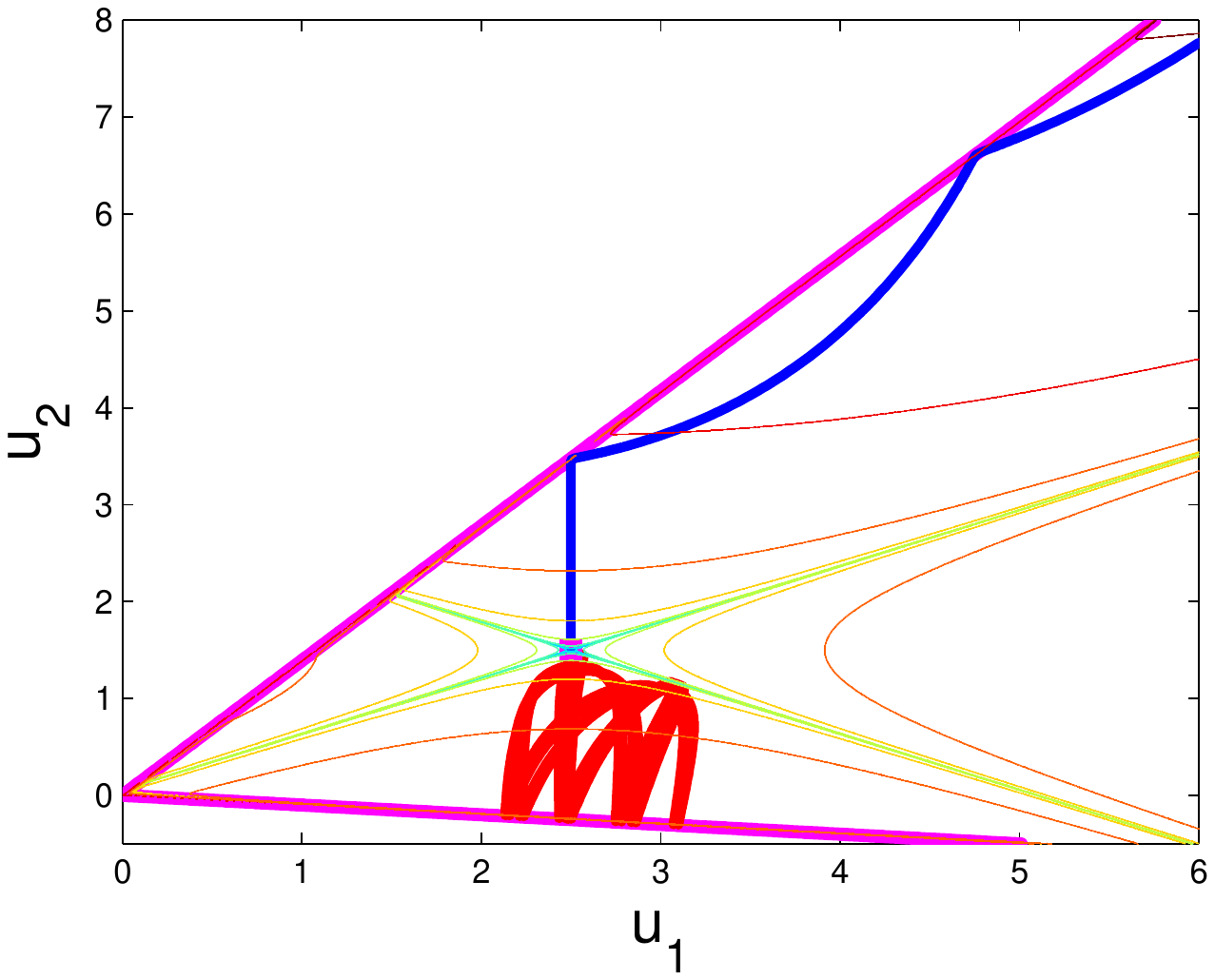}
(d)\includegraphics[width=0.4\textwidth]{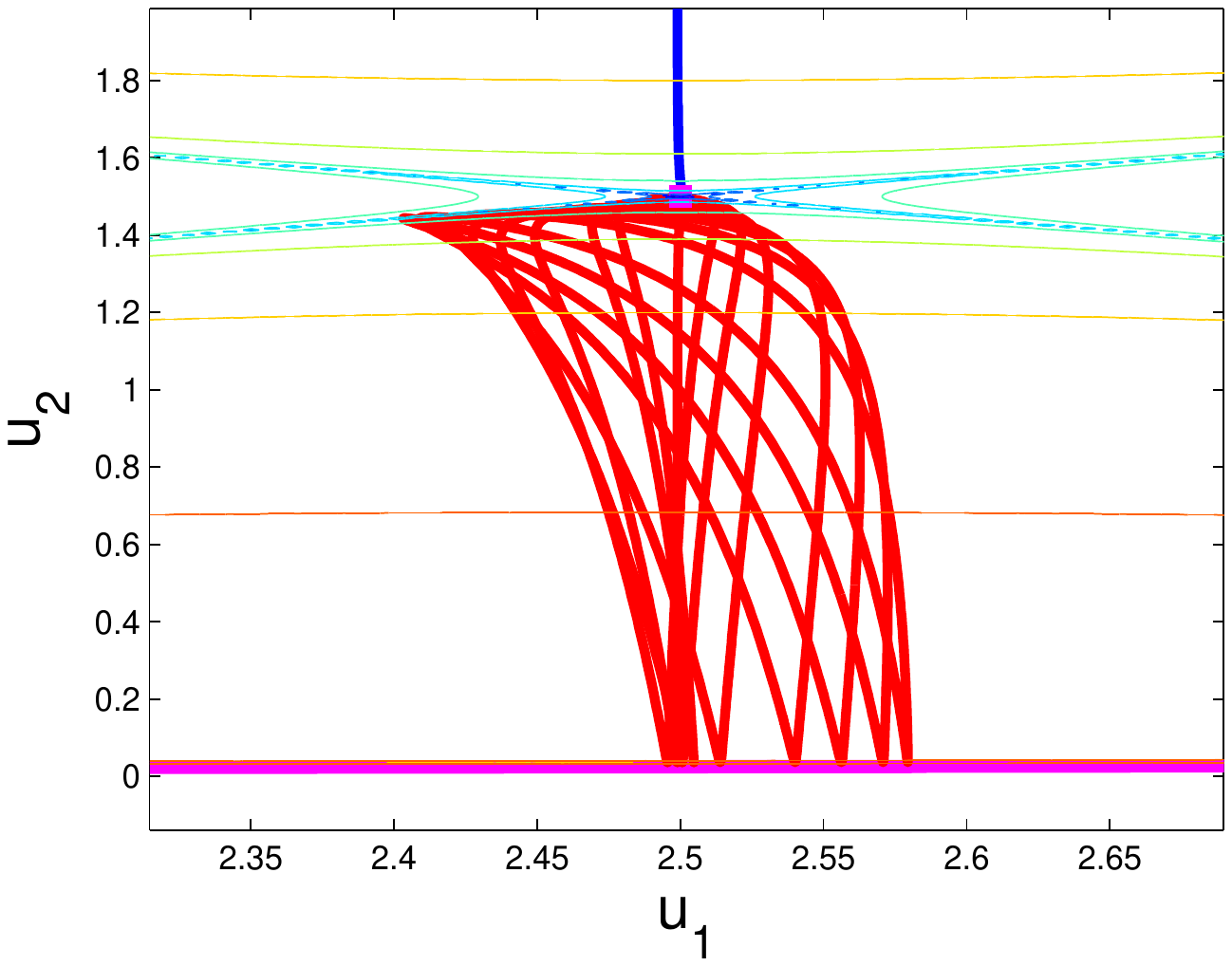}
\end{center}\caption{
%\begin{footnotesize}
Projections of the unstable manifold of \(P\)   onto the configuration space at small \(\theta\) values. \((a) \  \theta=0.1 \ (b) \ \theta=0.01 \ (c) \ \theta=-0.1\ (d)\)  Zoom-in of  (b). Thick blue (red) line: $W^{u}_+(P)$ ($W^{u}_-(P)$). Thin colored lines are the level sets of the geometric potential, with the exponential repulsion form (Eq. (\ref{eq:exppot})). Here, $b=10,$
 \(\varepsilon=0.01 , u_s=(2.5,1.5),\beta=\pi/3, \omega^2  =1,\lambda^2=3 . \ \)
 %\end{footnotesize}
 }%
\label{fig:smtheta}%
\end{figure}

Figure \ref{fig:smtheta} demonstrates that for small \(|\theta|\) and small \(\varepsilon\) (at \(c=0\)) the lower branch of the unstable manifold of \(P\)  indeed exhibits complicated behavior.

\subsection{Elliptic islands}

Theorem \ref{thm:bifurc} implies that for any small fixed $\mu=\tan\theta$,
the stable and the unstable manifolds of the Lyapunov periodic orbit
$\gamma_h$ with $h^{0} =\sigma\mu^2\approx(\lambda u_{2,s} \tan \theta)^2/{2}$ undergo a non-degenerate
tangent homoclinic bifurcation. Such a bifurcation leads, in particular, to
the creation of elliptic periodic orbits near the tangent homoclinic
orbit:
\begin{theorem}[Complicated Dynamics II]
At the semi-interval $h>h^0$, for sufficiently small \(\theta\),
the  limit system  (\ref{eq:linbilham})\(_{\varepsilon=c=0}\) has a countable set of
$h-$intervals $\Delta_n$ accumulating at $h=h^{0+}$ such that for
$h\in \Delta_n$ the limit flow has generic elliptic periodic
orbits  with their period tending to infinity as $n\to \infty. $
\end{theorem}

{\bf Proof}. We show that at \(\varepsilon=0\) one may construct a
smooth symplectic return map to a section close to the Lyapunov periodic orbit  $\gamma_{h^{0}}$. Then, the classical results regarding the emergence of elliptic
islands near homoclinic tangencies are applicable.

In order to formulate the classical
results precisely, let us recall some details
\cite{Newh75,Bir,MR,GSh,GG}. Suppose a family of smooth symplectic
diffeomorphisms $f_{\bar c}$ acts on a symplectic two-dimensional manifold
$(M, \omega)$. Let $p$ be a saddle fixed point for any parameter
value ${\bar c}$ near ${\bar c}=0$ with eigenvalues $0<\Lambda({\bar c}) < 1,
\Lambda^{-1}({\bar c}).$ Assume that  $f_{\bar c}$ undergoes a generic homoclinic bifurcation; at   ${\bar c}=0$ the stable and
unstable manifolds of $p$ have a quadratic intersection at some
point $q$ (different than $p$), creating a tangent homoclinic orbit $\Gamma$.
Furthermore, the unfolding $f_{\bar c}$ is generic, namely, in a small
neighborhood of $\Gamma$ there are no simple homoclinic orbits to $p $   at
${\bar c}<0$ and there are two transversal homoclinic orbits at ${\bar c}>0$ (or
vice versa). Near $p$ one can choose symplectic Darboux coordinates
$(x,y)$ in which the 2-form $\omega$ casts as $\omega = dx\wedge dy$. The local representation of the maps for any sufficiently small ${\bar c}$
is
$$
x_1 = [\Lambda + f_1(x,y)]x, \; y_1 = [\Lambda^{-1} + g_1(x,y)]y,
$$
with $f_1, g_1$ being of the first order at $(0,0).$ Recall that points of
$\Gamma$ tend to $p$ as $n\to \pm \infty$ ($n$ is the number of the
iterations of $f$).    Choose two points $q_+, q_-$ of $\Gamma$
such that $q_+ =(x_+,0)$ belongs to the segment $x > 0, y=0$ (a
piece of $W^s$) and $q_- =(0,y_-)$ belongs to the segment of $W^u$,
which we assume to be $y>0.$ Then one may take two small
neighborhoods $V_s$ of $q_+$ and $V_u$ of $q_-$ which belong both to
the coordinate chart. Then it is clear that points from the
semi-neighborhood $y>0$ of $V_s$ can reach under iterations of $f$ a
semi-neighborhood $x>0$ of $V_u$ for all iterations $n\ge n_0 \ge 1$
($n_0$ depends on $f,$ on the locations of $q_+, q_-$ and on the
size of $V_s, V_u$). The quadratic tangency of $W^s$ and $W^u$ along
$\Gamma$ means in these coordinates that there is some positive
integer $N$ such that the global map $f^N: V_u \to V_s$ can be
written in $V_u$ in the form $\bar{x} - x_+ = {\bar a}x+\bar b(y-y_-)+\cdots,$
$\bar{y}= -\bar b^{-1}x + \frac{1}{2}Ax^2 + Bx(y-y_-)+
\frac{1}{2}C(y-y_-)^2 +\cdots$ with nonzero $C.$ It is known that if
$\bar b>0,C<0$ then, at $\bar c=0$,  there are no orbits that
stay forever in $V_s\cup V_u$ under the action of the first return
maps $f^{n}\circ f^N$ (except of $\Gamma$). Suppose, to be definite, that two homoclinic orbits to $p$ appear
for small $\bar c>0$.
\begin{theorem}\label{thm:homtangen}\cite{Newh75,Bir,MR,GSh,GG}.
In the semi-interval $\bar c>0$ there is a countable set of $\bar c$-intervals
$\Delta_n$ accumulating at $\bar c=0^{+}$ such that for $\bar c\in \Delta_n$ the
map $f_{\bar c}$ has  generic elliptic periodic orbits and their periods
tend to infinity as $n\to \infty.$
\end{theorem}
Recall that the genericity of an elliptic fixed point or a
periodic orbit means that some of its Birkhoff coefficients in the related
normal form do not vanish.

For our case, the local map \(f_{\bar c}\) is obtained as follows. Let  $(I_1,\varphi)$
denote the symplectic polar coordinates on the $(u_1,v_1)$  plane
(see above). The semi-interval $\varphi = 0,$ $0< I_1 \le I_1^0$ is a
transverse cross-section. We fix
$h=h^{0}=\sigma\mu^2$, and choose, for sufficiently small $\delta, \zeta$,  the 3-disk $Q:$ $\varphi = 0,$
$|I_1 - I(\theta)|< \delta$, $|u_2 - u_{2,s}|<\zeta,$ $|v_2|<\zeta$,
where \(I(\theta)=h^{0}\) is the \(I_{1}\) action of the periodic orbit \(\gamma_{h^{0}}\). This $Q$ is a cross-section
to orbits that are close to the periodic orbit $\gamma_{h^{0}}.$
The generic family of symplectic maps in the 2-disks $Q_h$ is
defined by fixing the levels $H=h$ in $Q,$ and by defining the
bifurcation parameter $c=h-h^{0}.$ The coordinates on $Q_h$ are
$(u_2, v_2)$ and the map in these coordinates is linear in $u_2^0 -
u_{2,s}, v_2^0$: it is obtained from (\ref{eq:explicittraj}) if we set $t=
\frac{2\pi}{\omega }$ in $u_2(t), v_2(t).$ The quarter $D_2 \le
0,$ $u_2 \le u_{2,s}$ (recall $u_{2,s} >0$ for the case under
consideration) corresponds to those orbits of the flow which go from
$N^s_h$ to $N^u_h$ for positive $t.$ The tangent homoclinic orbit
$\Gamma$ cuts $Q_{h^{0}}$ transversely, when $t$ increases, along an
infinite sequence of points that lie on the semi-interval $D_2 =0,$
$u_2 < u_{2,s},$ $v_2 >0$ and accumulate to $(u_{2,s},0)$ -- the trace
of $\gamma_{h^{0}}$ in $Q_{h^{0}}.$ We fix some point $q_+$
of this sequence. This orbit also cuts transversely the cross-section
$N^s_{h^{0}}$ at some point $q^{s}_+$ on the trace of
$W^s(\gamma_{h^{0}}).$ Since the time of passage between these two points of transverse intersections is finite,
the flow defines a local symplectic map  $T_+: N^s_{h^{0}} \to Q_{h^{0}},$   in some small neighborhoods of $q^{s}_+$ and $q^{}_+$ respectively.

Similarly, the map $T_-$ is constructed. $T_-$  acts from some
neighborhood of $q_-$ on the unstable semi-interval $D_2 =0,$ $u_2 <
u_{2,s},$ $v_2 < 0$ to a neighborhood of the trace of $\Gamma$ on
$N^u_{h^{0}}.$ Thus, we have an analytic symplectic map (global
map) $T_{h}=T_+\circ S_{h}\circ T_-$ acting from a
small neighborhood of $q_-$ to a small neighborhood of $q_+$. The
global map \(T_{h^{0}}\) transforms  a small segment  of the unstable
manifold of the saddle fixed point $(u_{2,s},0) $   near $q_-\in Q_{h^{0}}$ to its image  in the
neighborhood of $q_+$. Since \(\Gamma\) is a non-degenerate tangent homoclinic orbit, this image is quadratically tangent to the stable
manifold of the saddle fixed point. Moreover, as the mutual position
of the two ellipses shows, this segment belongs to the quarter $D_2
> 0,$ $v_2
> 0$ and $u_2 < u_{2,s}$ for it. It means that we realize the case $\bar b >0, C<0$ of the above mentioned theorem, and we
get an infinite sequence of intervals in $h>h^{0}$
corresponding to the existence of elliptic periodic orbits of the
system.
\hfill $\blacksquare$
\bigskip

\begin{theorem}[Complicated Dynamics II-smooth]
Under the same conditions of Theorem \ref{conj:bifurcsm}, for sufficiently small \(|\theta|<\theta_{max}\),  \(\varepsilon \) and \(c\), at the semi-interval $h>h^{\varepsilon,c}(\theta)$,
the  system  (\ref{eq:linbilham}) has a countable set of
$h-$intervals $\Delta_n$ accumulating at $h=h^{\varepsilon,c}(\theta)^+$ such that for
$h\in \Delta_n$ the limit flow has generic elliptic periodic
orbits  with their period tending to infinity as $n\to \infty. $
\end{theorem}

\textbf{Proof:} For sufficiently small \(|\theta|\), near \({h^{0}}\), the gluing map \( S_{h}\)    corresponds to a regular reflection. Thus, under the same conditions as in Theorem \ref{conj:bifurcsm},  by \cite{KlRk11}, for sufficiently small \((c,\epsilon), \) the smooth version of the return map to $Q_h$, \(T^{\varepsilon,c}_{h}\), is \(C^{r}\) close to \(T_{h}\)  for  all \(|\theta|<\theta_{max}\) and    \(h\in[0,h^{\theta_{max}}_{crit-\gamma}] \).  Hence, by theorem \ref{conj:bifurcsm},  for \(h>h^{\varepsilon,c}(\theta)\), where \(h^{\varepsilon,c}(\theta)\) denotes the energy of the homoclinic tangent bifurcation of the smooth system,  theorem \ref{thm:homtangen} may be applied to the return map   \(T^{\varepsilon,c}_{h}\).
Moreover, this return map depends smoothly on \((\varepsilon,c)\), hence the theorem follows. Notice that the classical results
are clearly applicable for finite \(\varepsilon\) values for which
the homoclinic tangency  persists. Here, we see that one may
change the order of the limits, namely, even for arbitrarily small \(\varepsilon\) the stability islands appear.
\hfill $\blacksquare$
\section{\label{sec:upperstable}Criteria for simple dynamics }

\begin{figure}[t]
\begin{center}
\includegraphics[width=0.3\textwidth]{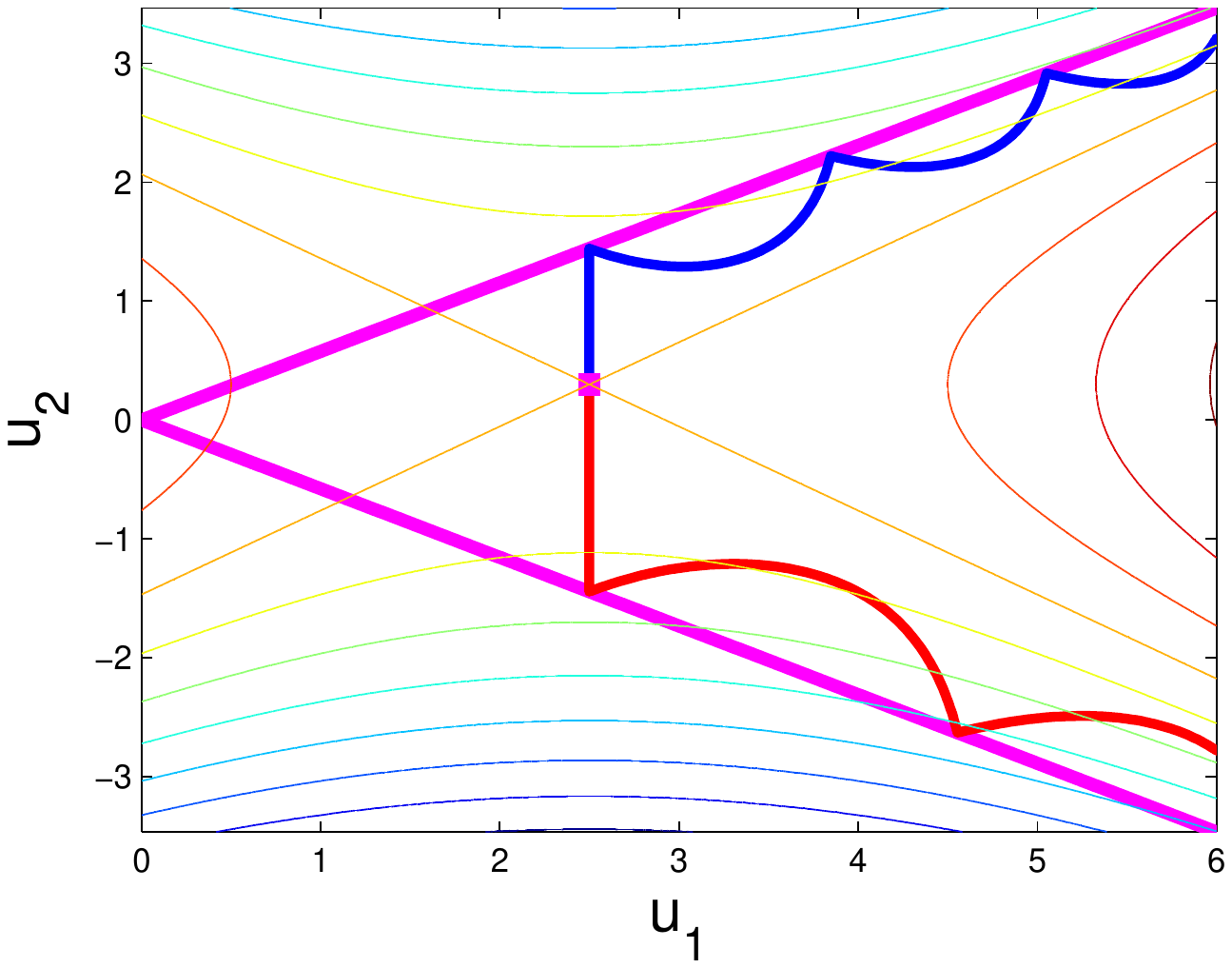}
\includegraphics[width=0.3\textwidth]{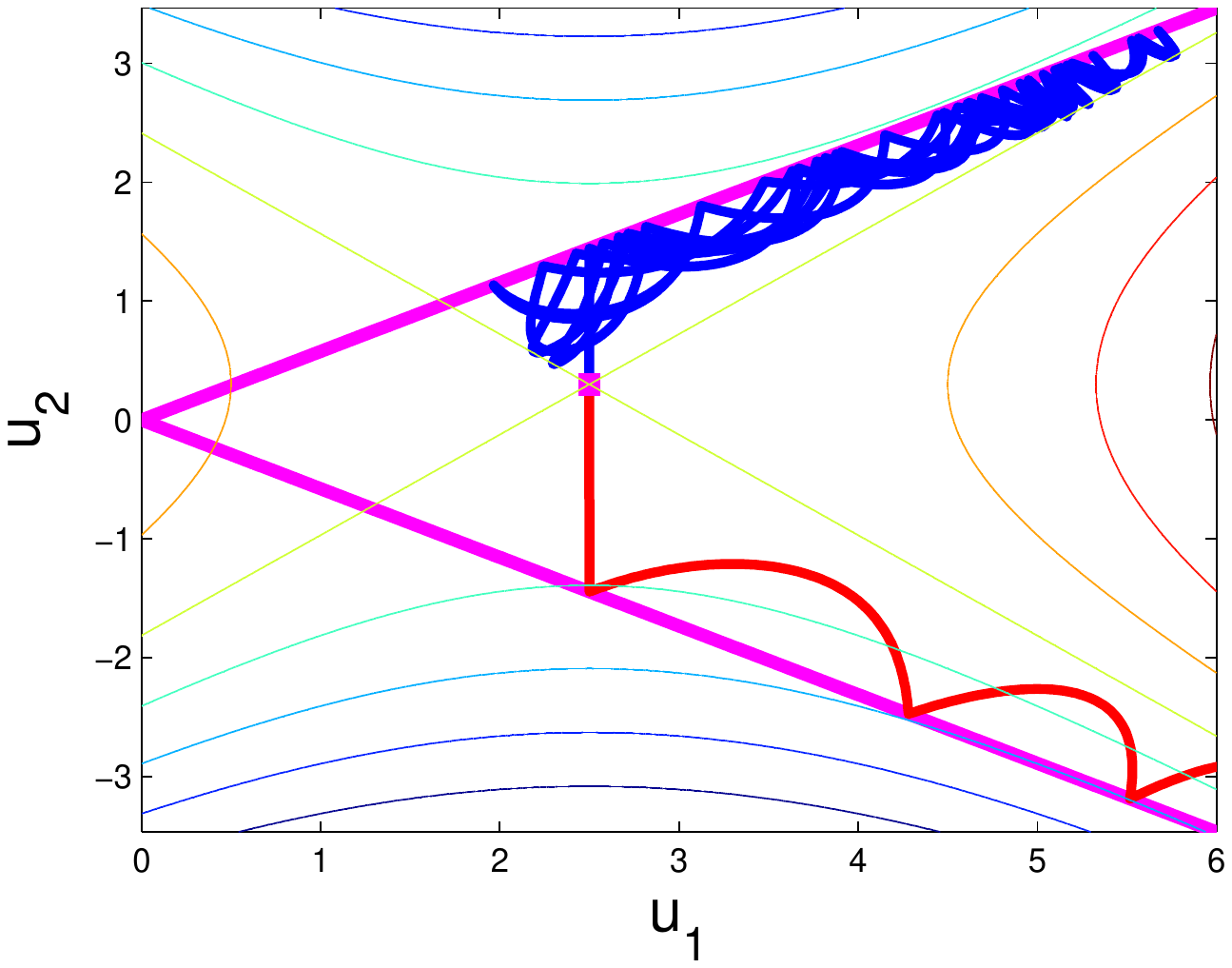}
\includegraphics[width=0.3\textwidth]{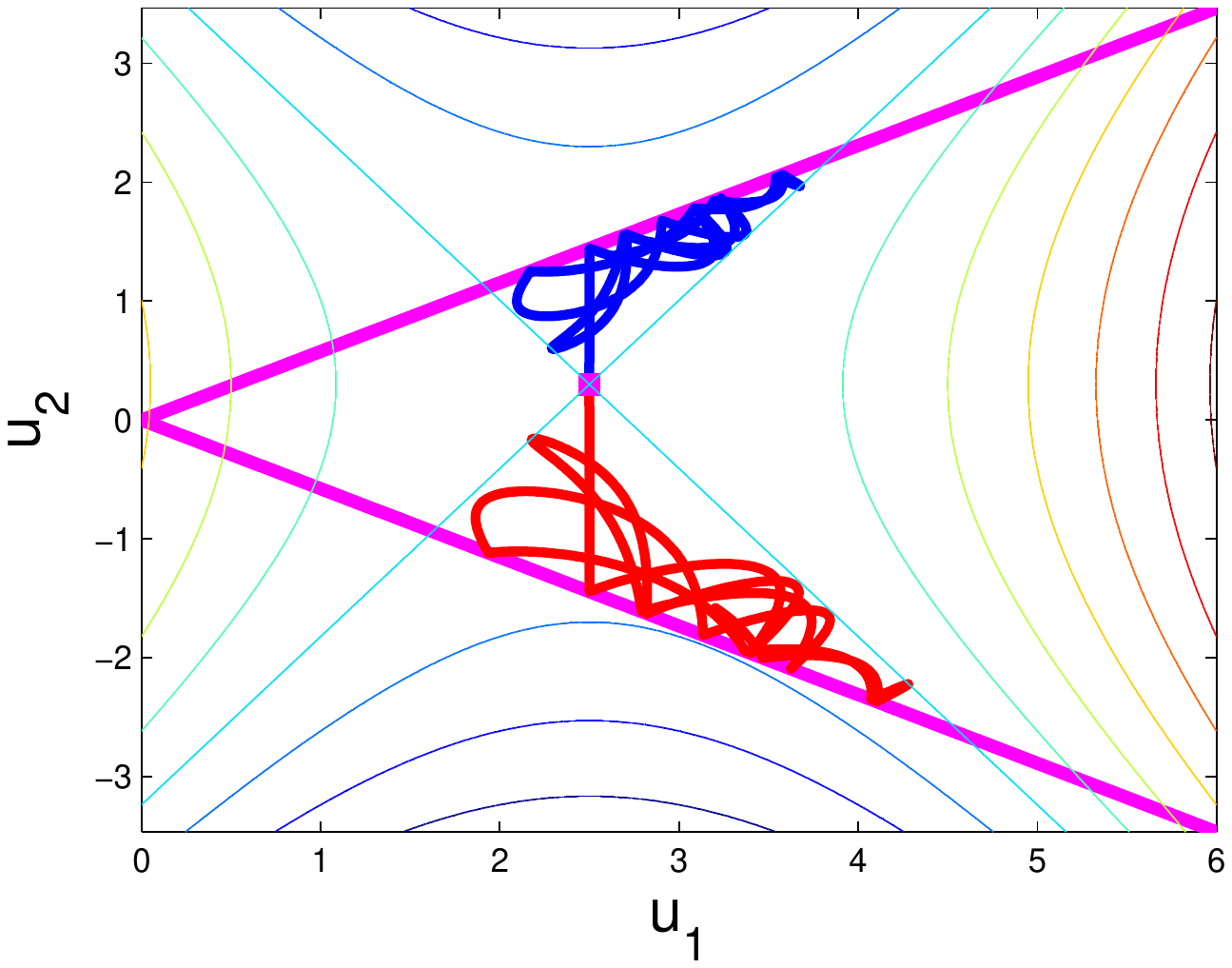}
\end{center}
\begin{center}
\includegraphics[width=0.3\textwidth]{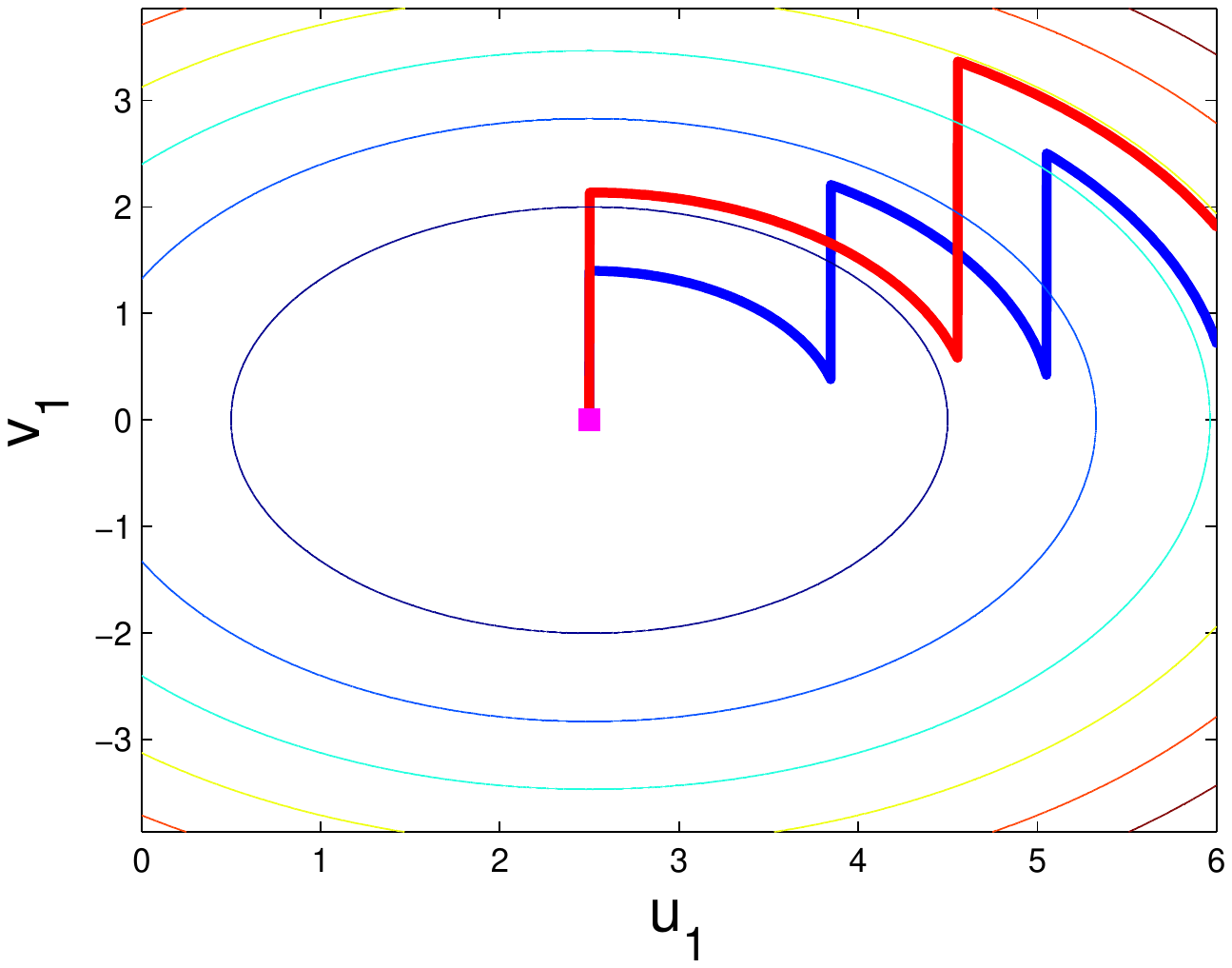}
\includegraphics[width=0.3\textwidth]{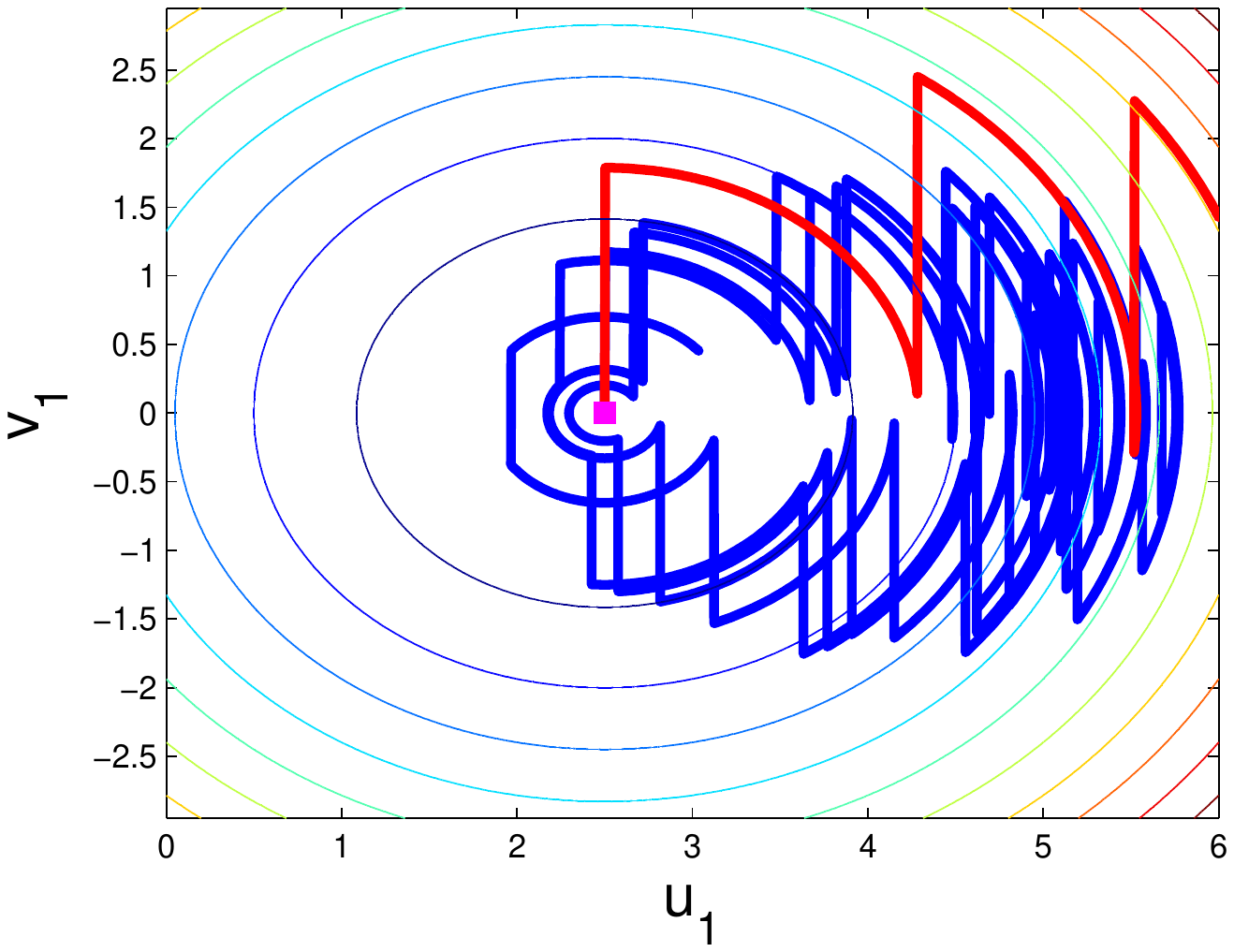}
\includegraphics[width=0.3\textwidth]{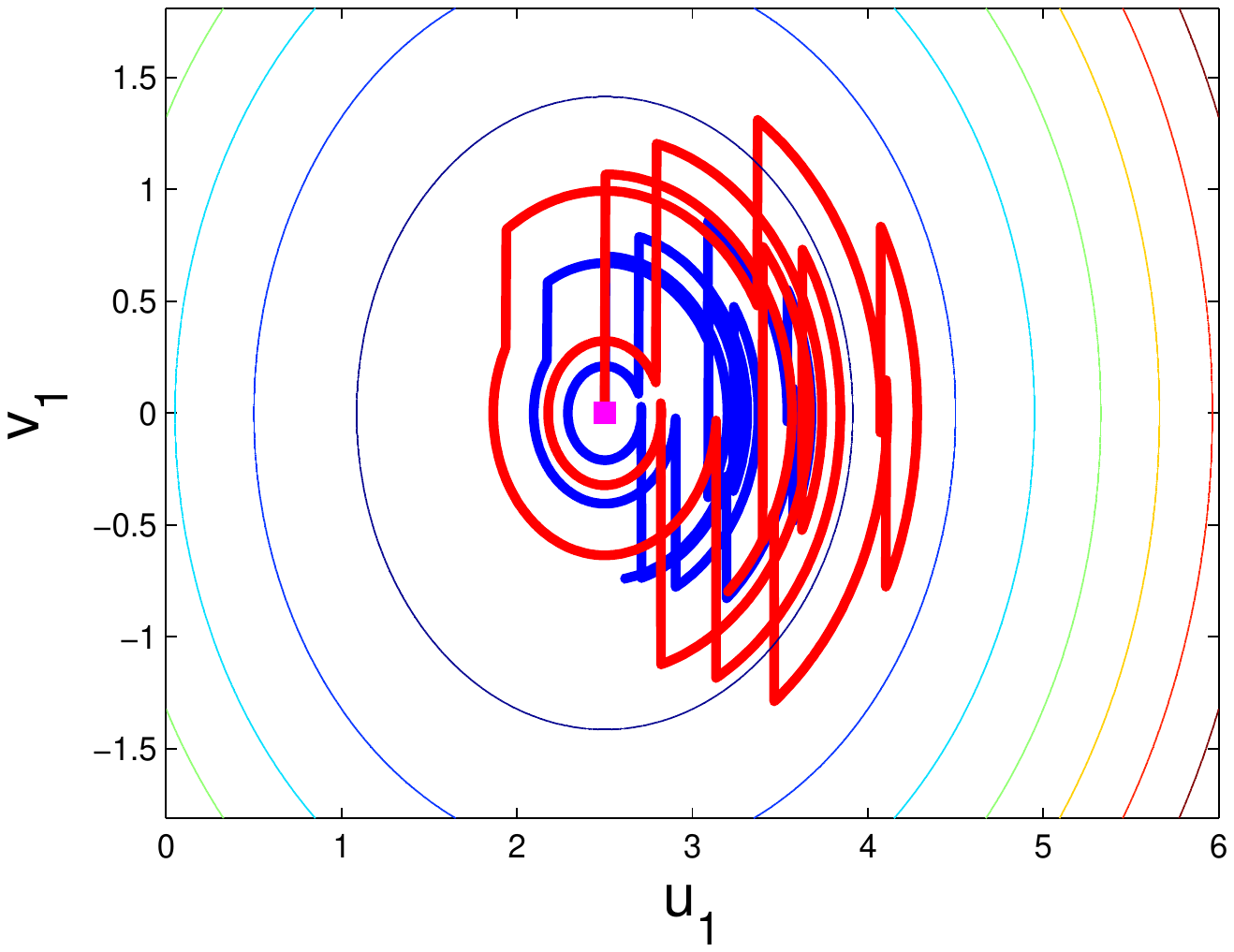}
\end{center}
\begin{center}
\includegraphics[width=0.3\textwidth]{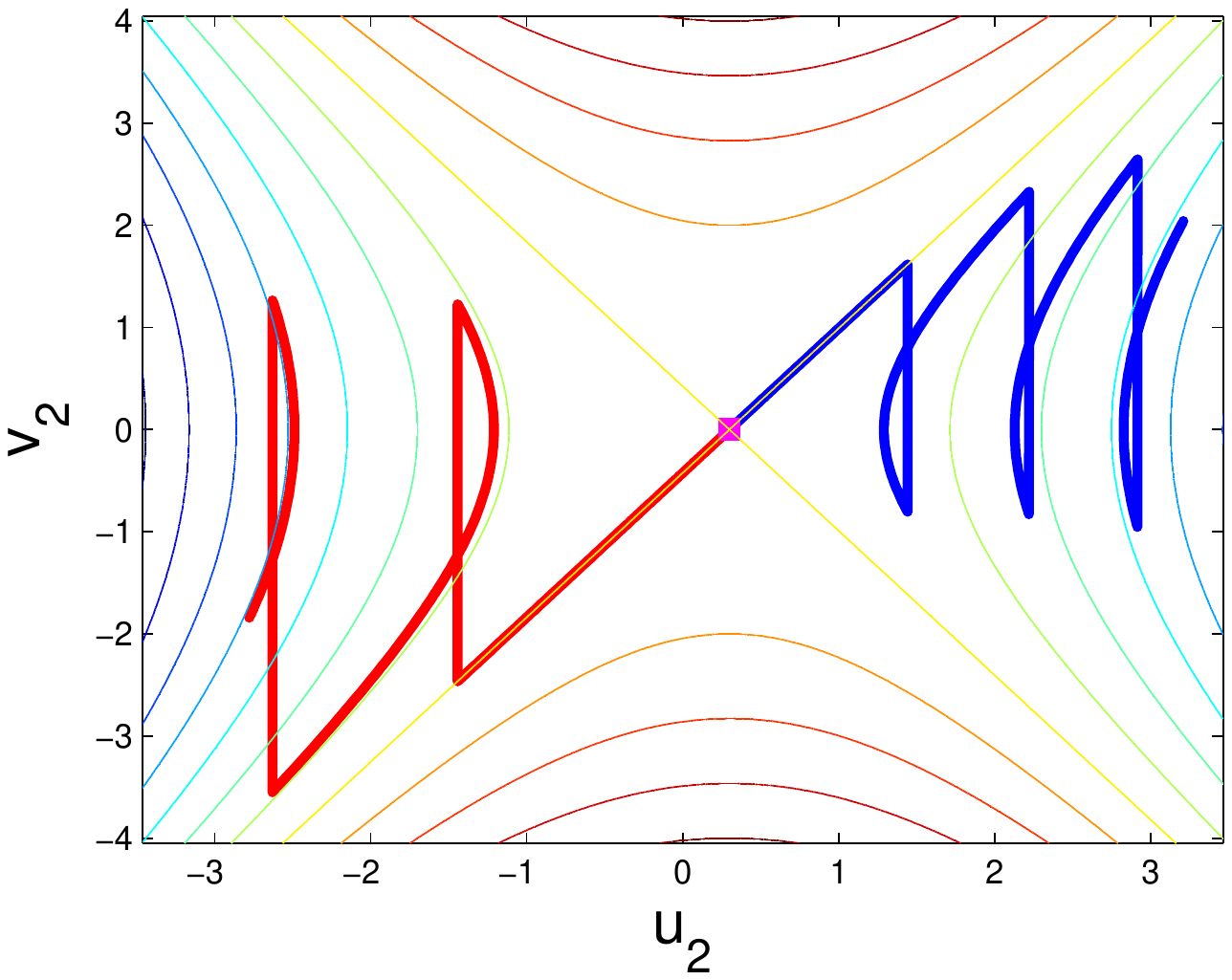}
\includegraphics[width=0.3\textwidth]{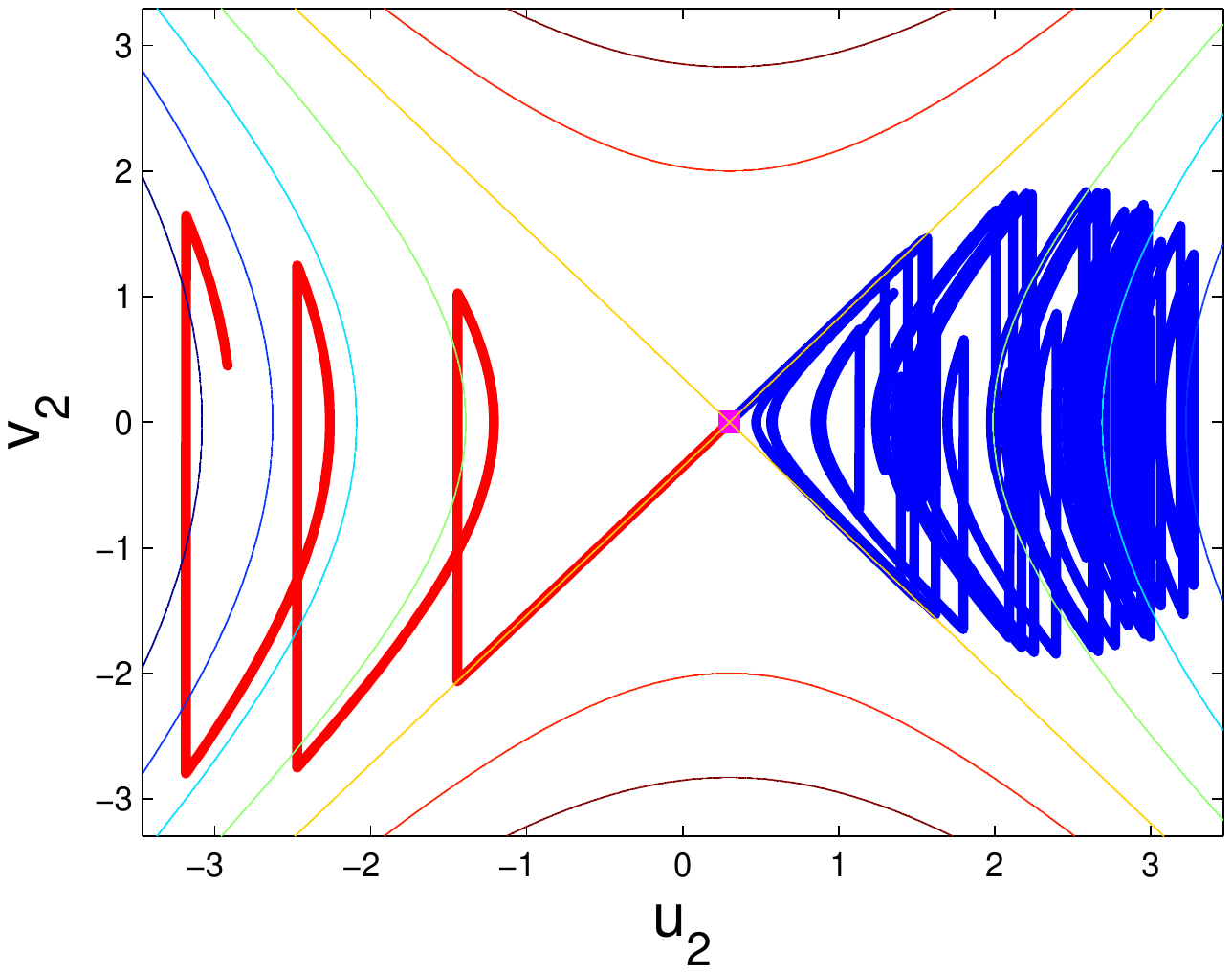}
\includegraphics[width=0.3\textwidth]{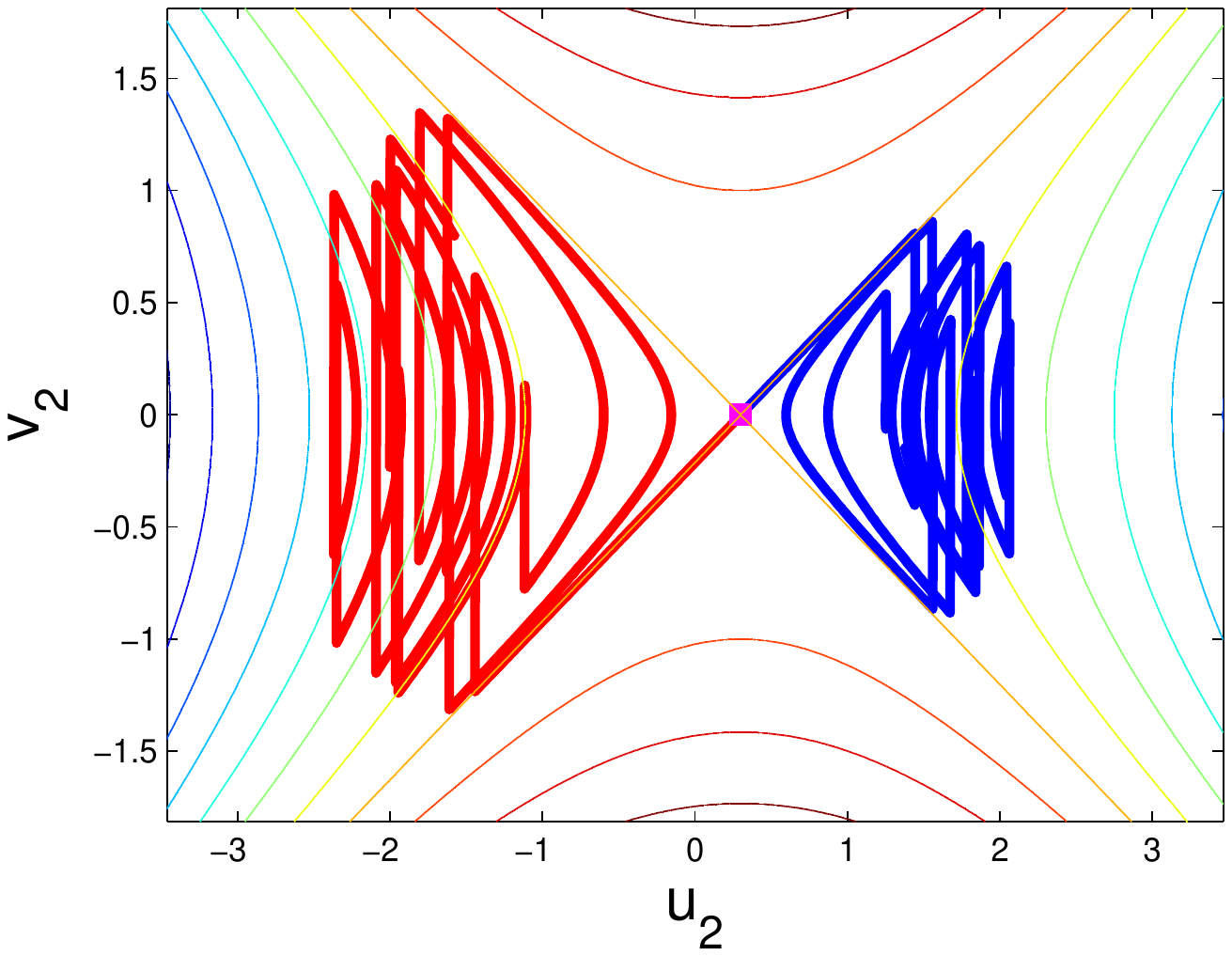}
\end{center}\caption{%\begin{footnotesize}
Simple and complicated dynamics for the limit system (\(\varepsilon=c=0\)).
The projections of $W^{u}_+(P)$ (blue) and $W^{u}_-(P)$ (red) to the configuration space (top), \((u_{1},v_{1}) \) space (middle) and \((u_{2},v_{2}) \) space (bottom) are presented for three different \(\lambda\) values: left, middle and right columns have \(\lambda^2=2.,1.4,0.5\) respectively. In all figures \(\omega^2  =1,u_s=(2.5,0.3),\beta=\pi/3,\theta=-\pi/6 \) and \(L=6\). The colored lines show the level curves of the quadratic potential \(V_{a}(u)\)  (top), the center constant \(I_{1}(u_1,v_1)\) (middle) and the hyperbolic constant \(D_{2}(u_2,v_2)\) (bottom). The magenta rays on the top panel indicate the impact surfaces and the magenta squares indicated the location of the stagnation point \((u_s,v_{s}=0)\).
%\end{footnotesize}
}%
\label{fig:bildynam}%
\end{figure}

The behavior near the lower branches of the stable and unstable
manifolds  of \(P\) was shown to be integrable when $\theta=\varepsilon=c=0$ and non-integrable for small non-zero $\theta$. Indeed,  we proved that the stable and unstable manifolds of \(\gamma_{h}\) intersect for a range of energy values, $h\in
(h^{0}\approx\frac{(\lambda u_{2,s} \tan \theta)^2}{2},h_{crit-\gamma})$.
Similar results apply
for the upper branches when
\((\beta+\theta)\) is small.

 Proving  analogous results regarding integrability or non-integrability of the dynamics near \(P\) for general
    \((\beta,u_{s},L,\theta,{\omega},{\lambda})\) is a difficult problem.  Rather then seeking the
general solution, we identify cases in which the nature of the
dynamics can be roughly predicted. More precisely, we classify the behavior of the upper/lower branches of the manifolds  as follows:
\begin{description}
\item[SD]\textbf{(Simple dynamics)} For small \(h\),  the upper (respectively lower) branches of the unstable and stable
manifolds are reflected to the regions of ``no-return" monotonically (with a monotone sequence of reflecting points exiting the corner region)  and thus, near these branches
there are no recurrent trajectories in the neighborhood of  $P$.
\item[PCBD (Possibly complicated bounded dynamics)] For small  \(h\) values the upper (resp. lower) branches of the manifolds are trapped in the upper (resp. lower) corner region: these manifolds cannot exit the corner region from the product (resp. reactant) channel.
\item[CD(Chaotic dynamics)] The branches of the manifolds intersect each other for a range of energy
levels that are close to the barrier energy.
\end{description}

The first and second rough criteria  check whether the manifolds reflect in a simple, monotone way out of the corner region or whether they are trapped. Clearly simple dynamics imply that no chaotic behavior is possible for the corresponding branches. Folding back of the manifolds is a necessary criterion for the creation of  homoclinic tangles\footnote{
Trapping of the manifolds implies the folding back of them but the opposite implication is not true in general - the lobes of homoclinic tangles may extend to infinity. }, yet it is not sufficient, as the special case $\varepsilon=\theta=0$  shows. Hence we call this behavior possibly complicated.

Notice that when both the upper and lower branches exhibit SD the linear structure near the saddle point governs the motion. On the other hand, when one set has CD and the other SD  we  have the ``open chaos" scenario in the reaction region. In particular, then the reactant/product region is strongly asymmetric.
When both branches intersect we have the classical double loop homoclinic tangles. If additionally the PCBD conditions are satisfied, the chaotic motion associated with this homoclinic tangle is limited to the reaction region (and then the implications regarding scattering have yet to be explored).

\begin{figure}[t]
\begin{center}
\includegraphics[width=0.5\textwidth]{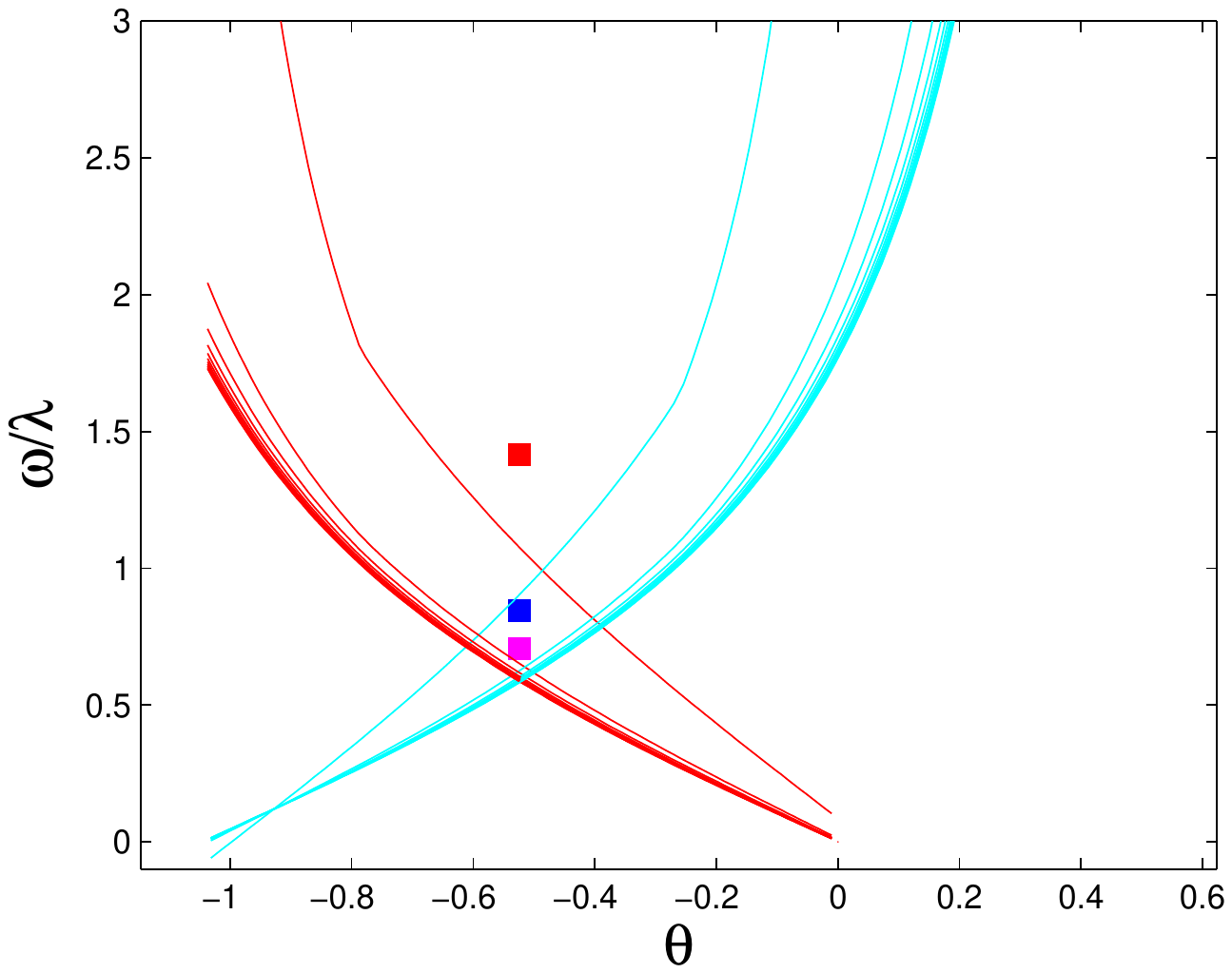}
\end{center}
\caption{
%\begin{footnotesize}
The dependence of the critical values of \(\omega/\lambda\) on \(\theta\) and \(L \). The critical values  \(\delta^{\pm}_c\) (see Theorems \ref{thm:compupper},\ref{thm:complower}) for the upper (blue) and lower (red) branches of the manifolds are plotted for increasing \(L\) values. When     \(\omega/\lambda>\delta^{\pm}\)
the upper/lower branches of \(P\) are trapped in the corner region. Here, \(\beta=\pi/3, L=6,26,46,..,206, \ u_s=(2.5,0.3)\). The convergence at large \(L\) values is apparent. Simple behavior is expected to appear at small
\(\omega/\lambda\), below both critical curves. The values of \(\omega/\lambda\) for the three $\lambda$ values of Fig. \ref{fig:bildynam} are indicated by squares (lowest,middle,upper squares correspond to left,middle,right columns of Fig  \ref{fig:bildynam}).
%\end{footnotesize}
}%
\label{fig:deltacrit}%
\end{figure}

Next, we show that there are some regions in the parameter space where we are able to determine that SD occurs and others where PCBD occurs. Figure \ref{fig:deltacrit} summarizes some of these results graphically.

First, we notice that independent of the saddle eigenvalues (i.e. of \({\omega},{\lambda}\)), when the upper and/or lower branches of \(W^{u,s}_{\pm}(P)\) do not reflect from the upper (respectively lower) boundary  ray we have simple dynamics:
\begin{theorem}  [Simple Dynamics I]\label{thm:sd1} For sufficiently small \(\epsilon,c\), if $\beta+\theta \ge \pi/2$ or if \(u_{1,s}\tan(\beta+\theta)>L, \) there exists \(h^{**}(u_{1,s},\beta+\theta,L;\epsilon,c)>0,\) such that for \(h\in[0,h^{**})\), $ W^{u,s}_+(\gamma_h)$, the upper  branches of  the stable and unstable manifolds of the Lyapunov orbit \(\gamma_h\) exit the corner region without intersecting each other. If  \(u_{1,s}\tan(\theta)<-L  \),  for   \(h<h^{**}(u_{1,s},-\theta,L;\epsilon,c)\) the same statement applies to  the lower branches,   $ W^{u,s}_-(\gamma_h)$.
\end{theorem}
{\bf Proof}: Consider the \(\epsilon=c=0 \) case. Here the projection onto the configuration space of the extensions of $ W^{u,s}_+(P) $ is a straight vertical segment that intersects the upper boundary of \( \mathcal A_{L}\) at \((u_{1},u_2)=(u_{1,s},L)\). The corresponding phase space intersection points of \(W^{u,s}_+(P)\) with the three dimensional cross section \(Q=\{(u,v)|u_{2}=L\}\) are  \( (u_{1},v_{1},u_2,v_{2}^{u,s})=(u_{1,s},0,L,\pm\lambda |L-u_{2,s}|)\), so, these points are well separated in phase space. Similarly, if \(h\) is not too large (see below), the traces of the stable and unstable manifolds of \(\gamma_h\) on \(Q\) are two planar ellipses, parallel to the \((u_{1},v_{1}) \) plane and separated by the finite distance \(2\lambda |L-u_{2,s}|\) along the \(v_{2}\) axis, so they do not intersect (this is simply the linear behavior). The restriction on \(h\) appears from the requirement that the manifolds should not hit the upper ray of the corner for any \(u_{2}\leqslant L\). An explicit bound on \(h  \) may be thus easily found from setting \(I_1=h \) and \(v_{1}=0\) in eq. (\ref{eq:i1def}). If $\beta+\theta \ge \pi/2$ then \(h<h_{obt}^{**}=\frac{\omega^2   }{2}(u_{1,s})^2\) (so that \(\gamma_{h}\) does not hit the upper ray). If \(u_{1,s}\tan(\beta+\theta)>L, \) we require \(u_1(t)<L/\tan(\beta+\theta)\) for all \(t\), hence  \(h<h_{acute}^{**}=\frac{\omega^2   }{2}(\frac{L}{\tan(\beta+\theta)}-u_{1,s})^2\). The same arguments apply to the lower branches when \(\theta<0\) (replacing \(L\) by \(-L\) and \((\beta+\theta)\) by \(\theta\)). These results are concerned with robust properties of trajectories within the corner region (with no impacts), hence, by the smooth dependence of trajectories within the corner region on parameters these are clearly  true for sufficiently small \(\epsilon,c\).
\hfill $\blacksquare$

\bigskip

The second geometrical observation is that the projection of  $W^{u,s}_+(\gamma_{h})$ onto the configuration space lies within the Hill region's of the energy level \(h\). In particular, for \(P=\gamma_0\), this Hill region boundaries are the potential zero-level lines \(V_a(u)=0\) together with the corresponding boundaries of the corner region \( \mathcal A_{L}\) (see below). We then notice that for sufficiently large \(\frac{\omega}{\lambda}\) these zero potential level lines always intersect the corners' boundary at \(u_{1}<L\), namely that  $W^{u,s}_+(\gamma_{h})$ must fold back inside \( \mathcal A_{L}\). We thus find the critical value of  \(\frac{\omega}{\lambda}\) above which the folding occurs:
\begin{theorem}   [Possibly Complicated Bounded Dynamics I] \label{thm:compupper} For any given geometrical parameters   \((u_{s},\beta+\theta,L)\) with  \(L>\tan(\beta+\theta)u_{1,s}\) and \(\beta+\theta<\pi/2\),  there exists \(\delta_{c}^+>\tan(\theta+\beta)\) as specified below, such that for all \(\frac{\omega}{\lambda}>\delta_{c}^+\),   for sufficiently small \(h\) and \((\varepsilon,c)\), the upper  branches of  the stable and unstable manifolds of the Lyapunov periodic orbit \(\gamma_h\) ($ W^{u,s}_+({\gamma_h})$) do not exit the corner region through the product channel (i.e. with \(u_{2}>u_{2,s}\)).
\end{theorem}
{\bf Proof}: Consider the region bounded by the corner and the upper segments of the zero level lines of the quadratic potential:\begin{equation}
\ \mathcal{C^{+}}=\mathcal{A}_L\cap\left\{ (u_1,u_2)|(u_2-u_{s,2})\geq\ \frac{\omega}{\lambda}|u_1-u_{s,1}|  \right\}.
\end{equation}
\(\mathcal{C^{+}}\) is  the Hill region for the energy level \(h=0\) for orbits of the limit system belonging to  $ W^{u,s}_+({\gamma_0}=P)$; since the kinetic energy must be non-negative, orbits belonging to the energy surface \(h=0\) must satisfy for all time \(V(u(t))\leq0\). Hence,at  \(\epsilon=c=0\),  $ W^{u,s}_+(P)$  is indeed trapped in \( \mathcal{C^{+}}\). Similarly, orbits belonging to   $ W^{u,s}_+({\gamma_h})$ for \(h>0 \) are restricted to reside in the configuration space region \(\mathcal{C}^{+}_h\)  at which \(V(u(t))\leq h\), and initially (eventually for the stable manifold) also satisfy  \(u_2 >u_{2,s} \).

The shape of the region \(\mathcal{C^{+}}\) depends on the geometrical parameters. When  \( \tan( \theta+\beta)<\frac{\omega}{\lambda} \), the upper ray of the corner $u_2 = u_1\tan(\beta+\theta)$ intersects the upper zero-potential energy level line
 $\{(u_{1},u_{2})|u_2 - u_{2,s} = \frac{\omega}{\lambda}(u_1- u_{1,s}),u_{2}>u_{2,s}\}$ at the vertex point \((u_1^+, u_2^+)\) with
  $$
  u_1^+ = \frac{u_{2,s}-\frac{\omega}{\lambda}u_{1,s}}{\tan(\beta+\theta)-\frac{\omega}{\lambda}}=
  u_{1,s} + \frac{\tan(\beta+\theta)u_{1,s}-u_{2,s}}{\frac{\omega}{\lambda}-\tan(\beta+\theta)},\;
  u_2^+ = u_1^+ \tan(\beta+\theta). $$
Let  \(\delta^{+}_{c}=\frac{\omega}{\lambda}\)  be the eigenvalue ratio for which \(L=\max( u_1^{+}, u_2^+)\):
\begin{equation}
\delta^{+}_{c}(L,u_{s},\tan(\beta+\theta))=\left\{\begin{array}{ccc}
 \tan(\beta+\theta)+ \frac{\tan(\beta+\theta)u_{1,s}-u_{2,s}}{L-u_{1,s}} & \text{for} & \beta+\theta<\pi/4\ \\
\tan(\beta+\theta)+ \frac{\tan(\beta+\theta)u_{1,s}-u_{2,s}}{L/\tan(\beta+\theta)-u_{1,s}} & \text{for} & \beta+\theta>\pi/4 \\
\end{array}\right.
\end{equation}
Notice that for  \(\frac{\omega}{\lambda}>\delta^{+}_{c}\)
   the vertex is inside the corner region,(\(\max( u_1^{+}, u_2^+)<L\)). Fig. \ref{fig:deltacrit} presents the typical dependence of \(\delta^{+}_{c}\) on \(\theta\) and \(L\) (blue curves).

 Similarly, for sufficiently small \(h\)\textgreater0, when
 \(\frac{\omega}{\lambda}>\delta^{+}_{c}\), the right boundary of the region   \(\mathcal{C}^{+}_h\)  (the upper part of the \(h\) Hill region), intersects the upper corner boundary at some finite value
\((u_1^*(h),u_2^* (h)= u_1^*(h)\tan(\beta+\theta))\). Then, the shape of the upper part of the Hill region is triangular like, and orbits in its upper part have \(u_1(t)<u_1^*(h)<L\). Thus, for small \(h\), trajectories belonging to  $ W^{u,s}_+({\gamma_h})$ cannot escape the corner region with \(u_2 >u_{2,s} \), namely, generically, the behavior is not simple. Finally,  for sufficiently small  \((\varepsilon,c)\), the restricted Hill regions are deformed into Hill regions with the same basic property:  when
 \(\frac{\omega}{\lambda}>\delta^{+}_{c}\), for sufficiently small \(h\geqslant0,\) for \(u_2 >u_{2,s} \), the \(u_1\) coordinate is limited by some \(u_1(t)<u_1^*(h,\varepsilon,c)<L\).
\hfill $\blacksquare$

 Note that the assumption that \(c\) is small is equivalent to requiring that the far-field terms remain small in \(\mathcal A_{L}\) - otherwise, these far-field terms may indeed change the shape of the Hill regions in  \(\mathcal A_{L}\).

 A similar statement for the lower branches is:
 \begin{theorem} [Possibly Complicated Bounded Dynamics II]\label{thm:complower}
For any given geometrical parameters \((u_{s},\theta<0,L)\) there exists \(\delta_{c}^->-\tan\theta\) as specified below, such that for all \(\frac{\omega}{\lambda}>\delta_{c}^-\),   for sufficiently small \(h\) and \((\varepsilon,c)\), the lower  branches of  the stable and unstable manifolds of the Lyapunov periodic orbit \(\gamma_h\) ($ W^{u,s}_-(\gamma_h)$) do not exit the corner region through the reactant channel (i.e. with \(u_{2}<u_{2,s}\)).
\end{theorem}
{\bf Proof}: As in Theorem \ref{thm:compupper}, we find the lower part of  the Hill region for \(h=\varepsilon=c=0\): \begin{equation}
\ \mathcal{C^{-}}=\mathcal{A}\cap\left\{ (u_1,u_2)|(u_2-u_{s,2})\leq\ \frac{\omega}{\lambda}|u_1-u_{s,1}|  \right\}.
\end{equation}
If $ \tan\theta>-\frac{\omega}{\lambda}$, the lower ray of the corner $u_2 = u_1\tan\theta$ intersects the lower-right zero-potential energy level line  $\{(u_{1},u_{2})|u_2 - u_{2,s} =- \frac{\omega}{\lambda}(u_1- u_{1,s}),u_{2}<u_{2,s}\}$ at the point \((u_1^-, u_2^{-})\) with
  $$
  u_1^- = \frac{u_{2,s}+\frac{\omega}{\lambda}u_{1,s}}{\tan\theta+\frac{\omega}{\lambda}}=
  u_{1,s} + \frac{u_{2,s}-u_{1,s}\tan\theta }{\tan\theta+\frac{\omega}{\lambda}},\; u_2^- = u_1^{-}\tan\theta. $$
  Setting, for \(\theta<0\),
  \begin{equation}
\delta_{c}^{-}(L,u_{s},\tan\theta)=\left\{\begin{array}{ccc}
 -\tan\theta+ \frac{u_{2,s}-\tan\theta u_{1,s}}{L-u_{1,s}} & \text{for} & -\theta<\pi/4\ \\
-\tan\theta+ \frac{u_{2,s}-\tan\theta u_{1,s}}{-L/\tan\theta-u_{1,s}} & \text{for} & -\theta>\pi/4 \\
\end{array}\right.
\end{equation}
we obtain that for all  \(\frac{\omega}{\lambda}>\delta^{-}(L,u_{s},\tan\theta)\)
the lower branches are trapped in the corner region,  see Fig. \ref{fig:deltacrit} for the typical dependence of \(\delta^{-}_{c}\) on \(\theta\) and \(L\) (red curves).
 Repeating the same arguments as for the upper branches (Theorem  \ref{thm:compupper}) the theorem is established.
  \hfill $\blacksquare$

Notice that for \(\theta>0\),  for all \(\frac{\omega}{\lambda}\),
\(u_1^- < \frac{u_{2,s}}{\tan\theta},  \) namely the Hill region is always bounded. Moreover, the manifolds in this case are reflected towards the corner region; Recall that the first hit of the lower boundary of trajectories belonging to  $ W^{u}_-(P)$     is with \(v_1^{-}=0,v_2^{-}<0\), hence, by  (\ref{lrl}),    \(v_1^{+}<0\). We conclude that these trajectories  reflect towards the corner region and hence the flow is not simple for \(\varepsilon =c=h=0\). Using continuity and the regular \(\varepsilon\) limit at regular reflection we see that a similar statement applies to the small \(h,\varepsilon,c\) case. Hence, we conclude:

 \begin{corollary} [Possibly Complicated Bounded Dynamics III]\label{cor:complowerthetneg} When \( \theta>0 \) the lower branches of the manifolds fold back and the flow is possibly complicated.
\end{corollary}

 Notice that when \(\varepsilon>0\), the appearance of closed level sets of the potential function (as established in the above theorems and corollary), implies that these regions also contain a  minimum point of the smooth potential. Thus,  we see that for large \(\omega  /\lambda\), even though the limit system has a single saddle fixed point in the corner, the smooth system has several fixed points and some of them are stable.

\begin{figure}[t]
\begin{center}
\includegraphics[width=0.3\textwidth]{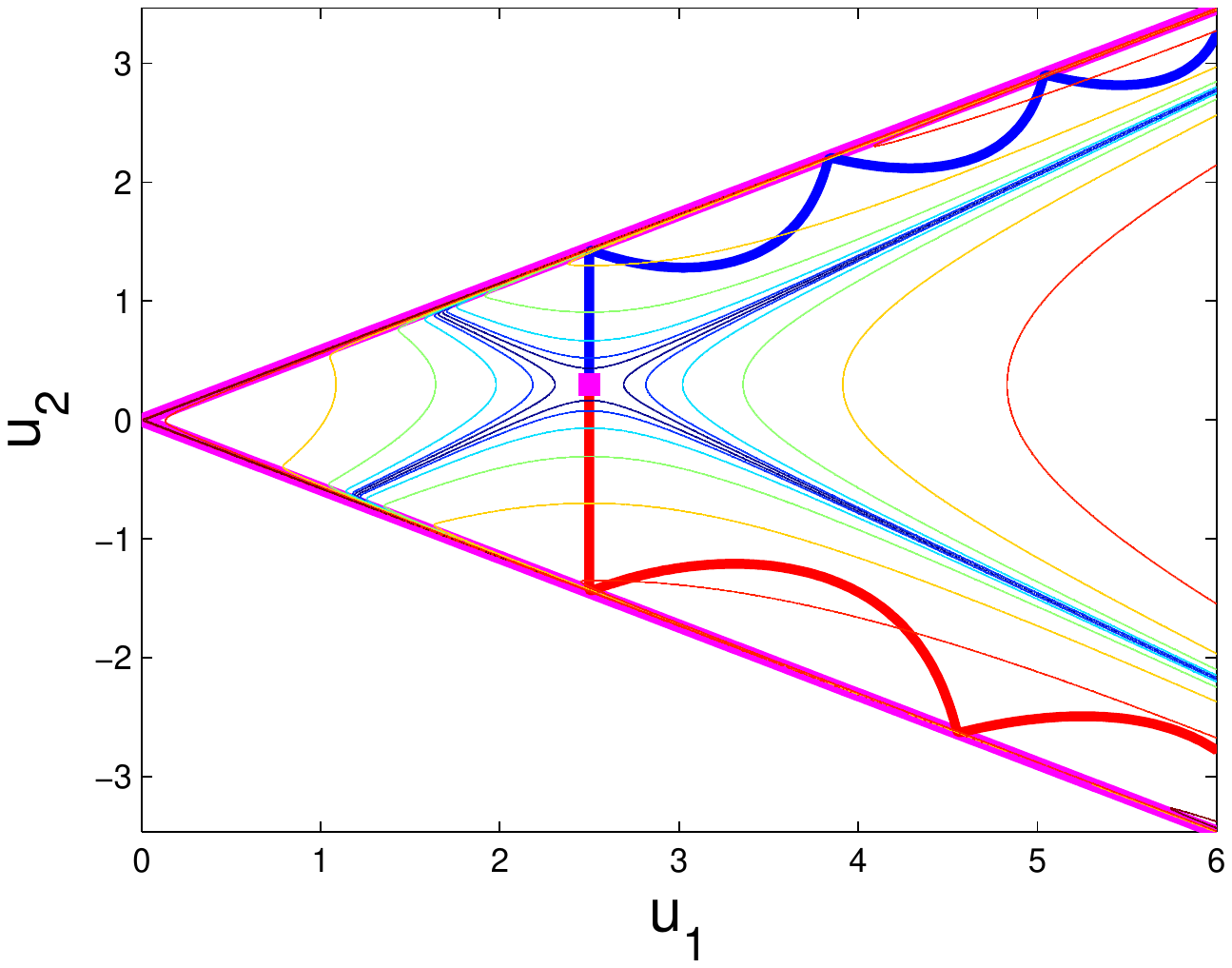}
\includegraphics[width=0.3\textwidth]{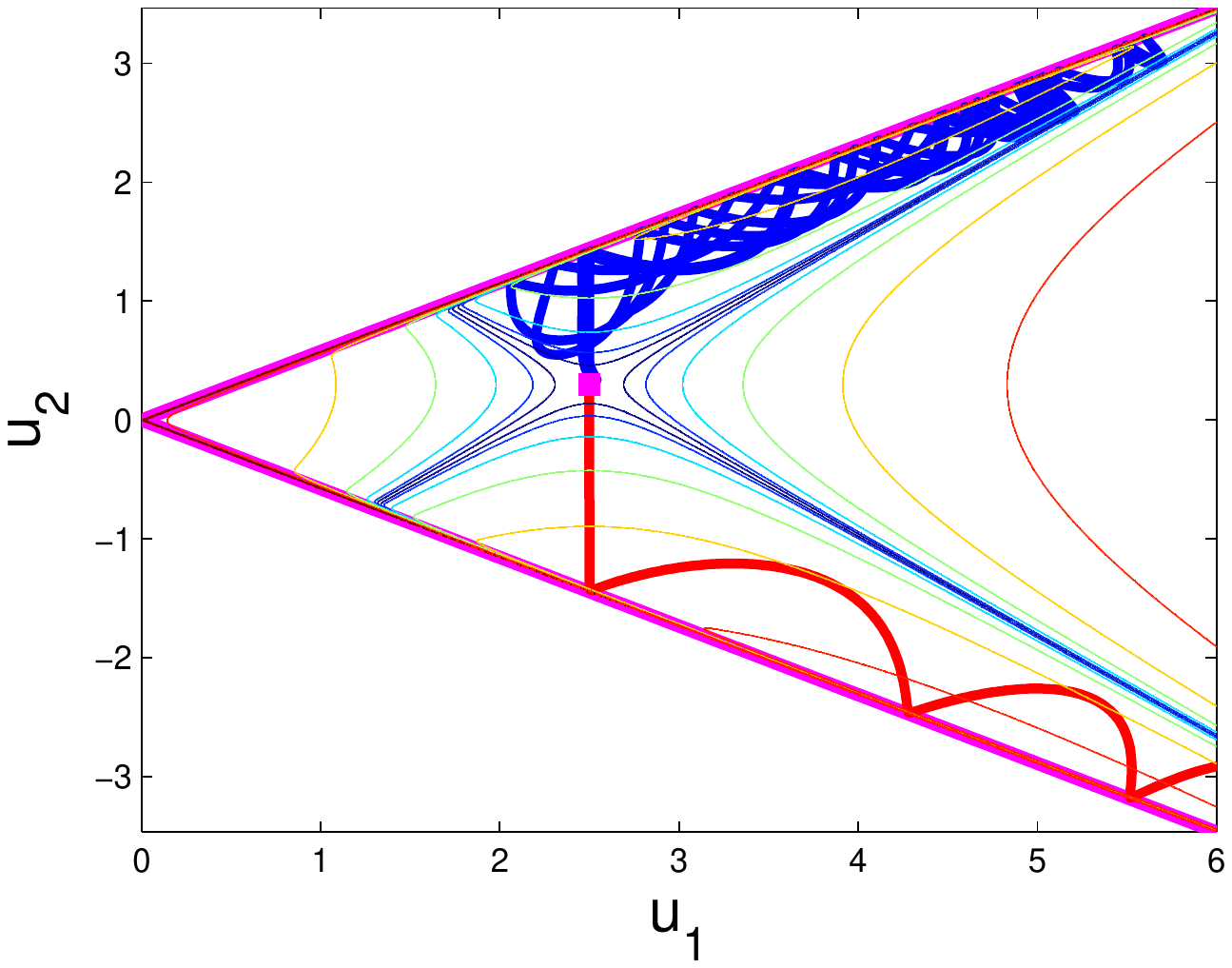}
\includegraphics[width=0.3\textwidth]{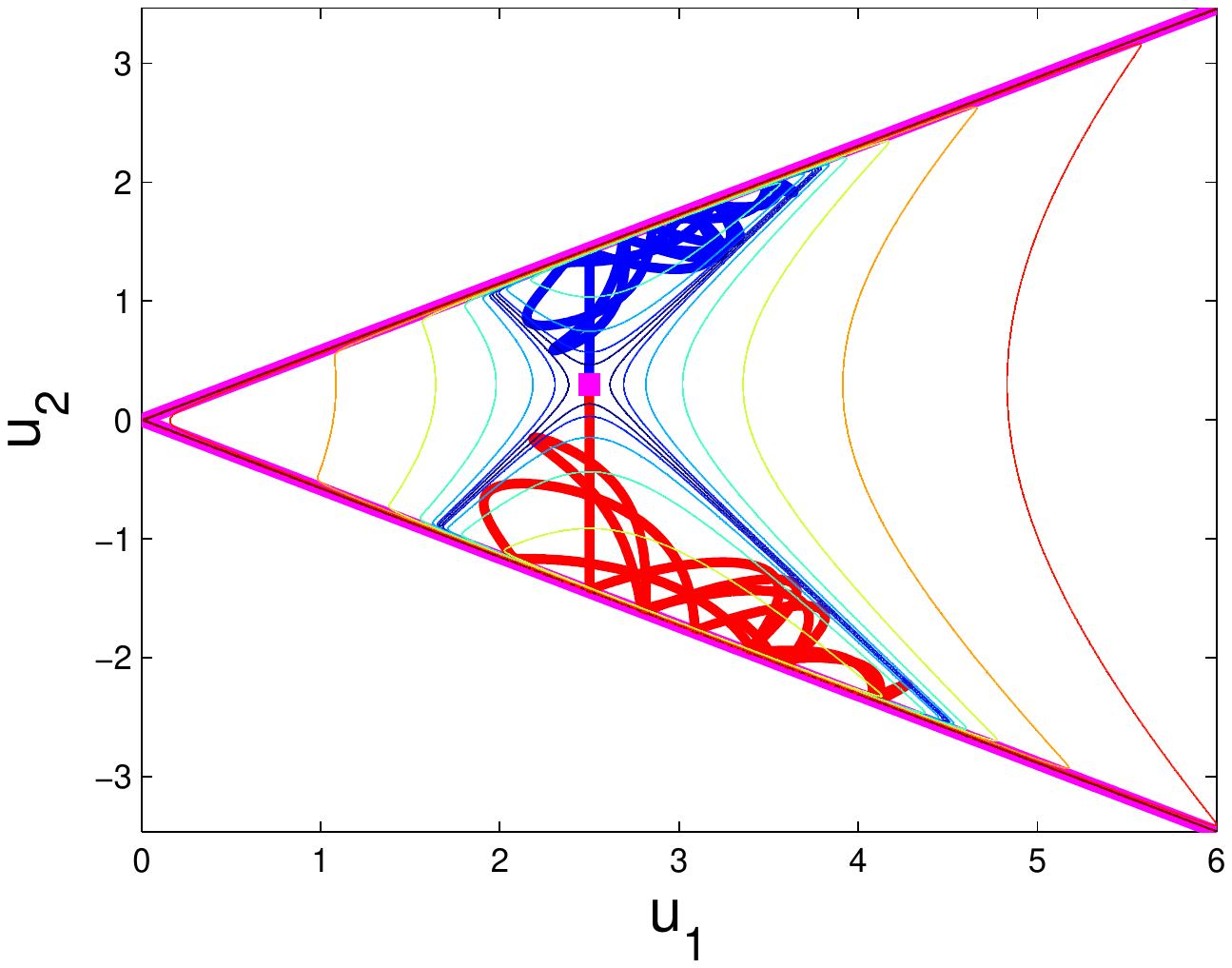}
\end{center}
\begin{center}
\includegraphics[width=0.3\textwidth]{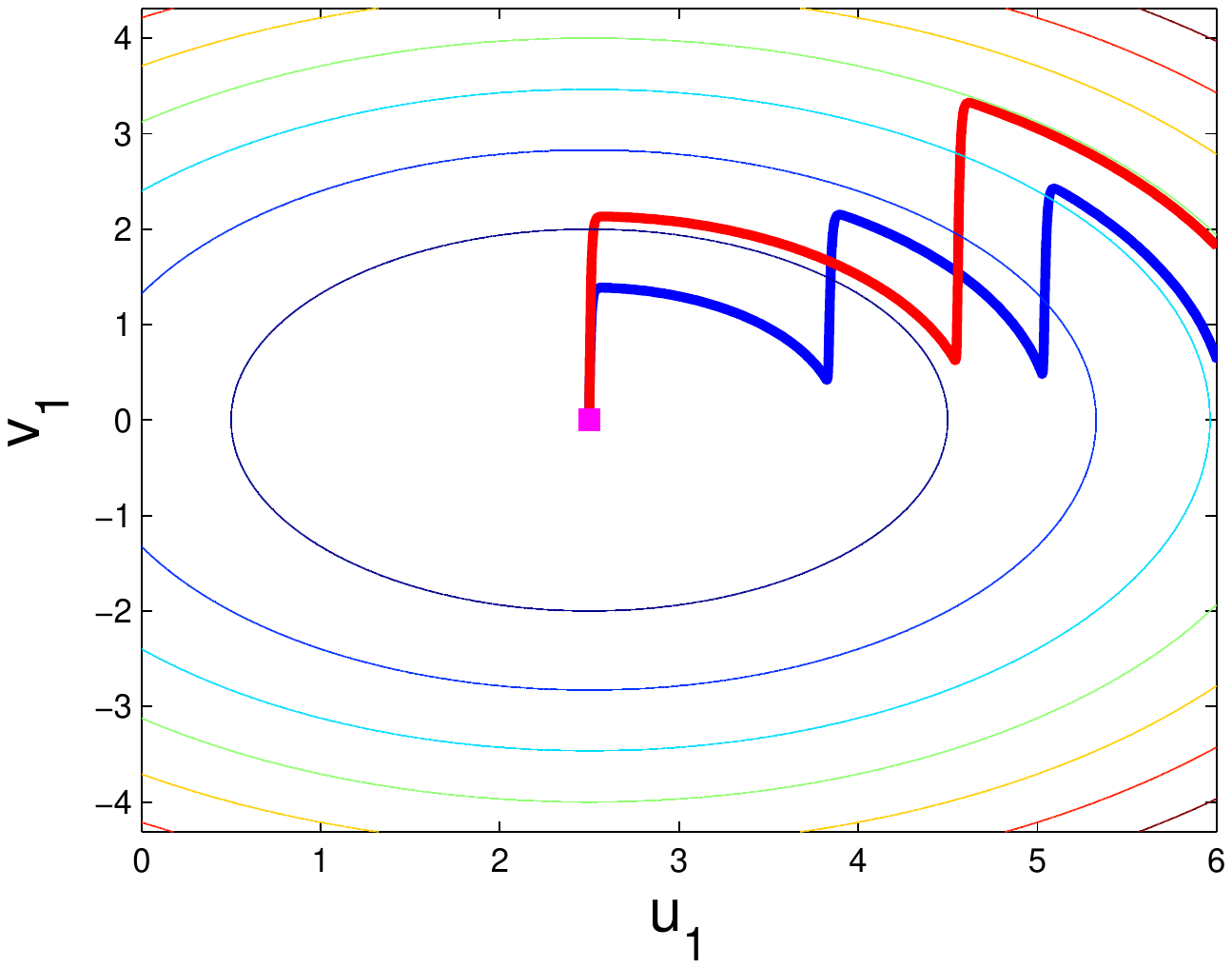}
\includegraphics[width=0.3\textwidth]{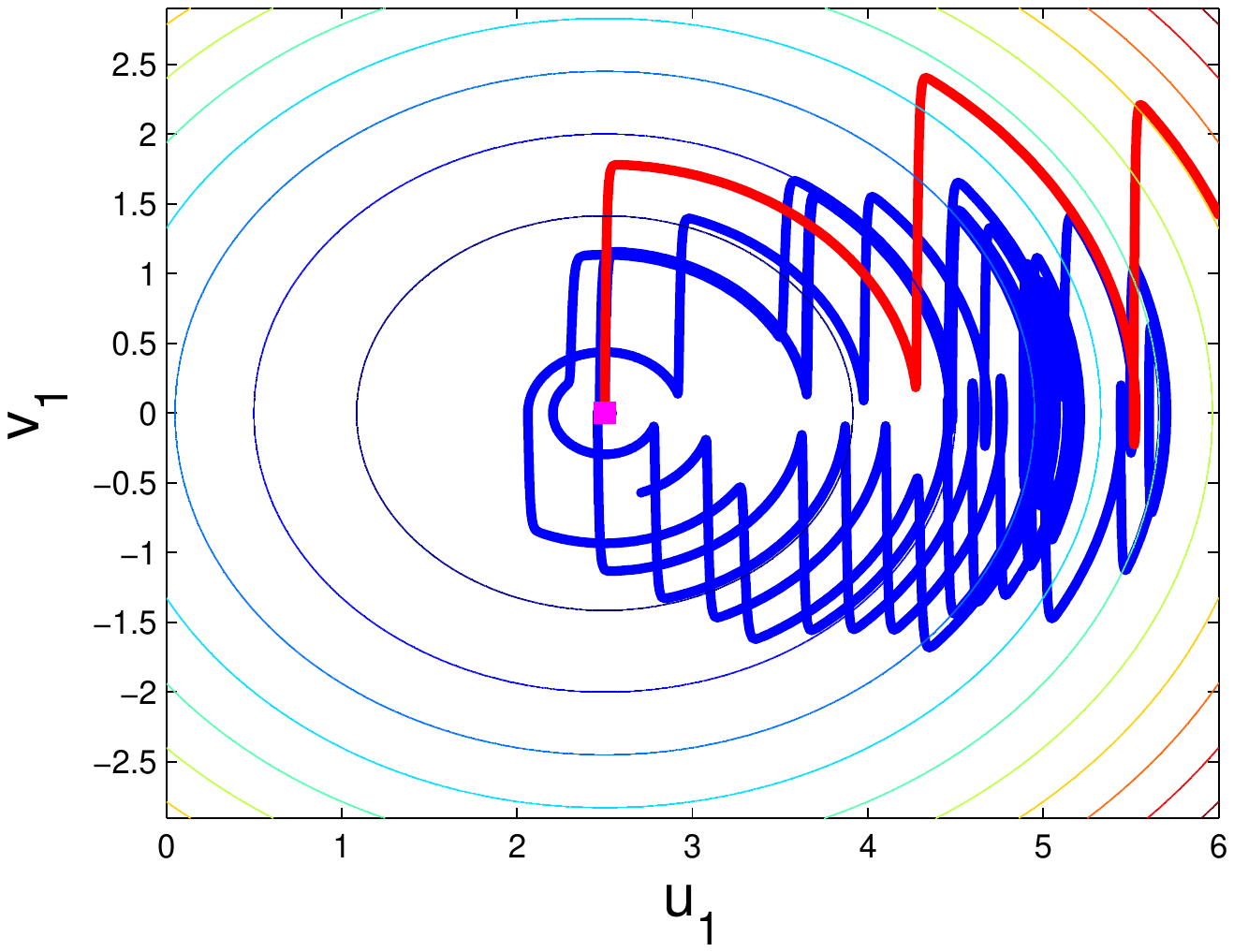}
\includegraphics[width=0.3\textwidth]{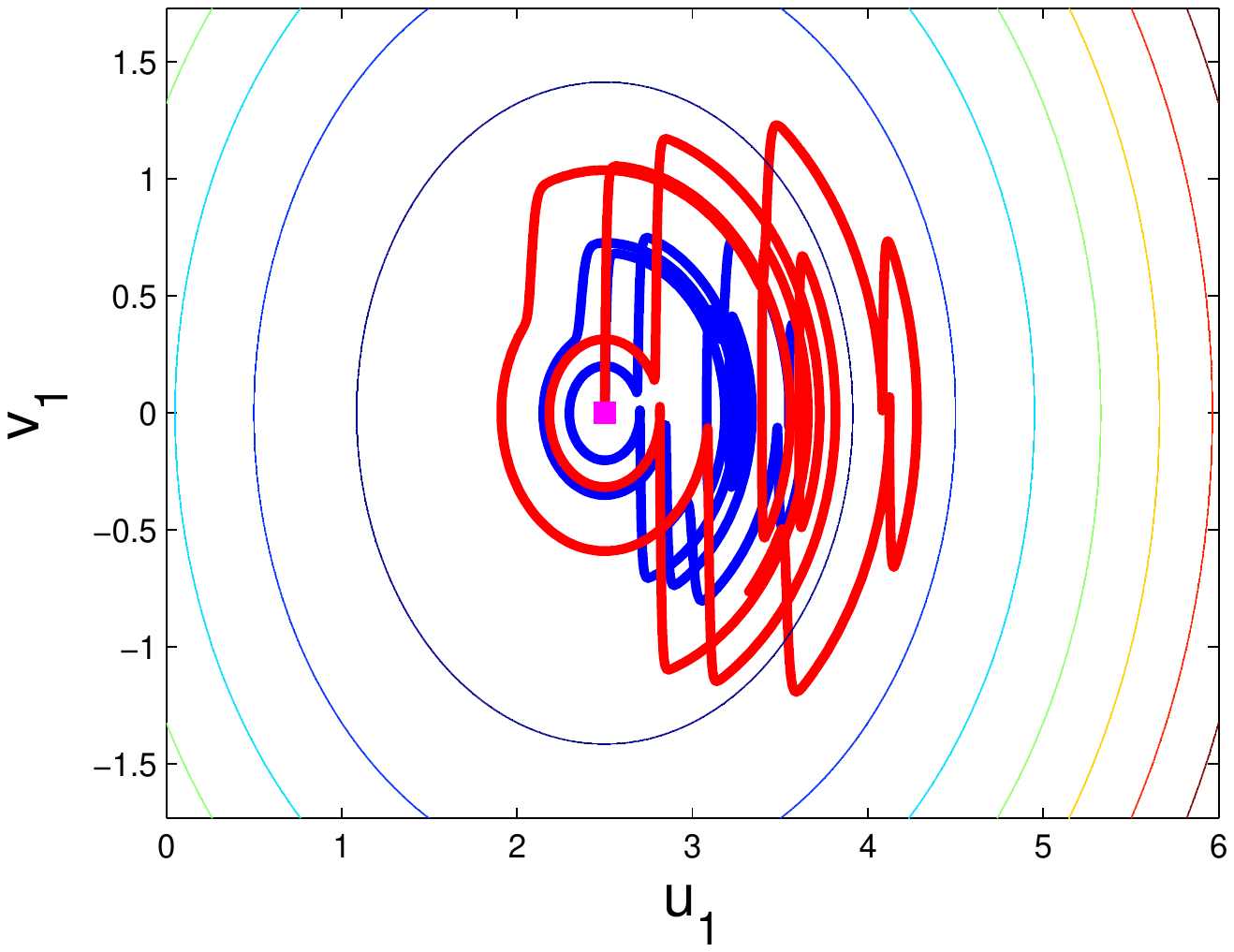}
\end{center}
\begin{center}
\includegraphics[width=0.3\textwidth]{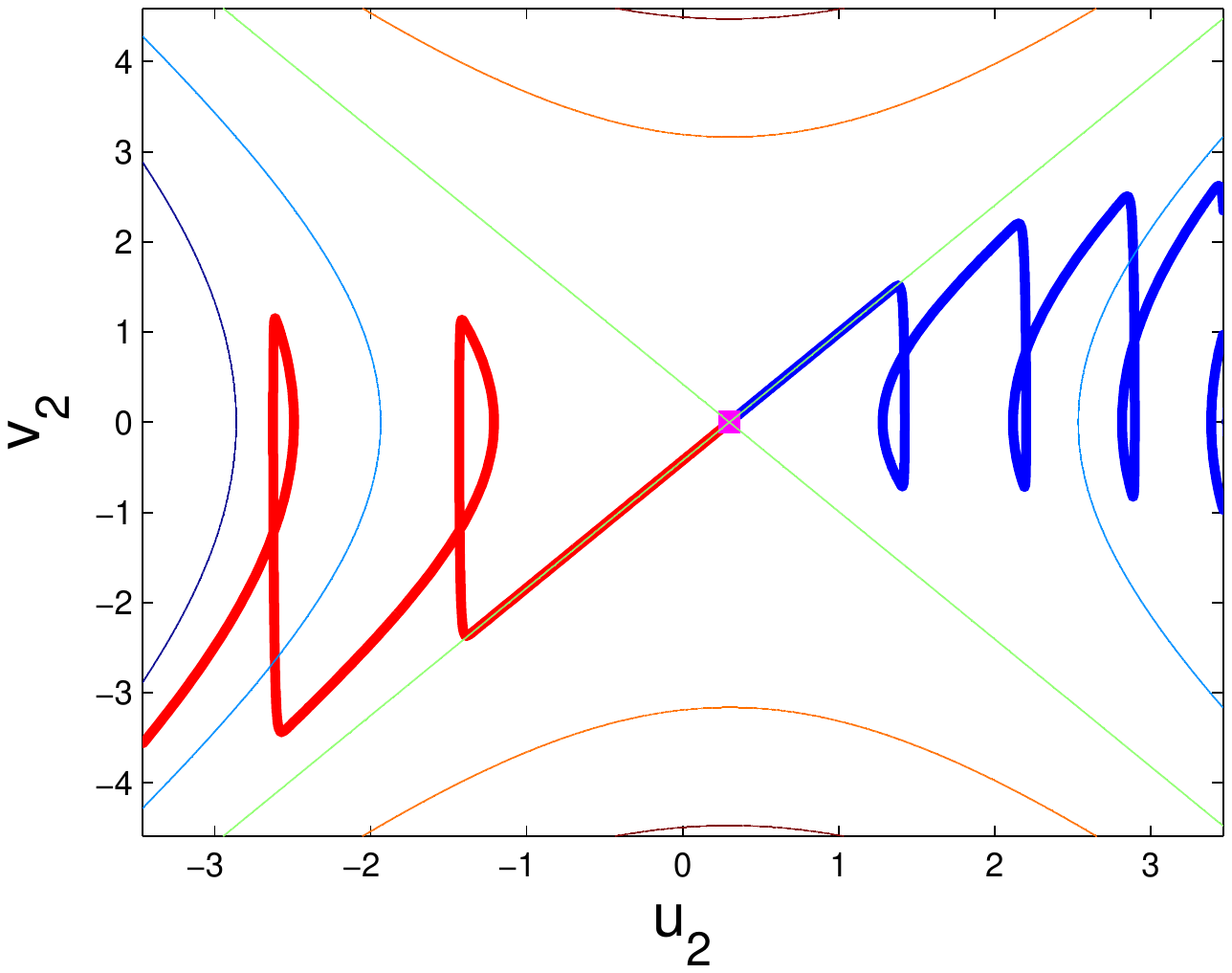}
\includegraphics[width=0.3\textwidth]{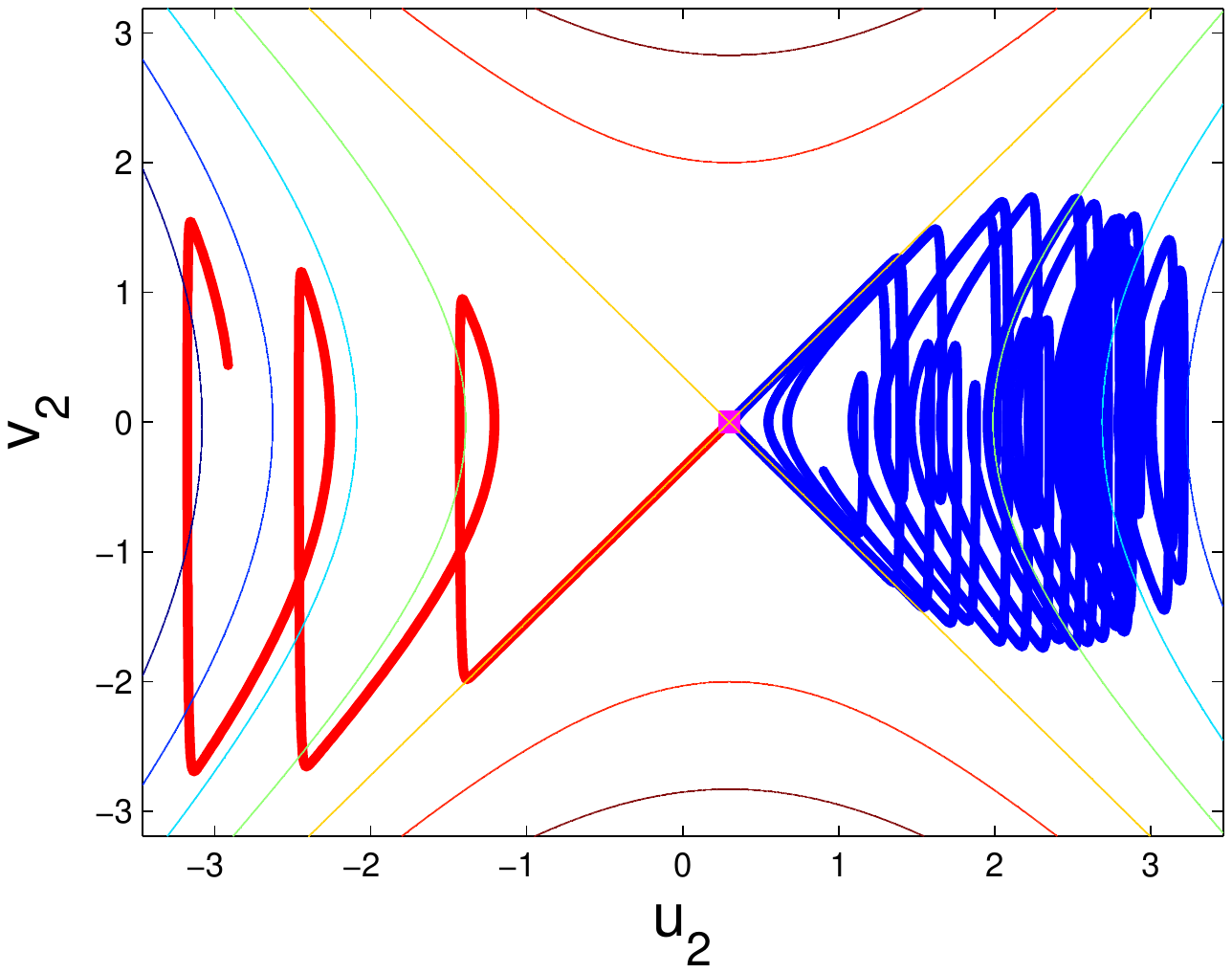}
\includegraphics[width=0.3\textwidth]{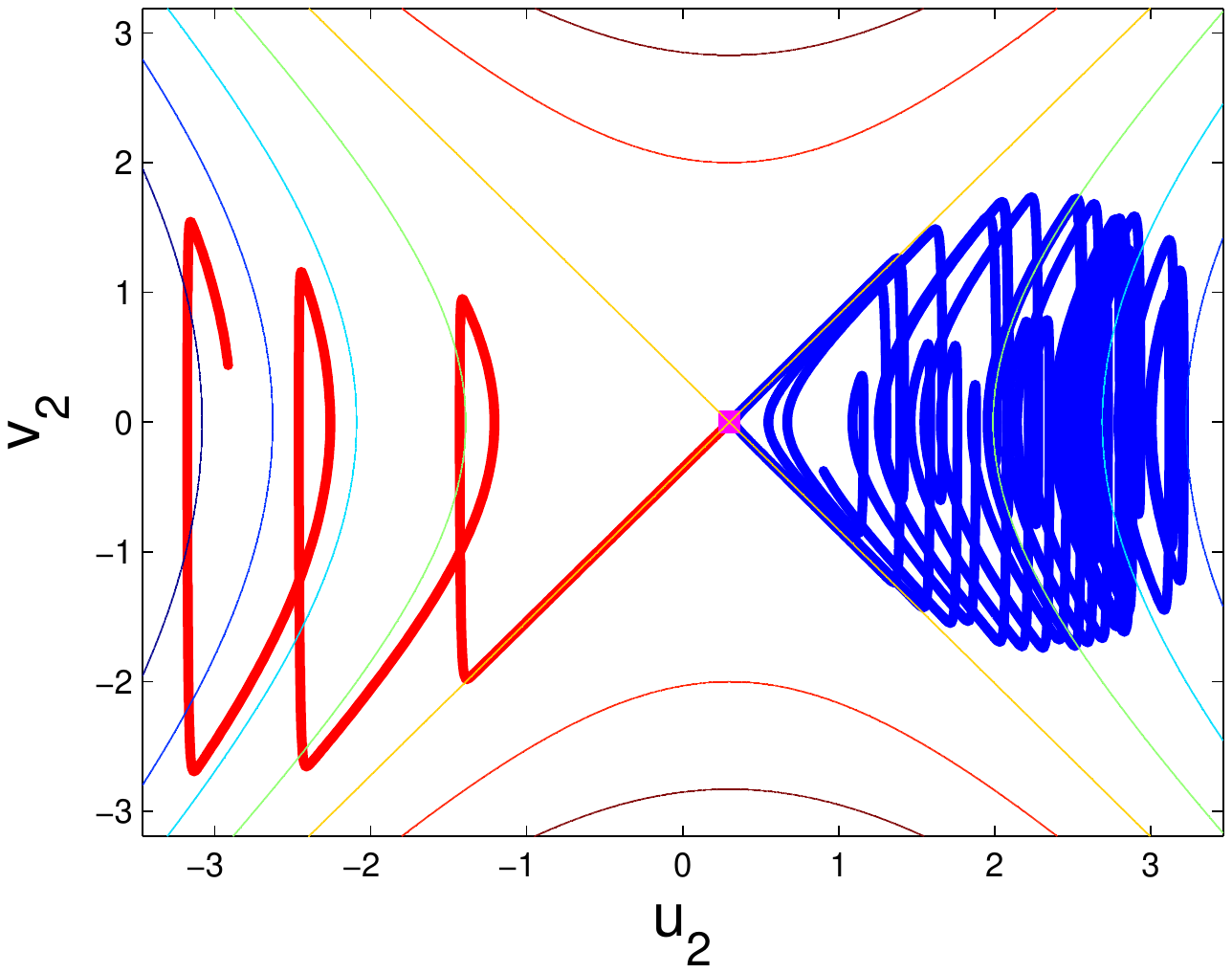}
\end{center}\caption{
%\begin{footnotesize}
Simple and complicated dynamics for the smooth system. The same parameters, initial conditions and projections  of Fig. \ref{fig:bildynam} are used, yet, the impact flow is replaced by a smooth exponential potential along the rays (Eqs. (\ref{eq:linbilham},\ref{eq:exppot})  with  \(b=10,\varepsilon=0.01)\). The colored lines on the top are the level curves of the smooth potential \(V_{a}(q)+bV_b(q,\varepsilon)\).
%\end{footnotesize}
 }%
\label{fig:smdynam}%
\end{figure}

\begin{figure}[t]
\begin{center}
\includegraphics[width=0.45\textwidth]{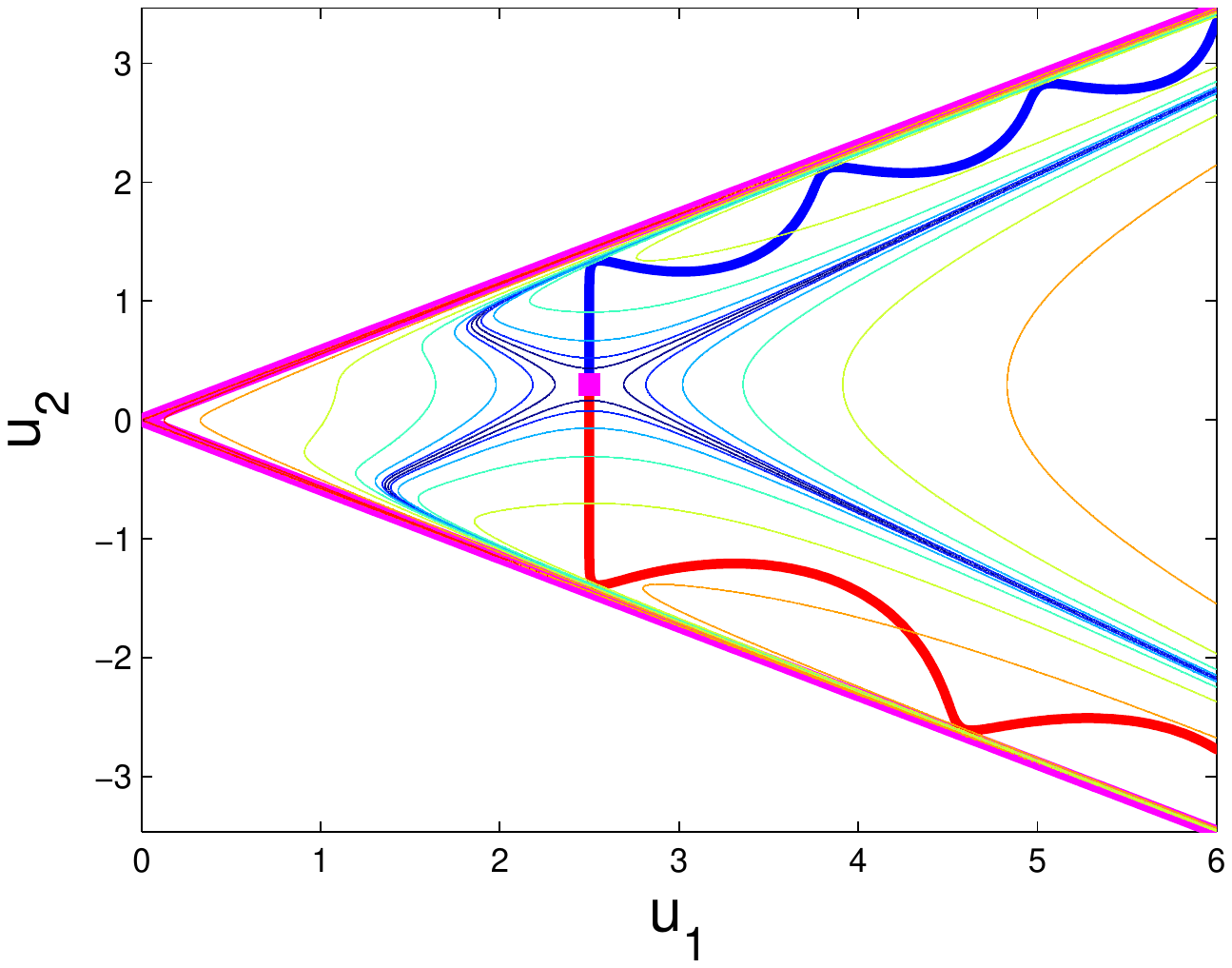}
\includegraphics[width=0.45\textwidth]{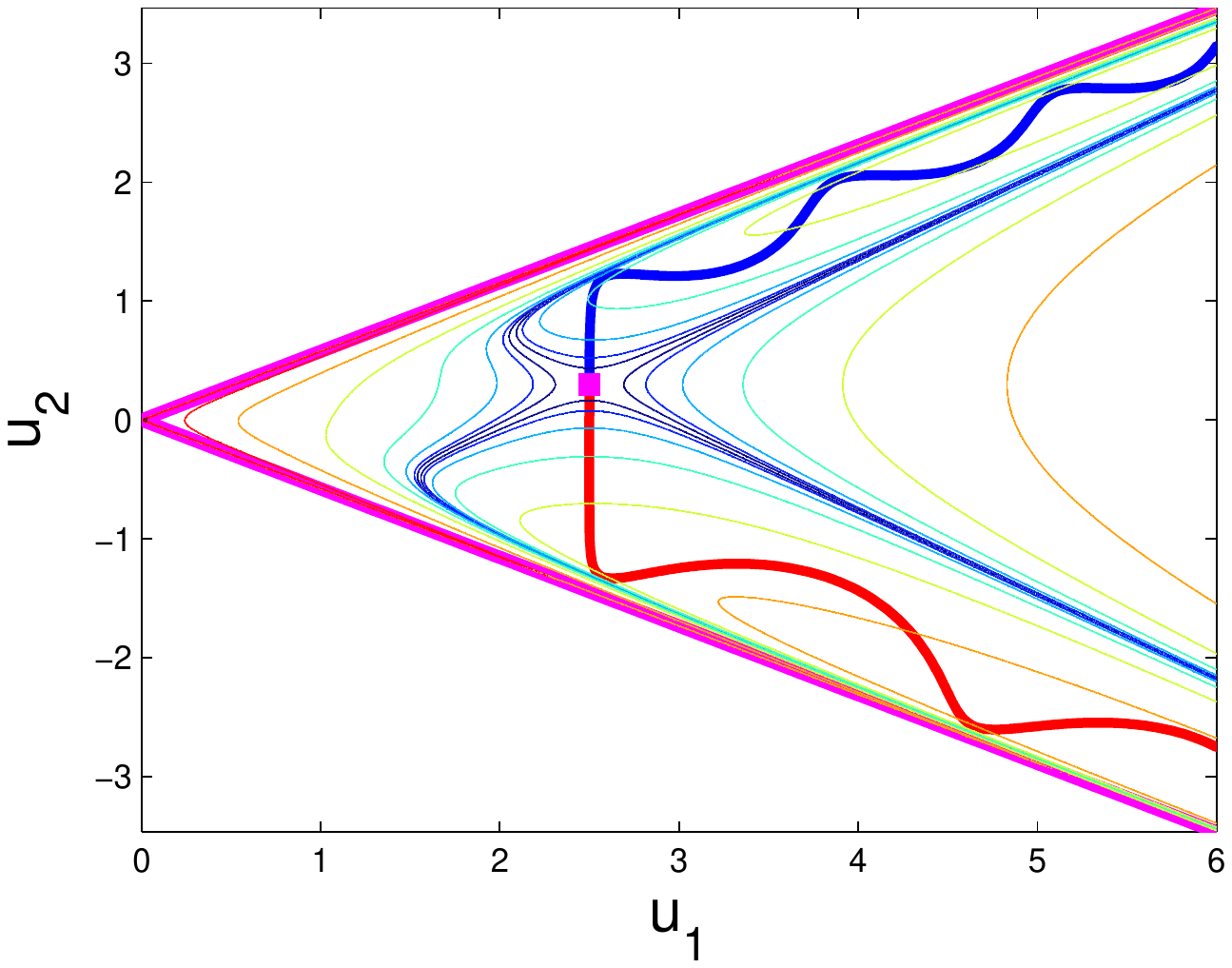}
\end{center}
\caption{ %\begin{footnotesize}
Simple dynamics for the smooth system: the effect of smoothing. The same parameters, initial conditions and projection of Fig. 6 top left panel are used with the smoothing parameter  increased to \(\varepsilon=0.05\)   (left) and  \(\varepsilon=0.1\)    (right). The smooth potential level curves still follow those of the linear potential quite closely in the region explored by  $W^{u}_\pm(P)$ .
%\end{footnotesize}
 }%
\label{fig:smeffect}%
\end{figure}

Finally, Figs. \ref{fig:bildynam},\ref{fig:smdynam},\ref{fig:smeffect} suggest that for   $\theta<0$     and  sufficiently large   \(\lambda \)     simple dynamics   are  realized:

\begin{conjecture} [Simple dynamics II] \label{thm:sd2} For any fixed geometrical parameter \( (u_s,\theta,\beta,L)\) satisfying \( -\beta<\theta<0\), for a fixed \(\omega \) and   sufficiently large \(\lambda \) (small   \(\delta= \frac{\omega}{\lambda} \)),  for small \(h\) and  sufficiently small \((\varepsilon,c)\), both the upper and lower  branches of  the stable and unstable manifolds of the Lyapunov orbit \(\gamma_h\) ($ W^{u,s}_{\pm}(\gamma_h)$) have simple behavior - these exit the corner region through the product (respectively reactant) channel without intersecting each other in the corner domain.
\end{conjecture}
 \textbf{Supporting evidence:}
 One may start by proving the claim  for \(h=\varepsilon=c=0 \), proving that trajectories belonging to $ W^{u,s}_+({P})$  reflect only a finite number of regular reflections before exiting the corner region through the right side of a box of size \(L\). Then, the claim follows by the smooth dependence of the manifolds on parameters (for  sufficiently small \(h,c\)) and by the closeness of the limit system to the smooth system at regular reflections (for sufficiently small \(\varepsilon\)).
Let \((u(t),v(t))\in W^{u}_+({P})\) hit the upper corner ray at times \(t_{i}, i\geq1\), so that \(t_1\) is its first hit.
Numerical simulations show that as   \(\lambda \) is increased the sequence of \(u_1(t_i)\) is indeed monotonically increasing. Moreover, these suggest that the gaps \(u_1(t_{i+1})-u_1(t_i)\)   approach a constant value:  in between hits the \(v_{1}\) velocity remains essentially constant since the time between reflections is roughly of order \(1/\lambda^2\) whereas the changes in \(v_1\) are on the much longer time scale \(1/\omega^2  \).
Proving the above statements  requires quite elaborate calculations of the asymptotic behavior at small \(\delta\), calculation that go beyond the scope of this manuscript.

Figures \ref{fig:bildynam} and \ref{fig:smdynam} demonstrate that the three different behaviors may be realized in the singular limit (\(\varepsilon=0\) in  Figure \ref{fig:bildynam})  and in a smooth case (\(\varepsilon=0.01\) in  Figures  \ref{fig:smdynam}) when we change \(\lambda\) and keep all other parameters fixed; The left columns show simple dynamics in both the upper and lower branches, the middle ones simple dynamics in the upper branch and possibly complicated dynamics in the lower branch and the right panels show possibly complicated dynamics in both branches. Fig \ref{fig:deltacrit} shows that these findings are consistent with the \(\delta_{c}^{\pm}\) bounds for \(\frac{\omega}{\lambda}\).
Figure \ref{fig:smeffect} shows the regularization effect that is achieved when \(\varepsilon\) is increased. Notice that even without introducing far-field potential terms, the geometrical potential level curves are reminiscent of Fig \ref{fig:potcon}b and are similar to other  PES appearing in  the Chemistry literature \cite{levinbook}-\cite{bugas94}.

\section{Discussion}

 The stable and unstable manifolds of unstable periodic orbits with energies that are slightly above the barriers' energy divide   the initial iso-energetic phase space region of incoming trajectories to reacting vs non-reacting regions   \cite{polak,davis,ujpyw02}. The structure of the manifolds is called simple if these manifolds do not intersect each other and simply extend to the reactant and product channels. Then, the phase-space transition state theory provides an accurate description of the transition rates from reactants to products. On the other hand, if the manifolds intersect each other or fold back into the reaction region, there is no cross-section which is crossed only once by all incoming trajectories. Then, the main assumption underlying the transition state theory fails   \cite{polak,davis,ujpyw02,WaWi10}.\\

 By introducing a geometrical model for the reaction dynamics  we find conditions under which the manifolds structure is simple and conditions under which it is complicated. Three qualitative observations emerge. First, we proved that a homoclinic bifurcation occurs when the manifolds are close to being perpendicular to one of the corner rays. Then, there are intervals of energies at which  stable periodic triatomic configurations emerge. Second, we found that when the projection of the unstable eigenspace  to the configuration space intersects the lower corner ray in an obtuse angle (so \(\theta<0\)), and, additionally, the saddle-center expansion rate is much larger than the oscillation frequency, the manifolds geometry is simple (Conjecture \ref{thm:sd2}).
Third, we established that provided the unstable eigenspace direction intersects the corner rays, the manifolds are trapped for sufficiently large oscillation frequency (Theorems \ref{thm:compupper},\ref{thm:complower}). \\ \\ We expect that similar qualitative statements may be formulated when nonlinear normal form and higher dimensional extensions are included. Namely, that conditions under which phase-space transition state theory   is adequate for describing the reaction for energies that are close to the saddle energy may be found by a similar methodology.
 More generally, while the effects of non-linear terms in the reaction region have been suppressed here for simplicity of presentation (by taking only quadratic terms of the integrable normal form and by considering small \(c\)), we {trust} that the main principles that were discovered hold for the nonlinear case as well; Indeed, once the fixed point \(P\) has the saddle-center structure, its stable and unstable manifolds may be computed. With a general non-linear smooth potential
their projection to the configuration space  will appear as curved lines that may or may not intersect the corner region boundary. The analysis presented here applies to the case where, in the limit system, the manifolds do hit the boundary and reflect back. Then, we expect to find similar behavior in terms of the Hill region dependence on the  ratio between the oscillatory and the hyperbolic eigenvalues.

 Finding the implications of the above observations for specific chemical reactions is an interesting and challenging endeavor - we believe the tools developed here may shed some light regarding the governing parameters.

Finally, we note that the analysis presented here applies to the traditional energy regime in which the behavior near the barrier is examined. The behavior for larger energies, or equivalently, for reactions that do not have a barrier, is expected to be quite different and will be described elsewhere.
Indeed, we hold that there are two distinct mechanisms that give rise to the observed sensitive dependence of the reaction rates on the energy and the initial conditions: those associated with the complicated structure of the manifolds as discussed here and those associated with the corner geometry of the nearly-billiard Hamiltonian.

\newpage

\section{Acknowledgement}

The authors thank M. Kloc and D. Tannor for their important comments and suggestions. The authors acknowledge a support from RFBR (Russia) and MSTI (Israel) under the grant 06-01-72023. L.M.L. also thanks RFBR for a partial support under the grant 10-01-00429a, RFBR and the administration of the Nizhny Novgorod region under the grant 11-01-97017 (regional-Povolzhye), the Ministry of Education and Science of the Russian Federation (the contract NK-13P-13, No.P945) and the Russian Federation Government grant, contract No.11.G34.31.0039.
V.R-K acknowledges the support of the Israel Science Foundation (Grant
273/07) and the Minerva foundation.

\section*{Appendix: The gluing map is symplectic.}
Let us check, for the reader convenience, that the map $S_h: N^u_h \to
N^s_h$ defined by the reflection law is a symplectic map w.r.t.
restrictions of the main 2-form on $N^u_h$ and $N^s_h,$
respectively. We shall verify it for the lower wall supposing
$\mu=\tan\theta$ is finite. We work in small neighborhoods of the
points $m_u$ and $m_s$ being the intersection points for the lower
branches of unstable and stable manifolds of the equilibrium $P$.
These cross-sections belong to the 3-plane given by the relation
$u_2 = \mu u_1$, the restrictions of 2-form $dv_1\wedge du_1 +
dv_2\wedge du_2$ to $N^u_h$, $N^s_h$ are the following:
$$
\begin{array}{l}
\hat \omega = d\hat v_1\wedge d\hat u_1 + \tan\theta d\hat v_2\wedge
d\hat u_1 = (1 - \tan\theta\frac{\hat v_1}{\hat
v_2})d\hat v_1\wedge d\hat u_1,\\
\omega = dv_1\wedge du_1 + \tan\theta dv_2\wedge du_1 = (1 -
\tan\theta\frac{v_1}{v_2})dv_1\wedge du_1.
\end{array}
$$
The symplecticity condition for $S_h$ means, as is known
$S^*_h\omega = \hat \omega$ \cite{Arn97}. Using the relations $u_1 =
\hat u_1,$ $v_1 = \hat v_1 \cos 2\theta + \hat v_2 \sin 2\theta$ and
expression for $\hat v_2$ on $N^u_h$ one gets:
$$
\begin{array}{l}
S^*_h\omega = \displaystyle{\left(1- \tan\theta\frac{\hat
v_1\cos2\theta + \hat v_2 \sin2\theta}{\hat v_1 \sin 2\theta -
\hat v_2 \cos 2\theta}\right) \left[d\hat v_1 \cos 2\theta + d\hat
v_2 \sin 2\theta\right] \wedge d\hat u_1} =
\\\displaystyle{\left(1- \tan\theta\frac{\hat v_1\cos2\theta +
\hat v_2 \sin2\theta}{\hat v_1 \sin 2\theta - \hat v_2 \cos
2\theta}\right)\left[\cos 2\theta - \sin 2\theta\frac{\hat
v_1}{\hat v_2}\right]d\hat v_1 \wedge d\hat u_1} =\\
\displaystyle{\left(1 - \tan\theta \frac{\hat v_1}{\hat
v_2}\right)d\hat v_1 \wedge d\hat u_1 = \hat \omega}.
\end{array}
$$
Thus we get that the Poincar\'e map $T_h\circ S_h$ is also symplectic.

\def\cprime{$'$} \def\cprime{$'$} \def\cprime{$'$} \def\cprime{$'$}

\end{document}